\documentclass[11pt,letterpaper]{article}
\usepackage{times}

\usepackage{graphicx}
\usepackage{subfigure}
\usepackage[justification=justified]{caption}
\usepackage{algorithm}
\usepackage{algorithmic}
\usepackage{amsmath}

\newcommand{\BigO}[1]{\ensuremath{\operatorname{O}\bigl(#1\bigr)}}
\newtheorem{example}{Example}
\newtheorem{definition}{Definition}

\begin{document}

\title{\Large Real-time Top-K Predictive Query Processing over Event Streams}
\author{\small Saurav Acharya$^+$, Byung Suk Lee$^+$, Paul Hines$^\pm$\\
      \small $^+$Department of Computer Science, University of Vermont, Burlington, VT, U.S.A.\\
      \small $^\pm$School of Engineering, University of Vermont, Burlington, VT, U.S.A.\\
      \small Emails: \{sacharya, bslee, phines\}@uvm.edu}
\date{}
\maketitle

\begin{abstract}
This paper addresses the problem of predicting the $k$ events that are most likely to occur next, over historical real-time event streams. Existing approaches to causal prediction queries have a
number of limitations. First, they exhaustively search over an \emph{acyclic} causal network to find the most likely $k$ effect events; however, data from real event streams frequently reflect \emph{cyclic} causality. Second,
they contain conservative assumptions intended to exclude all possible non-causal links in the causal network; it leads to the omission of many less-frequent but important causal links. We overcome these limitations by proposing a novel \textit{event precedence model} and a \textit{run-time causal inference} mechanism. The event precedence model constructs a first order absorbing Markov chain incrementally over event streams, where an edge between two events signifies a temporal precedence relationship between them, which is a necessary condition for causality. Then, the run-time causal inference mechanism learns causal relationships dynamically during query processing. This is done by removing some of the temporal precedence relationships that do not exhibit causality in the presence of other events in the event precedence model. This paper presents two query processing algorithms -- one performs \textit{exhaustive search} on the model and the other performs a more efficient \textit{reduced search with early termination}.
Experiments using two real datasets (cascading blackouts in power systems and web page views) verify the effectiveness of the probabilistic top-k prediction queries and the efficiency of the algorithms. Specifically, the reduced search algorithm reduced runtime, relative to exhaustive search, by $25-80\%$ (depending on the application) with only a small reduction in accuracy.
\end{abstract}



\section{Introduction} \label{sec:introduction}
Causal prediction (e.g.~\cite{Clark_TCS2003,GuyonACEP_SS08,SpiritesGS_book2000}) is emerging as an essential field for real-time monitoring, planning and decision support in diverse applications such as stock market analysis, electric power grid monitoring, sensor network monitoring, network intrusion detection, and web click-stream analysis. There is a need for active systems to continuously monitor the event streams from these applications to allow for the prediction of future effect events in real time. Specifically, given a sequence of potentially causal events, many applications would benefit from good algorithms to predict the next most likely (namely, top $k$) effect events. The potentially huge answer space, however, and the unknown dynamics as well as the streaming nature of data make such top-k prediction a challenging task.

Consider the following two scenarios as motivating examples.
\begin{example}{\textit{Web page click stream}: } \label{ex:click_stream}
\normalfont
Consider web-based online systems. A majority of them display the same content for everyone. However, the user experience can be more productive with a dynamic system where content is displayed based on real-time prediction of users' most likely activities, given historical data. One can use the results (i.e., the web pages/links most likely to be visited next) to display the most relevant links, content, and advertisements at each step of the user activity. Such an arrangement may help to retain the user longer by displaying the most relevant information, thereby increasing the content consumption (e.g., sales, page visits, ad clicks).
\hfill\fbox{}
\end{example}

\begin{example}{\textit{Electric power grid}: } \label{ex:power_grid}
\normalfont
Consider an electric power grid. When components of a power grid fail, as a result of a storm, malfunction or cyber-attack, a cascading sequence of subsequent component failures may result, which may lead to a very large blackouts (e.g.,~\cite{VaimanBCCDHPMZ_PS2012}). Thus, a timely prediction of the components that are most likely to fail next, given a list of a few components that have failed, may enable operators to take mitigating actions (like shutting down sections of the power grid) before a large-scale blackout occurs. Cascading blackouts typically progress slowly (minutes to tens of minutes) in the initial stages; a few seconds delay to compute and implement emergency controls is generally sufficient.
\hfill\fbox{}
\end{example}

In this paper, we address the challenge of continuous prediction of the \textit{top-k} most probable next effects in real time streams.
To the best of our knowledge, there are no existing top-k query processing mechanisms that are sufficiently efficient to support time-critical applications, such as Examples~\ref{ex:click_stream} and~\ref{ex:power_grid}. Moreover, most previous work on the prediction of effect events given one or more cause events is based on inefficient exhaustive search over a large search space of causal network (e.g.,~\cite{Akdere_DCP2010,Zhang_ECI1996}). A causal network represents the cause and effect relationships, called causality, in a directed acyclic graph. This \textit{traditional} causal network model (e.g.,~\cite{EllisW_JASA2008,Heckerman_UAI1995,LiL_PAKDD2009,MeganckLM_MDAI2006,Pearl_book2009,SpiritesGS_book2000,SpirtesGS_PACSS90}) has two major limitations to be used for causal prediction. First, since it is acyclic, it cannot have loops, such as $A \rightarrow B \rightarrow C \rightarrow A$ or bidirectional relationships such as $A \leftrightarrow B$, and consequently, does not support cyclic causality (e.g.,~\cite{FriedmanLNP_UBN2000,MargolinH_2013Reasoning}). The event streams from many applications, however, do show cyclic relationships. For example, a visitor to a news web site may visit the home page, proceed to read an article, and then return to the home page, creating a cyclic relationship between these vertices in the graph. Second, the causal Markov condition, often considered an essential property of traditional causal networks, is conservative in the causal inference, and as a result removes many infrequent but important causal relationships from the causal network~\cite{Pearl_B1995,Pearl_S1998,MargolinH_2013Reasoning,SpirtesGS_PACSS90}. That is, the causal Markov condition calls for the removal of those relationships which could potentially be independent in the presence of one or more events. The rationale for this is to avoid any suspicious and weak relationships.
However, this approach often backfires by removing rare but important causal relationships~\cite{MargolinH_2013Reasoning}. We call this limitation the \textit{causal information loss}.

Based on these facts, we identify three central research problems -- (1) how to model causal relationships among events in a stream to prevent causal information loss; (2) how to address cyclic causality in the causal model; (3) how to efficiently run a causal inference query on this causal model to continuously predict the top-k probable effects.

To address these problems, first, we propose an \textit{event precedence model} that captures temporal precedence relationship between every two 
event types into a first order absorbing Markov chain. We refer to the resulting model as an event precedence network (EPN), in which an edge signifies the temporal relationship between two events. This inclusion of \emph{all} temporal precedence -- hence likely causal -- relationships helps to avoid causal information loss. Note that EPN is a generative model of the observed event stream, which is built over a set of predefined event types instead of event instances. Second, we propose a \textit{run-time causal inference} method. Due to cyclic causality, causal inference cannot be performed until the cause event whose effects are being predicted is known, but a cause event is only observed at run time, hence run-time causal inference. 
EPN encodes all cyclic as well as non-cyclic precedence relationships from event streams on its edges, and therefore these edges are examined by run-time causal test (i.e., conditional independence test) to determine causality. Note that this run-time causal inference overcomes the two limitations of traditional causal model discussed earlier (i.e., lack of support for cyclic causality and loss of many important causal relationships). Third, we present two query processing algorithms -- the Exhaustive Search (ES) algorithm and the Reduced Search Early Termination (RSET) algorithm -- to continuously predict top-k event types with the highest scores based on the inferred causal relationships.
The ES algorithm formalizes the exhaustive search approach. The RSET algorithm is built upon the ES algorithm, and reduces the search space with possible early termination. As a result, it reduces the runtime with only marginal reduction in prediction accuracy.

We conduct experiments to evaluate the performance of the proposed \textit{ES} and \textit{RSET} algorithms using two real datasets. For each dataset we perform two sets of experiments to evaluate their accuracies and runtimes, respectively.  In each evaluation, there are two objectives. The first objective is to compare the run-time causal inference mechanism of the proposed algorithms (i.e., ES, RSET) against the state-of-the-art traditional causal inference mechanism called the Fast Causal Network Inference (FCNI) algorithm~\cite{AcharyaL_DaWaK2013}. The FCNI algorithm is essentially inapplicable to our problem due to its lack of ability to handle cyclic causality and run-time causal inference, but is the best available in the state of the art.  The second objective is to compare the query processing mechanisms between the ES algorithm and the RSET algorithm.

The contributions of this paper are summarized as follows.
\begin{enumerate}
    \item It presents an event precedence model to represent the temporal precedence relationships between event types and proposes an algorithm to construct an event precedence network incrementally over event streams.
    \item It introduces a run-time causal inference mechanism to infer the causal relationships in real time, and proposes two query processing algorithms: Exhaustive Search and Reduced Search Early Termination, to continuously predict the top-k next effects over event streams.
    \item It empirically demonstrates the advantages of the proposed run-time causal inference mechanism and the query processing algorithms in terms of the prediction accuracy and the runtime.
\end{enumerate}

The remainder of the paper is organized as follows. Section~\ref{sec:related_work} discusses the related work, and Section~\ref{sec:prelimiaries} presents some preliminary concepts. Sections~\ref{sec:event_precedence_model} and~\ref{sec:query_processing} describe the event precedence model and the query processing model, respectively. Section~\ref{sec:experiments} evaluates the proposed query processing algorithms. Section~\ref{sec:conclusions} concludes the paper and suggests future work.

\section{Related Work} \label{sec:related_work}
This section first discusses conventional causal inference techniques and then describes how this paper makes unique contributions relative to other work related to causal prediction.

There are two approaches for constructing a traditional causal network. The first approach, search and score based (e.g.,~\cite{EllisW_JASA2008,Heckerman_UAI1995,LiL_PAKDD2009,MeganckLM_MDAI2006}), performs greedy search (usually hill climbing) over all possible causal networks of the data to select the network with the highest score. This approach, however, has two limitations. First, the computational complexity increases exponentially as the number of variables in the causal network increases. Second, the problem of equivalence classes~\cite{Chickering_JMLR2002}, where two or more network structures represent the same probability distribution, makes the causal direction between nodes quite random and therefore unreliable. The second approach, constraint-based (e.g.,~\cite{ChengGKD_02,Pearl_book2009,SpiritesGS_book2000,SpirtesGS_PACSS90}), which performs a large number of conditional independence tests between variables to construct a causal network, does not have the problem of equivalence classes. The state-of-the-art Fast Causal Network Inference (FCNI) algorithm~\cite{AcharyaL_DaWaK2013} presents a traditional constraint-based causal network inference mechanism over event streams. (Thus, we consider the FCNI algorithm as the representative of the traditional causal network approach in this paper.) The FCNI algorithm learns temporal precedence relationships from the event stream and performs causal inference between only those event types which exhibit temporal precedence relationship. Such an approach helps to reduce the number of conditional independence tests required for causal network inference. However, this algorithm assumes acyclic causality in the data. (The idea of temporal precedence-based conditional independence test has been incorporated in the work presented in this paper.)

In addition, there has been some work (e.g.,~\cite{ElloumiZ_biological2013,KlopotekM_Studia2006,TulupyevN_MICAI2005}) to support \emph{cyclic} Bayesian network which aims to handle the cyclic causality in Bayesian networks. This work, however, still carries the drawbacks inherent in the Bayesian network approach -- that is, the ambiguity of equivalence classes and the inability to meet the requirement of a causal network that the parent node in the network should always represent the direct cause -- and hence is not useful in our work.

The existing body of work on prediction only addresses inference of the likelihood of occurrence of an effect variable given a cause variable (e.g., ~\cite{ChengM_JAIR00,ChengD_UAI01,MacKay_LIGM99,MoralDJ_SAC12,Shachter_UAI90,ZhangP_94}), while the prediction of top-k effects requires finding the most likely \textit{k} effects among all possible effect variables. Therefore, the only way to find the top-k next effects is to construct a traditional causal network, which ignores cyclic causality and suffers from causal information loss, over event streams and then infer the top-k effects of the cause exhaustively(e.g., ~\cite{Akdere_DCP2010,Henrion_UAI91,Zhang_ECI1996}). To the best of our knowledge, there is no solution to address cyclic causality, mitigate the causal information loss, and perform only necessary partial search to find the top-k effects of the given causes over event streams.

The well-established association rule mining algorithms (e.g.,~\cite{Jiang_DKE2005,RudinLSKM_COLT2011,VelosoAGMM_LRQ2008}) are extensively used for prediction and recommendation. However, association does not necessarily imply causation (e.g.,~\cite{BowesNGC_AAI2000,TsauYAC_Springer2008,Mazlack_ACI2004,MohammadN_SCI2010,SilversteinBMU_DMKD2000,YoungSK_SIGN2011}). Therefore, they are not useful to our problem due to the exclusion of the fundamental concept of causality. That is, two variables that are associated require stronger conditions, such as temporality and strength, to be considered causally related. {A few works on top-k query processing in the Internet domain, such as over social-tagging networks~\cite{Haghani_MA2010} and over web 2.0 stream~\cite{SchenkelCKMNP_2008}, have been published. Unlike our work, however, these works do not address causal prediction in an event-based environment at all.}

\section{Preliminaries} \label{sec:prelimiaries}
In this section, we introduce the concepts that are central to the techniques explained in the paper.

\subsection{Event Streams} \label{sec:event_streams}
An event stream is a discrete, indefinitely long sequence of event instances. An \textit{event instance} (or event) refers to a timestamped action which may have an effect. A prototype for creating events is called an \textit{event type}. Each event instance is created by one event owner. An event type can have many instances, and an event owner can create many instances of any type. Two events are related to each other if they share common attributes such as event owner, location, and time. These attributes are called \textit{common relational attributes} (CRAs). In Example~\ref{ex:click_stream} and Example~\ref{ex:power_grid}, the CRAs are the session id and the blackout id, respectively .

An event has the following schema: $\langle$\textit{timestamp}, \textit{type}, \textit{CRA}, \textit{attribute-set}$\rangle$. That is, an event has the timestamp at which it was created, the event type it belongs to, the CRA value, and a set of additional attributes called the attribute-set. An event is denoted as $e_{ij}$ where $i$ is the value of the CRA and $j$ (=$1,2,3,...$) is its event type id ($E_j$).

\begin{example}
\normalfont
Figure~\ref{fig:event-stream-sample} shows an illustrative example of events in a user click event stream of Example~\ref{ex:click_stream}. The first field in each line (e.g., $e_{21}$) denotes the actual event instance shown in the remainder of the line (e.g., $\langle$ 05/05/11 1:12 pm, 1, 2, [200s, ...]$\rangle$). The session id serves as the CRA and the webpage categories (e.g., frontpage, news, weather, sports, entertainment, tech, local, etc) are the event types. For instance, in the event instance $e_{32}$, 2 is the event type and 3 is the CRA. Note that the event type is represented by a numerical equivalent of the original event type {(e.g., frontpage = $E_1$, news = $E_2$, weather = $E_3$, sports = $E_4$, entertainment = $E_5$, tech = $E_6$, local = $E_7$).}
\begin{figure*}
\centering
\includegraphics[scale=0.32]{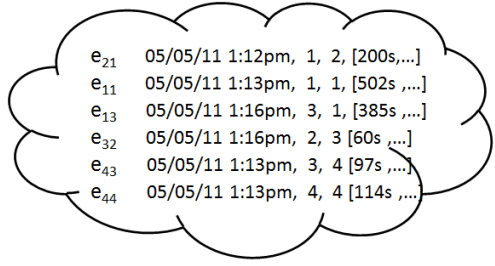}
\caption{Sample of event instances in a stream from Example~\ref{ex:click_stream}.}
\label{fig:event-stream-sample}
\end{figure*}
\hfill\fbox{}
\end{example}

\begin{figure*}
\centering
\subfigure[Raw Event Stream.]{\includegraphics[scale=0.27]{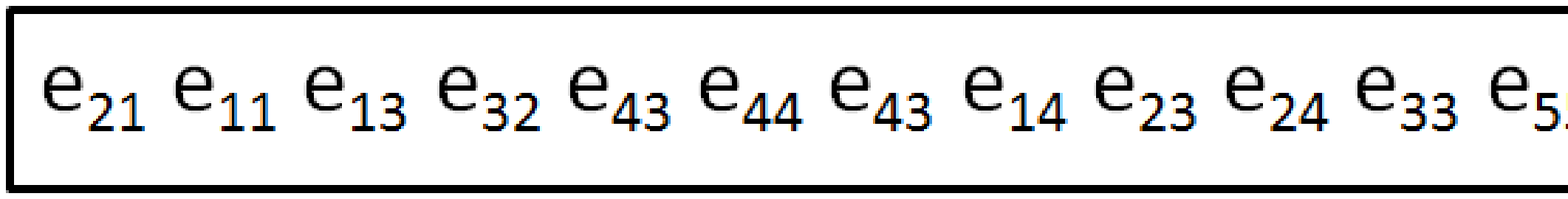}}
\quad
\subfigure[Partitioned Window.]{\includegraphics[scale=0.24]{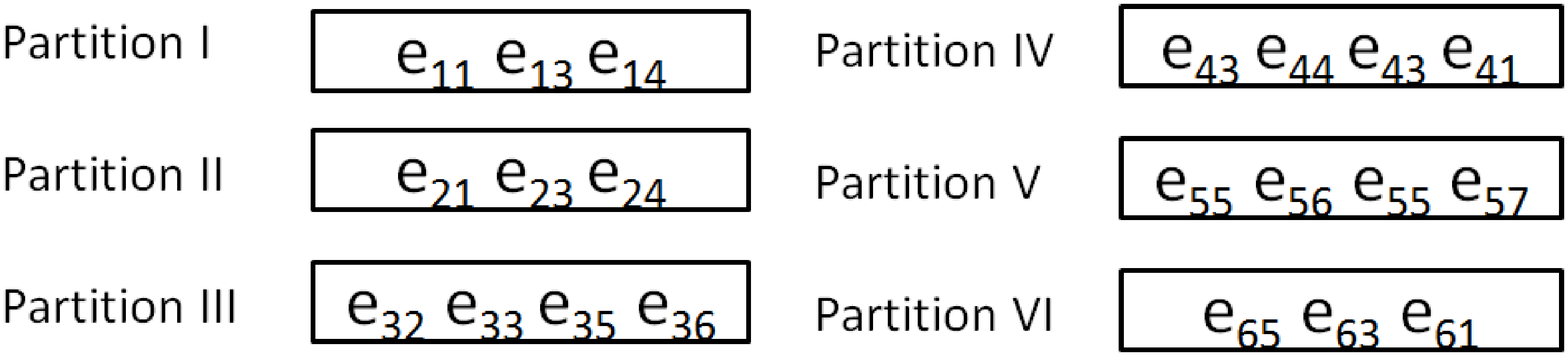}}\\
(\small{$e_{ij}$s are abbreviations of actual event instances such as shown in Figure~\ref{fig:event-stream-sample}.})
\caption{Event stream.}
\label{fig:event-stream}
\end{figure*}

We use a window, specifically called \textit{partitioned window}~\cite{AcharyaL_DaWaK2013}, to accumulate the events from the stream for a user-specified observation period \textit{T}. As a preprocessing step to group related events in the window, these events are partitioned by the CRA and then arranged in temporal order in each partition. Figure~\ref{fig:event-stream} shows what a partitioned window looks like for the event stream shown in Figure~\ref{fig:event-stream-sample}. Once the observation period expires, the window shifts to the next batch of events. The last event of one window overlaps the first event of the next window in order to ensure consistency in event precedence modeling across two consecutive windows.

\begin{definition}[Partition]
{\rm 
A partition $W_i$ in a partitioned window is defined as a set of observed events sharing the same CRA value $i$ and arranged in the temporal order over a time period $T$, that is
\begin{displaymath}
W_i = \{ e_{ij}(t) | t \le T, i \in \mathbf{A}, j \in [1,N] \}
\end{displaymath}
where $t$ is the timestamp, $\mathbf{A}$  is the set of all possible CRA values, $j$ is the event type id, and $N$ is the number of event types.
}
\hfill \fbox{}
\end{definition}

The events which are being predicted are \textit{effect events} while the events which are used for prediction are \textit{cause events}. We denote the cause event type and the effect event type as $C_i \in \mathbf{E}$ and $T_j \in \mathbf{E}$, respectively, where $i$ and $j$ are the positions of the events in each sequence. Note that $C_i$ and $E_i$ are not necessarily the same, and nor are $T_j$ and $E_j$. Table~\ref{tbl:notations} summarizes the key notations used in this paper.
\begin{table*}
  \centering
  \begin{tabular}{ l l }
    \hline
    Symbols & Definitions \\
    \hline
    $N_p$ & Number of partitions \\
    $N_{ei}$ & Number of event instances in the $i$-th partition \\
    $N_e$ & Total number of event instances in all partitions \\
    $E_i$ & Event type with id $i$ \\
    $e_{ij}$ & Event instance of type $j$ and CRA $i$\\
    $N$ & Number of event types\\
    $C_i$ & Cause event type at position $i$ \\
    $T_i$ & Effect event type at position $i$ \\
    $O$ & Causal search order \\
    $S_i$ & Event type at the $i$-th position in $O$ \\
	$C_{\delta}$ & The most recent cause event type\\
    $\mathbf{E}$ & Set of N event types in the data $\{E_1, E_2, ..., E_N\}$\\
    $R_k$ & Ranked list of the \textit{top-k} event types \\
    $N_{instances}$ & Number of event instances \\
    $N_{CRA}$ & Number of common relational attributes\\
    $f(E_i, E_j)$ & Number of observations of the instances of type $E_i$ followed \\
                  & by the instances of type $E_j$\\ \hline
  \end{tabular}
  \caption{Definitions of main symbols.}
  \label{tbl:notations}
\end{table*}

\subsection{Causal Networks} \label{sec:causal_networks}
A causal network (or causal \emph{Bayesian} network) is a directed acyclic graph \textit{G} = (\textit{V, $\Xi$}) to encode causality, where \textit{V} is the set of nodes (representing event types) and $\Xi$ is the set of edges between nodes. For each directed edge, the parent node denotes the cause, and the child node denotes the effect.

The joint probability distribution of a set of $N$ event types $\mathbf{E} \equiv {\{E_{1},...,E_{N}\}}$ in a causal network is specified as
\begin{displaymath}
    P(\mathbf{E}) = \prod_{i=1}^N P(E_{i}|\mathbf{Pa_{i}})
\end{displaymath}
where $\mathbf{Pa_{i}}$ is the set of the parent nodes of event type $E_{i}$.

\begin{figure*}[t!]
\centering
\subfigure[{Causal network with event type names.}]{\includegraphics[scale=0.32]{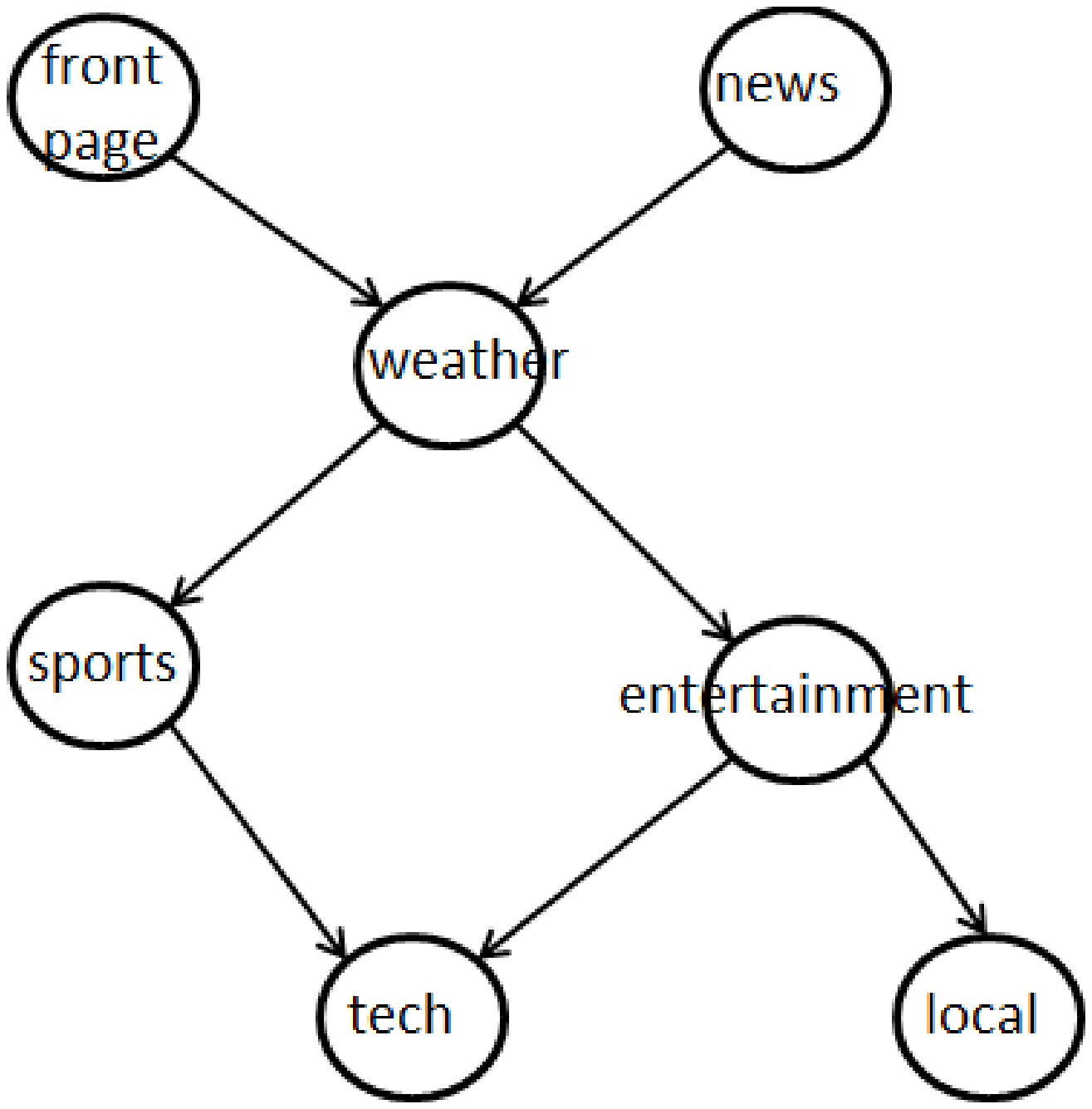}}
\quad
\subfigure[Causal network with event type notations.]{\includegraphics[scale=0.32]{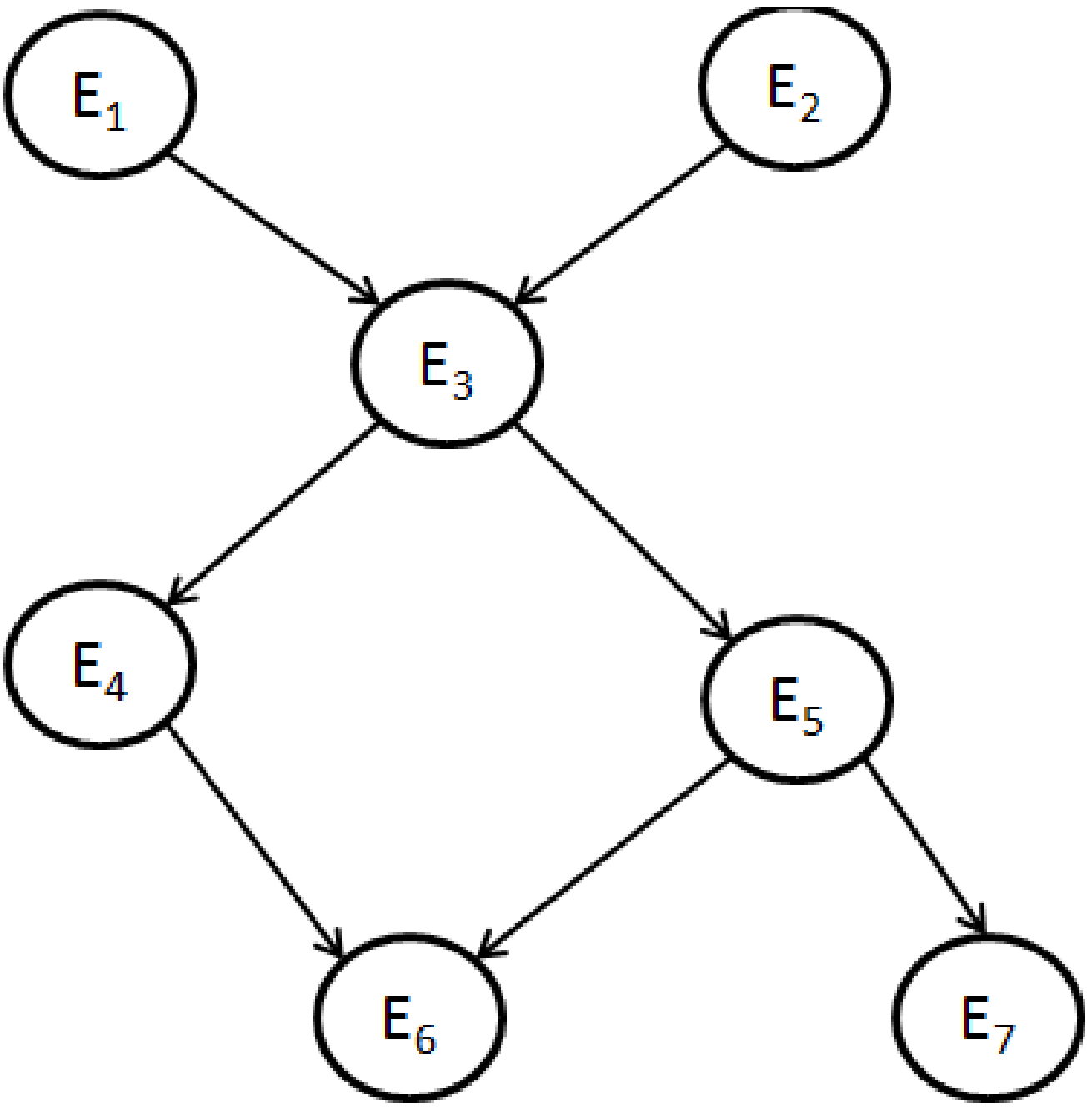}}\vspace*{1em}\\
\begin{minipage}{0.75\linewidth}
{\small (The event type names in the subfigure a are from the web page click stream in Figure~\ref{fig:event-stream}, and the event type notations in the subfigure b are symbols used to represent the real event type names.)}
\end{minipage}
\caption{Causal network.}
\label{fig:causal-network}
\end{figure*}

Consider the event stream of Figure~\ref{fig:event-stream}. The causal relationships among the event types in the stream may be modeled as a causal network like the one shown in Figure~\ref{fig:causal-network}.

\subsection{Conditional Independence Tests} \label{sec:conditional_independence_tests}
A popular approach for testing the conditional independence (CI) between two random variables X and Y in a set of random variables, $\mathbf{C}$, is \emph{conditional mutual information} (CMI) (e.g.,~\cite{ChengGKD_02,deCampos_JMLR2006}).
\begin{displaymath}
    \mbox{CMI}(X,Y|\mathbf{C}) = \sum_{x\in X}\sum_{y\in Y}\sum_{c\in \mathbf{C}} P(x,y,c) log_{2}\frac{P(x,y|c)} {P(x|c) P(y|c)}
\end{displaymath}
where $P$ is the probability mass function calculated from the frequencies of variables. CMI gives the strength of dependency between variables in a measurable quantity, which helps to identify the weak (or spurious) causal relationships.

In the traditional CMI, two variables $X$ and $Y$ are said to be independent if $\mbox{CMI}(X,Y | \mathbf{C})$ = $0$, and dependent otherwise. This criterion itself offers no distinction between weak and strong dependencies. With a higher value of $\mbox{CMI}(X,Y | \mathbf{C})$, the dependency between X and Y should be considered stronger. Thus, to prune out weak dependencies, we need a threshold CMI value, below which we consider the evidence ``too weak''. To do so, we relate CMI with the $G^2$ test statistics~\cite{BishopFH_DMA75,SpiritesGS_book2000} as below.
\begin{displaymath}
   G^2(X,Y|\mathbf{C}) = 2 \cdot N_s \cdot log_e2 \cdot \mbox{CMI}(X,Y|\mathbf{C})
\end{displaymath}
where $N_s$ is the number of samples (i.e., event instances).

Under the independence assumption, $G^2$ follows the $\chi^2$ distribution~\cite{Kullback_book1968} with the degree of freedom \emph{df} equal to $(n_x - 1) (n_y - 1)\prod_{s\in S}n_s$, where $n_x$, $n_y$, and $n_s$ are the number of possible distinct values of X, Y, and S, respectively. So, we perform the test of independence between X and Y given \textbf{C} by using the calculated $G^2$ test statistics as the $\chi^2$ test statistics in a $\chi^2$ test, which provides the threshold based on \emph{df} and significance level ${\alpha}$, to validate the result. We set $\alpha$ as the generally accepted value of $95\%$.

We define a Boolean function \textit{IsIndependent}$(X,Y,\mathbf{C})$ to test the conditional independence between two variables \textit{X} and \textit{Y} given a condition set of variables $\mathbf{C}$ using the $G^2$ test statistics. It returns true if these two variables are conditionally independent; otherwise, it returns false.

The unbounded and continuous nature of event streams of interest makes it infeasible to store all of the historical data. Therefore, we use an incremental approach such that when a new batch of events is processed, we only update the record of the frequency of observations without storing the old events.

\section{Event Precedence Model} \label{sec:event_precedence_model}
In this section, we introduce the proposed incremental mechanism to model the precedence relationships between events in a network structure.

\subsection{Model} \label{sec:model}
 To overcome the limitations of the existing causal models described in Section~\ref{sec:introduction}, we propose the \textit{event precedence model} (EPM). It represents the temporal precedence relationships in a first order absorbing Markov chain, called \textit{event precedence network (EPN)}, over which further analysis is done to predict the probable effect events based on the observed cause events. Note that the temporal precedence is a required criterion of a causal relationship. To avoid any information loss, evidence of the precedence between every two events in the stream is preserved. EPM takes the partitioned window (collected from the event stream) as an input and incrementally builds a model to reflect all precedence relationships in the input data. The actual data is discarded once a new batch of events arrives. Such an adaptive approach is essential for a streaming environment with continuous and unbounded data.

We make the following assumptions in the event precedence model.
\begin{itemize}
\item Given an ordered sequence of cause event types \{$C_{0}, C_{1}, ...., C_{\delta} $\}, an instance of the effect event type $T_{0}$ cannot occur without the occurrence of an instance of the most recent cause event type $C_{\delta}$. Moreover, while all past events influence the future events, the strongest influence is exerted by the immediately preceding event of the effect event. With this in mind, the precedence relationships only between every two consecutive events are considered.
\item There cannot be a causal relationship between events of the same type and, therefore, such precedence relationships are ignored.
\item The cause and effect events should share the same CRA value. As described in Section~\ref{sec:event_streams}, the events are grouped into partitions based on their \textit{CRA} values (e.g., session id and blackout id in Examples~\ref{ex:click_stream} and~\ref{ex:power_grid}, respectively). In other words, two events are not related to each other if they have different CRA values.
\end{itemize}

The proposed EPM is a \textit{first order absorbing Markov chain}~\cite{KemenyS_FMC1969} where an observation is independent of all previous observations except the most recent one and every state can reach an absorbing (a.k.a., terminating) state. Thus, the probability of occurrence of an effect event given past cause events is given as follows.\\
\[
    P(T_0 | C_{0}, C_{1}, ... , C_{\delta}) = P (T_0 | C_{\delta}).
\]
$P (T_0 | C_{\delta})$ can be rewritten as below.
\[
   P (T_0 | C_{\delta})  =  \frac{P(T_0, C_{\delta})}{P(C_{\delta})}
\]
\noindent which, then, can be estimated as
\begin{equation} \label{eq:cond_prob}
P (T_0 | C_{\delta}) = \frac{f(T_0, C_{\delta})}{\sum_{E_{j} \in children(C_{\delta})} f( E_{j}, C_{\delta})  }
\end{equation}
where $f(E_i, E_j)$ denotes the number of observations in which instances of the type $E_i$ precedes instances of the type $E_j$.

\begin{figure}[t!]
\centering
\includegraphics[scale=0.25]{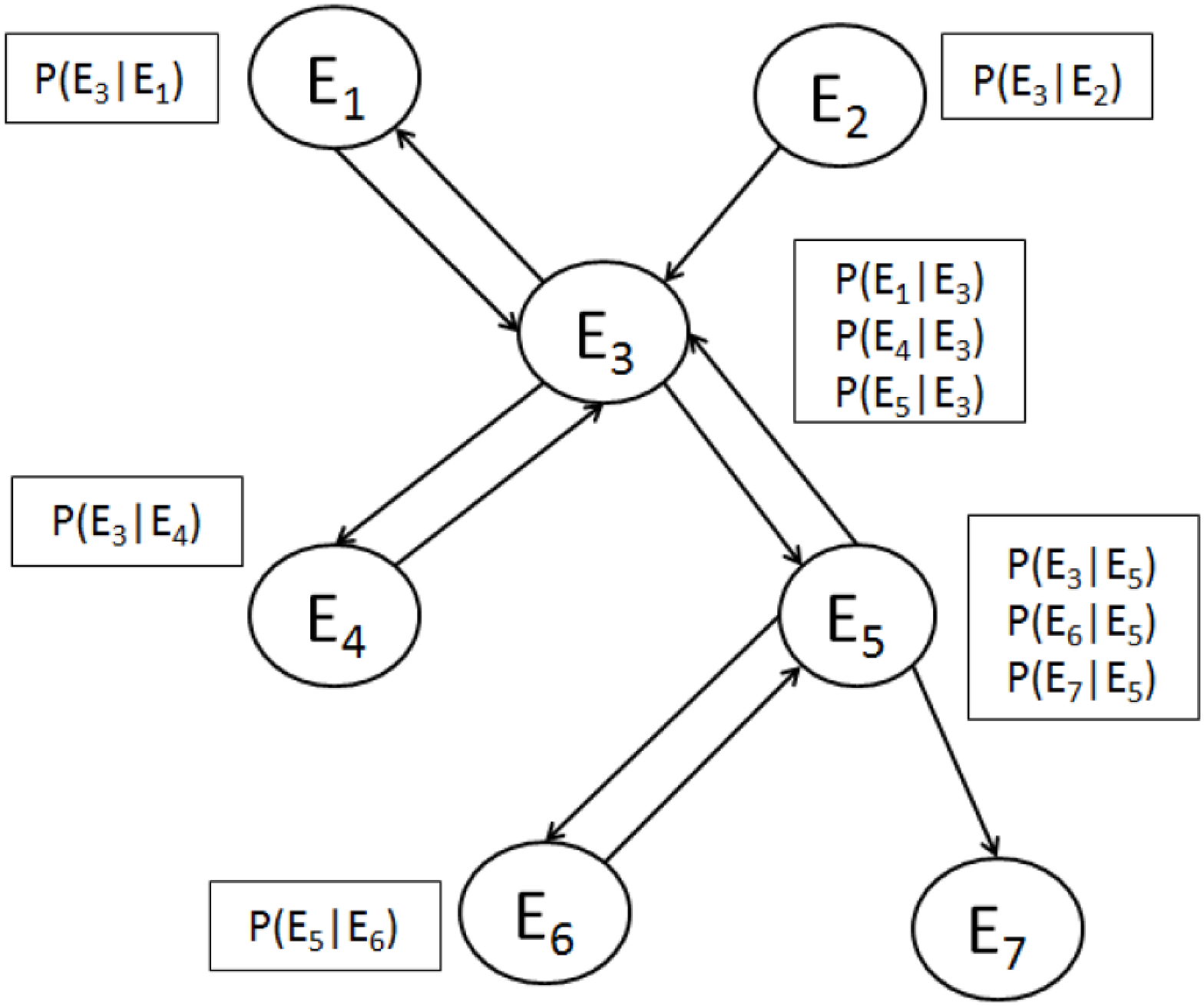}
\caption{Illustration of event precedence network construction from the event stream in Figure~\ref{fig:event-stream}.}
\label{fig:events-precedence-model}
\end{figure}

In summary, \textit{EPM} allows us to automatically build a tractable probabilistic graphical model from the events, discovering the existing dependencies among the event types in the event stream. These dependencies are represented by a graph, as illustrated in Figure~\ref{fig:events-precedence-model}, where the conditional probabilities are stored at the node level.

\begin{algorithm}[!ht]
    \floatname{algorithm}{Algorithm}
    \caption{\textit{Event Precedence Model}}
    \label{algorithm:EPM}
    \begin{algorithmic}[1]
        \REQUIRE A partitioned window $P$ for a batch of new events.\\ %
        \textbf{Observation: }
        \FOR{ each partition $W_k \in P$ where \textit{k} is the CRA value}
            \FOR{ each pair of consecutive events (of type $E_i$ and $E_{j}$ such that $i \neq j$) in $W_k$}
                \STATE  $f(E_i, E_j)$++ i.e., increase the observed frequency by 1;

            \ENDFOR
        \ENDFOR

        \textbf{Graph Generation: }
        \STATE Construct an edgeless network $G = (V, \Xi)$;

        \FOR{ each pair of event types, $E_i$ and $E_j$ such that $i \neq j$,}
            \IF {$f(E_i, E_j) > 0$}
                \STATE $\Xi \leftarrow \{ \Xi \cup \{ E_i \rightarrow E_j \}\} $ i.e., add an edge $E_i \rightarrow E_j$;
                \STATE $P(E_j | E_i) \leftarrow \frac{f(E_i, E_j)}{\sum_{E_k \in children(E_i)}{f(E_i, E_k)}} $;
            \ELSE
                \STATE $P(E_j | E_i) \leftarrow 0$;
            \ENDIF
            \IF {$f(E_j, E_i) > 0$}
                \STATE $\Xi \leftarrow \{ \Xi \cup \{ E_j \leftarrow E_i \} \} $ i.e.,  add an edge $E_j \rightarrow E_i$;
                \STATE $P(E_i | E_j) \leftarrow \frac{f(E_j, E_i)}{\sum_{E_k \in children(E_j)}{f(E_j, E_k)}} $;
            \ELSE
                \STATE $P(E_i | E_j) \leftarrow 0$;
            \ENDIF
        \ENDFOR
    \end{algorithmic}
\end{algorithm}

\subsection{Algorithm} \label{sec:algorithm}
Algorithm~\ref{algorithm:EPM} outlines the event precedence network construction algorithm. It has three steps: \emph{observation}, \emph{graph generation}, and \emph{evidence inscription}. These steps are discussed below.
\begin{enumerate}
    \item \textit{Observation : } This step observes adjacent neighbor events in each partition of the window to learn the precedence relationships {and update the frequency matrix \textit{f}}.
        Note that, based on the assumptions stated earlier, the precedence relationships should be between events in the same partition and between events of different types. Suppose $E_i$ and $E_j$ are the event types of two adjacent events. Then, their count f($E_i$, $E_j$) is increased by 1.  {Additionally, note that the frequency matrix \textit{f} is updated incrementally for each new partition of events.}

    \item \textit{Graph Generation : } This step starts with an edgeless graph $G = (V, \Xi)$ where $V$ is the set of nodes (event types) and $\Xi$ is an empty set of edges. Then, for any evidence of the precedence relationship between event types $E_i$ and $E_j$ (i.e., f($E_i, E_j$) $> 0$), an edge is added between the two nodes representing these event types. Note that the graph supports anti-parallel edges between nodes; in addition, a cyclic loop of edges is also supported. Thus, the graphical model offers the flexibility to incorporate all possible types of relationships, unlike in the traditional systems where only directed edges are supported.

    In addition, for every edge added in the graph, the probability of an event type given its parent event type is calculated using Equation~\ref{eq:cond_prob}. The calculated probabilities are then stored in the nodes.
\end{enumerate}

The running time complexity of the algorithm is polynomial with the total number of events that have arrived thus far and the number of event types. The observation step counts every pair of consecutive events in every partition of the window. Clearly, for each partition, the number of the counts is always one less than the number of events in it. If $N_e$ and $N_p$ are the number of events and the number of partitions, respectively, then the running time complexity is given as \BigO{\sum_{i=1}^{N_p} (N_{ei}-1)} $\approx$ \BigO{N_e}, where $N_{ei}$ is the number of events in the $i$-th partition. The graph generation phase checks for the evidence of the precedence relationships between every pair of event types. In the worst case, the event precedence network is completely connected (including cyclic edges and self referencing edges) and has $N^2$ edges. So, the running time complexity of this step is proportional to $N(N-1)$ or \BigO{N ^ 2}. Hence, the total running time is given as \BigO{N_e} + \BigO{N^2} = \BigO{N_e + N^2}.

\section{Top-K Predictive Query Processing} \label{sec:query_processing}
In this section, we first describe the predictive query processing model and its run-time causality test and then present the two top-k continuous predictive query processing algorithms -- \textit{Exhaustive Search} (ES) algorithm and the more efficient \textit{Reduced Search Early Termination} (RSET) algorithm.

\subsection{Predictive Query Processing Model} \label{sec:predictive_query}
The predictive query processing problem can be formulated as a search problem to find the possible effects of a given set of observed events in a causal network. However, the traditional causal network cannot be used for query processing due to the causal information loss and its lack of support for cyclic causality. To address this issue, since we already know that every causal relationship is always a temporal precedence relationship, we propose to infer causality during query processing from the event precedence network to determine the possible effects.

\begin{figure*}[t!]
    \centering
    \subfigure[$E_3$ as EOP]{\includegraphics[scale=0.195]{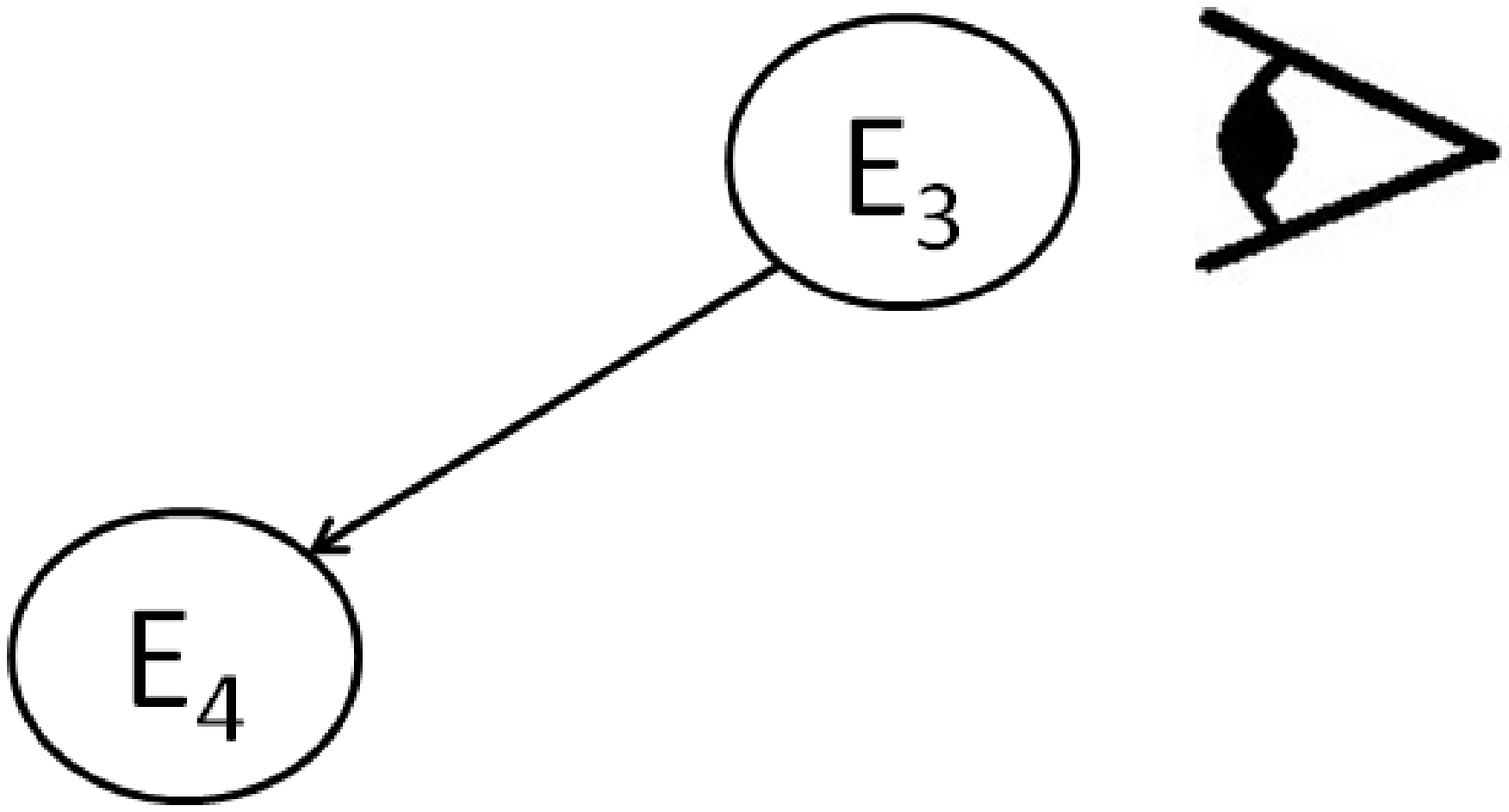}}
    \quad
    \subfigure[$E_4$ as EOP]{\includegraphics[scale=0.195]{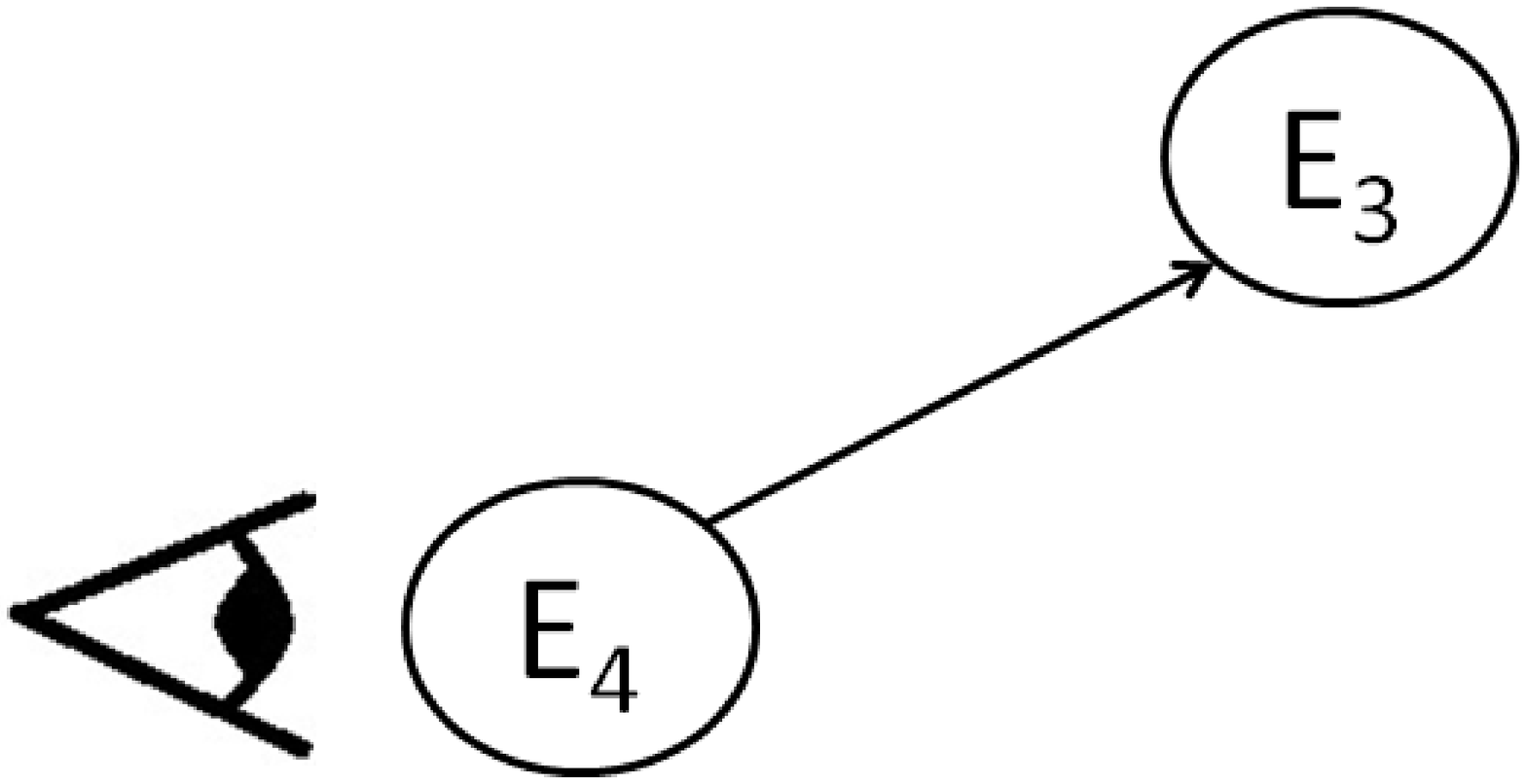}}
    \caption{Views from the EOPs for Figure~\ref{fig:events-precedence-model}.}
    \label{fig:eop}
\end{figure*}

In our work, the predictive query is a standing continuous query, so the ranked result list may change every time a new event is observed. The idea is to explore the event precedence network (EPN), which represents all the precedence relationships (including cyclic precedence relationships) among the event types, to answer the predictive queries when evidence is available. Indeed, the effect events are always the descendants of the cause events. Therefore, an outward breadth first search on the EPN is required to find the effect events. In situations where a visited node is encountered again, as EPN is cyclic, we ignore it. As discussed in Section~\ref{sec:event_precedence_model}, the next effect events cannot occur without the existence of the most recent cause event. Therefore, the starting point for exploring the EPN is always the event type $C_{\delta}$ of the most recent cause event. We call this event type the \textit{effect observation point} (EOP). For instance, in Figure~\ref{fig:events-precedence-model}, consider the two event types $E_3$ and $E_4$. $E_3$ is the effect of $E_4$ when $E_4$ is the EOP whereas $E_4$ is the effect of $E_3$ when $E_3$ is the EOP, as illustrated in Figure~\ref{fig:eop}.

However, there are two issues which make EPN not directly usable to answer a predictive query. First, a precedence relationship is not necessarily a causal relationship. So, we have to remove from the precedence relationships those that are not causal. Second, two variables may have causal relationship in the absence of other variables, but may not exhibit causality in the presence of certain condition variables. For example, rain and wet ground are dependent variables, as rain causes wet ground. However, they are independent in the presence of a roof over the ground (which is a conditional variable), as rain does not cause wet ground, given the existence of a roof. Therefore, we test causal relationships to resolve these issues during query processing for finding possible effects. To determine causality, the conditional independence tests, as described in Section~\ref{sec:conditional_independence_tests}, are performed between two event types with an edge in the EPN.

The ranking score of the predicted effect event type $E_i$ is calculated as $P(E_i|C_{\delta})$ given its EOP $C_{\delta}$. An EPN node stores the conditional probability of every child node given the current node as the parent  node. Scores across a chain of event types, $E_g \rightarrow E_p \rightarrow E_i$ (where $E_p$ is a parent of $E_i$ and $E_g$ is a parent of $E_p$), in EPN is calculated using the multiplicative property of conditional probability $P(E_i|E_g)$ = $P(E_i|E_p) \cdot P(E_p | E_g)$. 


\subsection{Exhaustive Search Algorithm} \label{sec:es}

\subsubsection{Approach} \label{sec:es_approach}
The most straightforward solution to the top-k prediction problem is to search for all possible effects exhaustively during the run-time causal inference over EPN and then sort them, according to their scores, in non-increasing order to determine the \textit{k} effects with the top scores. We call this solution the \textit{Exhaustive Search} (ES) approach.

\begin{figure*}[t!]
    \centering
    \includegraphics[scale=0.34]{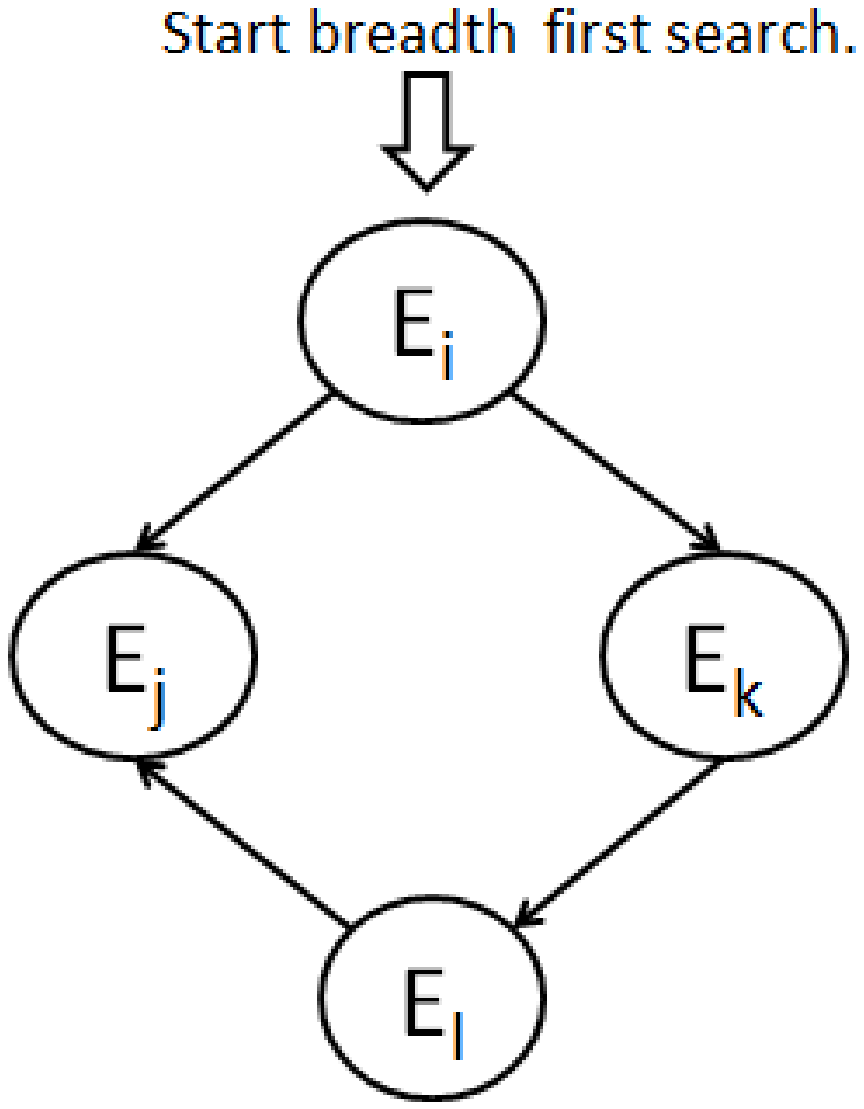}\vspace*{2ex}\\
    \begin{minipage}{0.75\linewidth}
    {\small
    (Score of $E_j$ is $P(E_j|E_i) + P(E_j|E_l) \cdot P(E_l|E_i)$. Note that $E_l$ is explored only after $E_j$. Therefore, $P(E_l|E_i)$ is unknown during the calculation of $E_j$'s score.)
    }
    \end{minipage}
    \caption{Illustration of the need for a causal search order.}
    \label{fig:CSOneed}
\end{figure*}

The ES approach should have a robust strategy for exploring the EPN to infer effects. As discussed earlier, an outward breadth first search may be run over the EPN for run-time causal inference.
However, the score calculation of the effects is not straight-forward. To apply multiplicative property of conditional probability described in Section~\ref{sec:predictive_query},
the scores of the parents of an event type should be known before its score is calculated, which is not always possible as demonstrated in Figure~\ref{fig:CSOneed}. Therefore, we define a
search order, called \textit{causal search order}, before exploring the EPN for run-time causal inference.

\begin{definition}{(\textbf{Causal Search Order}).}
{\rm 
The causal search order $O$ is an ordered set of event types \{$S_1, S_2, ...., S_N$\} observed during the outward breadth first search of the EPN such that $S_{i+j}$ is never an ancestor of $S_i$, where $j > 0$ and $i+j \le N$.
}
\hfill \fbox{}
\end{definition}

In addition to guiding the search for possible effects, the causal search order provides us with an effective strategy to calculate the ranking score. It gives an order of the event types such that the probabilities of parents are always known before calculating the probabilities of their children.

\begin{example}
\normalfont
Let us illustrate the \textit{causal search order} considering the EPN shown in Figure~\ref{fig:events-precedence-model}. As described earlier, we run outward breadth first search in EPN from the EOP. Suppose $E_3$ is the EOP. Initially, $E_3$ is added to $O$ and is explored. Then, the children of $E_3$ are added, so $O$ becomes $\{E_3, E_1, E_4, E_5\}$. Then, since $E_3$ has already been explored, the next unexplored node in $O$, $E_1$, is explored. However, no new nodes are added to $O$ as $E_1$ has no child. Then, we consider the next unexplored node ($E_4$) in $O$. Similar to $E_1$, no new nodes are added to $O$ as $E_4$ has no child. Now, the only remaining unexplored node in $O$ is $E_5$. So, the children of $E_5$ are added to $O$, which then becomes $\{E_3, E_1, E_4, E_5, E_6, E_7\}$. The recently added unexplored nodes $E_6$ and $E_7$ do not have any children and, therefore, no new nodes are added to $O$. So, the final causal search order $O$ is $\{E_3, E_1, E_4, E_5, E_6, E_7\}$. These steps are shown in Figure~\ref{fig:cso}(a).
Similarly, when EOP is $E_5$, the causal search order $O$ may be determined to be $\{ E_5, E_7, E_6, E_3, E_1, E_4\}$, as shown in Figure~\ref{fig:cso}(b).
\begin{figure*}[t!]
    \centering
    \subfigure[$E_3$ as EOP]{\includegraphics[scale=0.25]{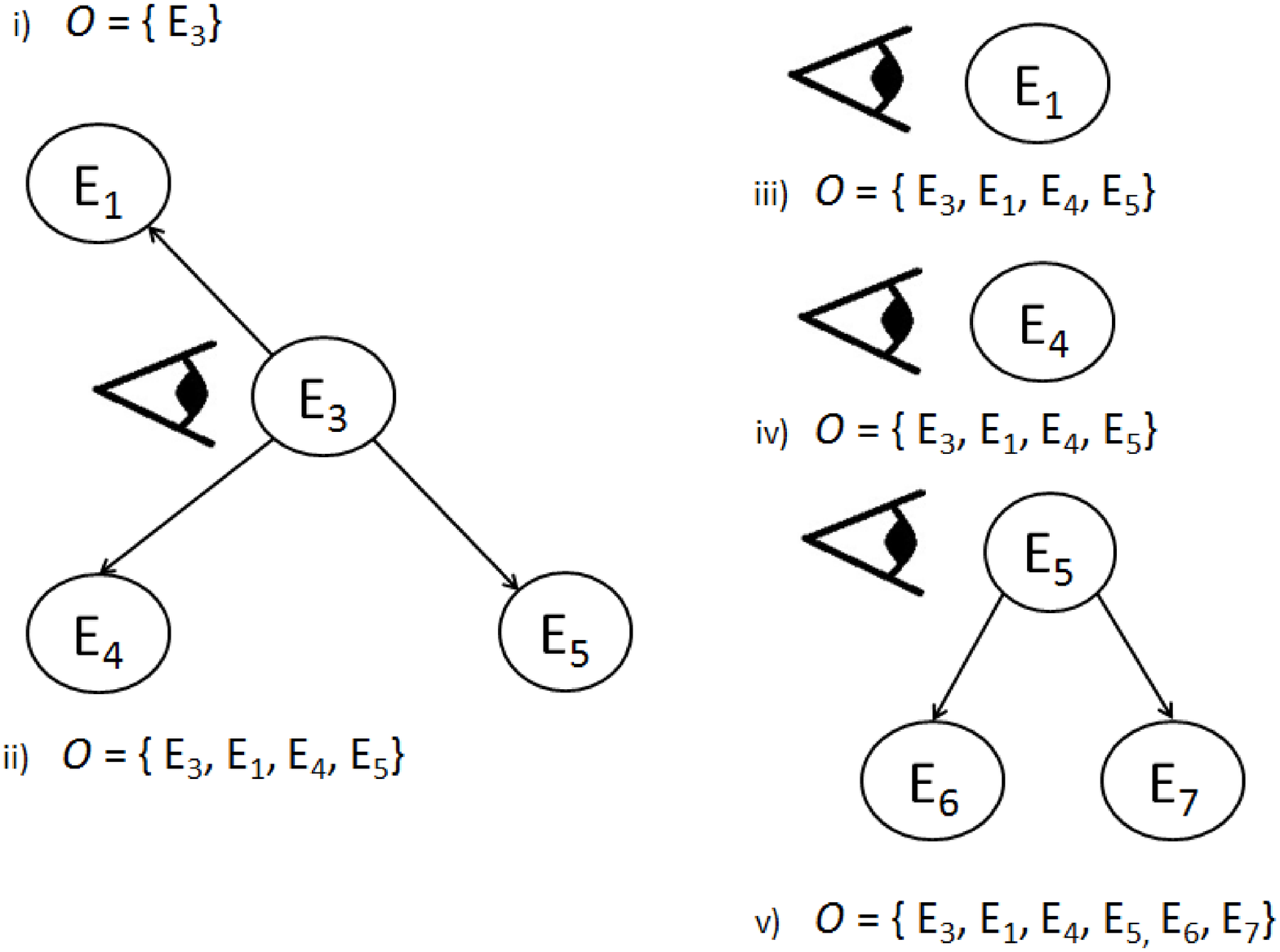}}
    \qquad
    \subfigure[$E_5$ as EOP]{\includegraphics[scale=0.25]{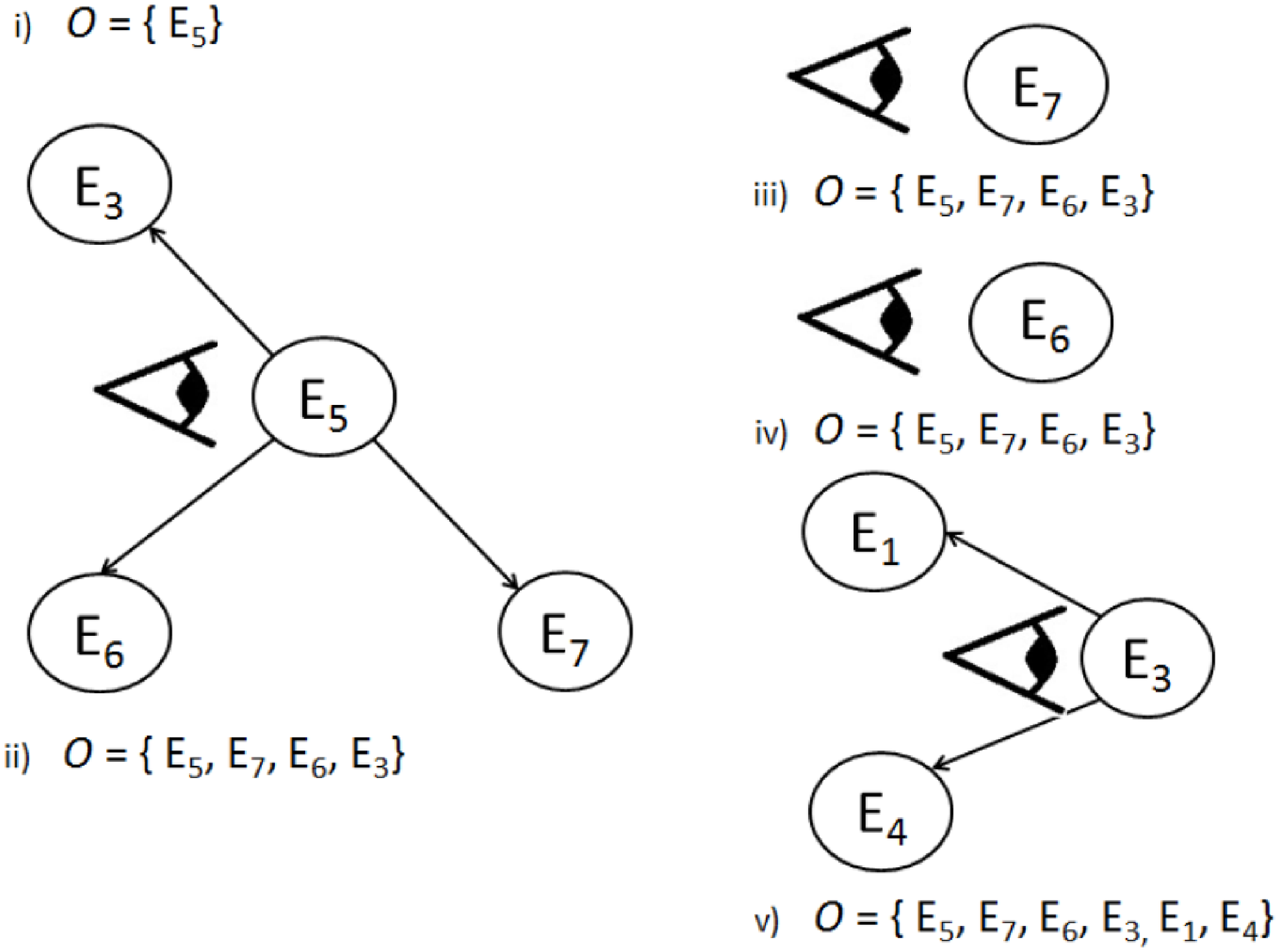}}
    \caption{Illustration of the steps to determine causal search order for the event precedence network of Figure~\ref{fig:events-precedence-model}.}
    \label{fig:cso}
\end{figure*}
\hfill \fbox{}
\end{example}

\subsubsection{Algorithm} \label{sec:es_algorithm}
Algorithm~\ref{algorithm:ex_approach} outlines the ES algorithm. It has a two-pass strategy for exploring EPN to infer effects. In the first pass, breadth-first search is run over the EPN to determine causal search order. In the second pass, EPN is explored with this search order for run-time causal inference. The input to the algorithm is the size of the result \textit{k}, the event precedence network \textit{G}, and the set of the recently observed $\delta$ cause event types arranged in temporal order $\{C_1, C_2, ...., C_{\delta}\}$. The four main steps of the algorithm are given as follows.
\begin{enumerate}
\item First, an outward breadth first search over $G$ from EOP (the most recent cause event type $C_{\delta}$) is run to determine the \textit{causal search order O}. Line 1 shows this step.

\item Second, the \textit{marginal independence tests} are performed between the event types during the search to remove any weak relationships (lines 2-7). {(A marginal independence test disregards the effect of other event types; in other words, it is equivalent to a conditional independence test with an empty condition set.)}

\item Third, $G$ is searched to find the effects of every unexplored node $E_j$ based on the ordering of the event types in $O$. Lines $8$ - $19$ shows this step.
    The CI tests between $E_j$ and each of its parents are performed as only the parents of $E_j$ can have effect on it. These tests are required to make sure that the event types being considered are not independent in the presence of other event types. In case of independence between $E_j$ and its parent, the edge representing their precedence relationship in G is removed. Lines $9$ - $16$ shows this step.

    Then, the score of the node $E_j$ is calculated and stored, as shown in lines $17$ and $18$.

\item Finally, the ranking scores of all event types explored in non-increasing order are sorted and then the \textit{k} event types with the top scores are selected (line 20).

\end{enumerate}

\begin{algorithm}[!ht]
    \floatname{algorithm}{Algorithm}
    \caption{\textit{ES Algorithm}}
    \label{algorithm:ex_approach}
    \begin{algorithmic}[1]
        \REQUIRE A temporally ordered set of recently observed $\delta$ cause event types $S = \{C_1, C_2,...., C_{\delta}\}$, the size of the result \textit{k}, an empty buffer $B_T$ to store the effect event types and their scores, and the event precedence network $G = (V, \Xi)$.

        \STATE Determine \textit{causal search order, O}, with the outward breadth first search in $G$ from the $EOP$ ($C_{\delta}$).\\
        \FOR{ every edge $E_i \rightarrow E_j \in \Xi$}
            \STATE 
            \textit{isIndependent} $\leftarrow$ \textit{IsIndependent}$(E_i, E_j , \phi )$);
            \IF{\textit{isIndependent} is true}
                \STATE $\Xi  \leftarrow \Xi - \{E_i \rightarrow E_j\}$ (//remove the weak relationship.)
            \ENDIF
        \ENDFOR

        \FOR{ every node $E_j \in (O-C_{\delta})$}
            \STATE $S_{parents} \leftarrow$ parents of $E_j$;

            \FOR{ every parent node $E_i \in S_{parents}$}
                \STATE \textit{isIndependent} $ \leftarrow I_{CMI}(E_i, E_j, S_{parents} - \{E_i\})$
                \IF{\textit{isIndependent} is \textit{true}}
                    \STATE $\Xi \leftarrow \Xi - \{E_i \rightarrow E_j \}$;
                    \STATE $S_{parents} \leftarrow S_{parents} - \{E_i\}$;
                \ENDIF
            \ENDFOR

            \STATE $P(E_j | \mathbf{C}) \leftarrow \sum_{E_p \in S_{parents}} P(E_j | E_p) P(E_p | \mathbf{C})$;
            \STATE Insert the pair $(E_j, P(E_j | \mathbf{C}))$ into $B_T$;

        \ENDFOR

        \STATE Sort the nodes in $B_T$ in the non-increasing order of the score and return the top-k results from $B_T$.

    \end{algorithmic}
\end{algorithm}

\begin{example}\label{example:ES}
\normalfont
Let us illustrate the ES algorithm considering the EPN shown in Figure~\ref{fig:events-precedence-model}. Suppose $\mathbf{C}$ is $\{E_2, E_3\}$ and \textit{k} is 2.
\begin{enumerate}
    \item First, the causal search order \textit{O}, from the EOP (i.e., $E_3$), is determined as $\{E_3, E_1,$ $E_4, E_5, E_6, E_7\}$.
    \item Then, the marginal independence tests are performed on each edge in EPN. For simplicity in illustration, we assume that these tests fail to remove any edges.
    \item Now, the score of each event type in \textit{O} is calculated and stored into the buffer $B_T$ based on their ordering in \textit{O}.
     \begin{enumerate}
        \item The score of the first unexplored event type $E_1$ is calculated and updated in $B_T$ as follows.
            \begin{itemize}
                \item Determine the parents of $E_1$, $S_{parents} = \{E_3\}$
                \item Perform CI tests between every parent of $E_1$ (i.e., $S_{parents}$) and $E_1$. Suppose the CI test succeeds, and thus no edge is removed.
                \item Calculate the score of $E_1$: $P(E_1 | \mathbf{C})$ =  $P(E_1 | E_3) \cdot P(E_3 | \mathbf{C})$ = 0.333.
                \item Update $B_T$ as $\{(E_1, 0.333)\}$.
            \end{itemize}
        \item The score of the next unexplored event type $E_4$ is calculated similarly as above, and $B_T$ is updated to $\{(E_1, 0.333), (E_4, 0.50)\}$.
        \item Following the same step as above, $B_T$ is updated to $\{(E_1, 0.333),$ $(E_4, 0.50),$ $(E_5, 0.167)\}$ for the next event type $E_5$.
        \item Next, $B_T$ is updated to $\{(E_1, 0.333), (E_4, 0.50), (E_5, 0.167), (E_6,$ $0.0835)\}$ for the event type $E_6$.
        \item $B_T$ is updated to $\{(E_1, 0.333), (E_4, 0.50), (E_5, 0.167), (E_6, 0.0835),$ $(E_7, 0)\}$ for the event type $E_7$.  Note that
                the score $P(E_7 | \mathbf{C})$ of $E_7$ is assigned as zero. For the parent event type $E_5$, the probability of its child $E_7$ is much lower (half) than the probability of its other child $E_6$. Thus, for this illustration, we assume the CI test between $E_7$ and $E_5$ fails. Consequently, there is no edge between them.
     \end{enumerate}

    \item Finally, $B_T$ is sorted in non-increasing order of the score as $\{(E_4, 0.50), (E_1,$ $0.333), (E_5, 0.167)$, $(E_6, 0.0835), (E_7, 0)\}$. Then, the top-2 predicted next event types are selected from $B_T$ as $\{(E_4, 0.50),$ $(E_1, 0.333)\}$.

\hfill \fbox{}
\end{enumerate}
\end{example}

\subsection{Reduced Search Early Termination Algorithm} \label{sec:rset}

\subsubsection{Approach} \label{sec:rset_approach}
The ES algorithm searches exhaustively in the EPN during query processing to determine the top-k results and, therefore, may well be slow. More importantly, it scales very poorly and, therefore,
this approach is likely to be intractable for a large network. Naturally there is a need for an alternative method that is faster. There are two issues to deal with for that purpose.
\begin{enumerate}
    \item \textit{Running time} : With the exhaustive search of ES on EPN to find all possible results, the search space increases with the number of variables and it performs unnecessary computations. 
        A good query processing method should avoid redundant and unnecessary computations to reduce the running time.
    \item \textit{Accuracy} : Usually, the tradeoff for reducing running time is the loss in the accuracy of the results. The running time is decreased by reducing the search space, which may skip the correct effect events. The query processing method should avoid such an incident as much as possible.
\end{enumerate}

To achieve faster running time while predicting accurate and consistent results, we propose a new algorithm called the Reduced Search Early Termination (RSET) algorithm. It is built upon the ES algorithm.
The following strategies reduces the runtime with only marginal reduction in prediction accuracy.
\begin{enumerate}
    \item \textit{To reduce the running time. } There are two ideas to reduce the running time. First, we reduce the search space in EPN by exploring only the descendants of the nodes in the current top-k during query execution. The nodes not in the top-k or their descendants have lower scores, due to multiplicative property of conditional probability described in Section~\ref{sec:predictive_query}, thereby disqualified from being top-k candidates. Second, we use a priority-based breadth-first search with an early termination criterion such that the query execution is stopped as soon as it is certain that the top-k results have been found. The priority-based breadth-first search always chooses the unexplored descendant node with the highest score to explore its children. The early termination criterion is met if there is no change in the list of event types in the top-k; that means, there is no more descendant node whose score can be greater than those in the current top-k. Consequently, the search space is only partially explored. For this reason, even though the causal inference is done at runtime, it incurs only a small overhead.
    \item \textit{To predict accurate results. } In an effort to achieve the same level of accuracy as the exhaustive search, we employ two ideas. First, we calculate the ranking score with the evidence from all explored nodes. Second, we perform the breadth-first search of the EPN in such a way that the events with greater score are processed earlier. It is worth noting that the ancestors always have higher or the same (in the worst condition) score as their descendants.
\end{enumerate}

\subsubsection{Algorithm} \label{sec:rset_algorithm}
Algorithm~\ref{algorithm:sm_approach} outlines the RSET algorithm and can be described as follows.

\begin{enumerate}
    \item First, two empty buffers, $B_C$ and $B_k$, are created to store the event types explored during query processing and the top-k effect event types computed, respectively. $B_k$ can hold maximum \textit{k} event types.  Line 1 states this step.

    \item Second, we add the $EOP$, $C_{\delta}$, with 1 as its score to both these buffers. $C_{\delta}$ has the probability of 1 since the event type has already been observed. Line 2 states this step.

    \item Then, the algorithm employs the following strategies to (a) reduce the search space, (b) predict accurate results, and (c) terminate as early as possible.

        \begin{itemize}
        \item \textbf{To reduce the search space.} Two ideas are employed to reduce the search space. First, it explores only the children of EOP or the event types in the buffer $B_k$ for further computation. Line 3 reflects this strategy. For any event type $E_c$ not in $B_k$ (i.e., top-k ), $E_c$ and its descendants are ignored, thus reducing the search space. The probabilities of the children of $E_c$ are much lower than that of $E_c$ due to multiplicative property of conditional probability, or equal even in the (rare) best case. Second, the buffer $B_k$ is always sorted in non-increasing order of the score after a new event type is added to it (lines 18--19 and 23--24). So, the priority for network exploration is always given to the \textit{unvisited} node with the highest score. This means that we consider the fact that its children might have higher ranking scores than the unvisited nodes already in $B_k$. Consequently, the nodes with the lowest probability ($E_{lowest}$) in $B_k$ can be removed if $B_k$ is full, which further reduces the search space. Lines 3 and 23 reflect this mechanism.

        \item \textbf{To predict accurate results. } It keeps in $B_c$ the record of all event types visited. Basically, the event types in EPN are explored in the breadth-first search. Therefore, the parents of a node have already been explored by the time the algorithm considers the child node. While doing so, the causal relationships are tested in lines 7 -- 14. Although the first strategy above reduces the search space, the algorithm still keeps all ancestors in $B_c$ by recording all visited nodes so far. Due to this search strategy, only the nodes with lower scores are not visited. So, the ranking score calculation is not affected by the reduced search space. Consequently, the accuracy of prediction result is not degraded.  Lines 7, 15, and 16 make use of this strategy.

        \item \textbf{To terminate as early as possible. } An early termination is possible if there is no change in the list of event types in $B_k$ after exploring their children. It means that there are no more event types further down the current level of exploration that can have higher probability than those already in $B_k$. Checking only unvisited nodes in line 3 reflects this strategy.
        \end{itemize}

\end{enumerate}

\begin{algorithm}[!ht]
    \floatname{algorithm}{Algorithm}
    \caption{\textit{RSET Algorithm}}
    \label{algorithm:sm_approach}
    \begin{algorithmic}[1]
        \REQUIRE A temporally ordered set of recently observed $\delta$ cause event types $\mathbf{C} = \{C_1, C_2, ....., C_{\delta}\}$, the size of the result \textit{k}, and event precedence network $G = (V, \Xi)$.
        \STATE $B_C\leftarrow\{\}$, $B_k\leftarrow\{\}$ i.e., create two buffers $B_C$ and $B_k$;
        \STATE $B_C \leftarrow B_C \cup \{(C_{\delta}, 1)\} $;
        \FOR{ every \textit{unvisited} node $E_c \in (\{C_{\delta}\} \cup$ (set of event types in $B_k))$ with the highest score}
            \STATE Mark $E_c$ as \textit{visited}.
            \STATE $S_{children} \leftarrow$ children of $E_c$;

            \FOR{ each node $E_j \in S_{children}$}

                \STATE ``\textit{Visited}'' parents $S_{parents} \leftarrow $ (set of parents of $E_j$) $\cap B_C $;

                \FOR{ every node $E_i \in S_{parents}$} 
                    \STATE \textit{isIndependent} $ \leftarrow I_{CMI}(E_i, E_j, S_{parents} - \{E_i\})$
                    \IF{\textit{isIndependent} is \textit{true}}
                        \STATE $\Xi \leftarrow \Xi - \{E_i \rightarrow E_j \}$;
                        \STATE $S_{parents} \leftarrow S_{parents} - \{E_i\}$;
                    \ENDIF
                \ENDFOR

                \STATE  $P(E_j | \mathbf{C}) \leftarrow \sum_{E_p \in S_{parents}} P(E_j | E_p) P(E_p | \mathbf{C} )$;
                \STATE $B_C \leftarrow B_C \cup \{(E_j, P(E_j | \mathbf{C} ))\}$;

                \IF{$|B_k| \leq k $}
                    \STATE $B_k \leftarrow B_k \cup \{(E_j, P(E_j | \mathbf{C} ))\}$;
                    \STATE Sort the event types in $B_k$ in non-increasing order of their scores;
                \ELSE
                    \STATE Find the entry with the lowest score, ($E_{lowest}, P_{lowest}$), in $B_k$.
                    \IF{ $P(E_j | \mathbf{C} ) > $ the lowest value in $B_k$}
                        \STATE $B_k \leftarrow (B_k - \{(E_{lowest}, P_{lowest})\}) \cup \{(E_j, P(E_j | \mathbf{C}) )\} $;
                        \STATE Sort the event types in $B_k$ in non-increasing order of their scores;
                    \ENDIF
                \ENDIF

            \ENDFOR
        \ENDFOR

    \end{algorithmic}
\end{algorithm}

{The computational complexities of the RSET and ES algorithms are dominated by the number of conditional independence tests, which is exponential in the \emph{worst} case as shown in our prior work on the FCNI algorithm~\cite{AcharyaL_DaWaK2013}, and thus both the RSET and ES algorithms have exponential computational complexity in the worst case. In practice, however, the RSET algorithm achieves a significant reduction in the computational complexity due to pruning by reduced search space and early termination.}

\begin{example}\label{example:RSET}
\normalfont
Let us illustrate the RSET algorithm considering the EPN shown in Figure~\ref{fig:events-precedence-model}. Suppose $\mathbf{C}$ is $\{E_2, E_3\}$ and \textit{k} is 2.
\begin{enumerate}
    \item First, two empty buffers $B_C$ and $B_k$ are created to store all the event types explored so far and to store the current top-k predicted event types, respectively.
    \item Then, the search starts with the EOP ($E_3$) and updates $B_C$ to $\{(E_3, 1)\}$.
    \item Now, every \textit{unvisited} event type is explored as follows. For simplicity in illustration, we assume that the CI tests return false and hence the edges are not removed.
        \begin{enumerate}
            \item The first unvisited event type, $E_3$, is marked as visited and its children, $S_{children}$ (= $\{E_1, E_4, E_5\}$), are explored.
                \begin{enumerate}
                    \item The score of the first unexplored child, $E_1$, is calculated and added to the two buffers as follows.
                        \begin{itemize}
                            \item Determine the parents of $E_1$; $S_{parents}$ i set to $\{E_3\}$.
                            \item Perform CI test of the edge between $E_1$ and $E_3$ given $S_{parents}$.
                            \item Calculate the score of $E_1$ as $P(E_1 | \mathbf{C}) = P(E_1 |E_3) P(E_3 | \mathbf{C})$ = 0.333.
                            \item Update $B_C$ to $\{(E_3, 1) ,(E_1, 0.333)\}$.
                            \item Update and sort $B_k$ to $\{(E_1, 0.333)\}$.
                        \end{itemize}

                    \item The same steps as above are followed for the next unexplored child $E_4$. The two buffers $B_C$ and $B_k$ are updated to $\{(E_3, 1) ,(E_1, 0.333),$ $(E_4, 0.50)\}$ and $\{(E_4, 0.50), (E_1, 0.333)\}$, respectively.
                    \item $B_C$ and $B_k$ are updated to $\{(E_3, 1) ,(E_1, 0.333), (E_4, 0.50), (E_5,$ $0.167)\}$ and $\{(E_4, 0.50), \\(E_1, 0.333)\}$, respectively, for the next unexplored child $E_5$.

                  \end{enumerate}

            \item Now, the next unvisited event type, $E_4$ in $B_k$, is marked as visited and its children are explored. However, there are no child of $E_4$, i.e., $S_{children}$ is empty. Therefore, there is no computation to be performed.
            \item The same result is seen for the next event type, $E_1$, as well. As it has no child (i.e., $S_{children}$ is empty), there is no computation to be performed.

        \end{enumerate}
    \item The top-k result in $B_k$ is obtained as $\{(E_4, 0.50), (E_1, 0.333)\}$.
\end{enumerate}
For the same $\mathbf{C}$, the RSET algorithm considered only four event types -- $E_3 , E_1, E_4,$ $E_5$ -- and the ES algorithm considered all event types -- $E_3 , E_1, E_4, E_5, E_6, E_7$. Note that in this example, RSET produced the same result (i.e., $B_k = \{(E_4, 0.50),$ $(E_1, 0.333)\}$) as ES (which is typical unless the value of k is significantly large). It shows the merit of the early terminating reduced search approach of the RSET algorithm against the exhaustive search approach of the ES algorithm.
\hfill \fbox{}
\end{example}

\section{Performance Evaluation} \label{sec:experiments}
We conduct experiments to evaluate the run-time causal inference model and the top-k query processing mechanism in the proposed \textit{RSET algorithm} and the \textit{ES algorithm}. One evaluation is with respect to the accuracy of the top-k results, and the other evaluation is with respect to the runtime. Section~\ref{sec:settings} describes the experiment setup, including the evaluation metrics, datasets and the platform used, and Section~\ref{sec:results} presents the experiment results.

\subsection{Experiment Setup} \label{sec:settings}

\subsubsection{Evaluation Metrics} \label{sec:evaluation_criteria}
Intuitively, the performances of the \textit{top-k query processing} algorithms are best evaluated by examining two important evaluation criteria -- accuracy and runtime.

\begin{enumerate}
\item \textbf{Accuracy. } Suppose $R_k$ is the ranked list of the top-k effects predicted from the set of cause event types, $\textbf{C}$, observed so far in the test sequence.
    To measure accuracy, the next event type observed in the test sequence, $E_o$ ($o \in [1, N]$), is checked against the predicted ranked list $R_k$. If $E_o$ exists in $R_k$, we say the prediction is correct (hit); otherwise we say the prediction is incorrect (miss).

    There are two methods for deciding the correctness of a top-k predicted result. First, in some scenarios, a hit may be a sufficient condition for correctness. In such a case, we say the accuracy is $100\%$ for a hit and $0\%$ for a miss. We call this perspective \textit{hit}-\textit{or}-\textit{miss} (\textit{non}-\textit{weighted}). Second, in other scenarios, the rank of the predicted result may also play an important role in determining the accuracy. Clearly, if the algorithm predicts $E_o$ as the most likely effect (i.e., at the top of $R_k$ with the highest probability), then the accuracy is $100\%$. As we go down the list in $R_k$, the accuracy decreases. Therefore, to reflect this point, we take the probability of each event in $R_k$ into consideration when calculating the prediction accuracy. We call this perspective \textit{weighted}.

  \textbf{\textit{Hit}-\textit{or}-\textit{miss accuracy}. } Let $n_{hits}$ and $n_{misses}$ be the number of hits and the number of misses, respectively out of $n_{tests}$ tests. Then, the hit-or-miss accuracy of the result, $\alpha_{hm}$, is calculated as follows.
            \begin{equation}\label{eq:hit_or_miss_accuracy}
                \alpha_{hm} = \frac{n_{hits}}{n_{tests}} = \frac{n_{hits}}{n_{hits} + n_{misses}}
            \end{equation}

  \textbf{\textit{Weighted accuracy}. } Suppose $P(E_o)$ is the score of $E_o$ in $R_k$. As discussed earlier, the rank of $E_o$ in $R_k$ contributes towards the calculation of the prediction accuracy. The rank is based on the score; therefore, we normalize the score such that the prediction accuracy decreases gradually with the decrease in the rank of $E_o$ in $R_k$. The accuracy of $E_o$, in the case of a miss, is $0\%$ whereas the accuracy of $E_o$, in the case of a hit, is $\frac{P(E_o)}{max\{ P(E_j) | E_j \in R_k \}}$, where the denominator is the highest probability among all event types in $R_k$. (Note that this measure gives the top event type the accuracy of $100\%$.)

        Let $n_{hits}$ and $n_{misses}$ be the same as above. Then, the weighted accuracy of the result, $\alpha_{weighted}$, is computed as follows.
            \begin{displaymath}\label{eq:1_weighted_accuracy}
                \alpha_{weighted} = \frac{\sum_{i = 1}^{ n_{tests}}\frac{P(E_{o_i})}{max\{ P(E_{j_i}) \mid E_{j_i} \in R_{k_i} \}}}{n_{tests}}
            \end{displaymath}
            where $E_{o_i}$ is the $i$-th observed event type in the test data.

        As mentioned earlier, the accuracy of a miss is $0\%$. Therefore, we can consider only the hits in the numerator.
             \begin{equation}\label{eq:2_weighted_accuracy}
                \alpha_{weighted} = \frac{\sum_{h = 1}^{ n_{hits}}\frac{P(E_{oh})}{max\{ P(E_{j_h}) \mid E_{j_h} \in R_{k_h} \}}}{n_{hits} + n_{misses}}
            \end{equation}
             where $E_{o_h}$ is the $h$-th observed event type which has the result hit in the test data.

\item \textbf{Runtime. } The \textit{runtime} is the CPU time taken during query processing. Note that the event precedence network construction is not part of the query processing mechanism. Therefore, we do not include it in the measurement of the runtime. In the query processing with the RSET algorithm and the ES algorithm, there is an overhead of run-time causal inference and it is included in the runtime. In contrast, the query processing with the traditional causal inference (i.e., FCNI algorithm) does not include the causal network construction time in the runtime as the causal inference is performed only once (during the causal network construction) prior to query processing. In our work, latency is the interval between the arrival of a new event and the identification of its top-k effect events. However, as the time for EPN update is insignificant (see the polynomial runtime in Section~\ref{sec:algorithm}) compared to the time for query processing (which is exponential), latency is essentially the query processing time.
\end{enumerate}

\subsubsection{Datasets} \label{sec:datasets}
Experiments are conducted using two real-world datasets (summary in Table~\ref{tbl:datasets}) to evaluate the proposed algorithms.
\begin{table*}
\centering
\begin{tabular}{|c|c|c|c|c|} \hline
 \textbf{Dataset} &  $\mathbf{N}$ &  $\mathbf{N_{instances}}$ &  \textbf{CRA} &  $\mathbf{N_{CRA}}$ \\ \hline
MSNBC & 17 & 4698795 & session id & 989818\\ \hline
Power grid & 565 & 94339 & blackout id & 4492\\ \hline
\end{tabular}
\vspace*{1em}\\
\small ($N_{\textit{CRA}}$ is the number of different CRA values. $N_{\textit{instances}}$ is the number of event instances in the dataset.)
\caption{Profiles of the datasets.}
\label{tbl:datasets}
\end{table*}

\paragraph{\textbf{Electric power grid dataset}}
This dataset contains simulated temporal sequences of cascading electric power grid component outages, such as those that can lead to very large blackouts (e.g.,~\cite{Blackout_2004d}). The sequences were generated using a model of the Polish power network, which is described in~\cite{EppsteinH_PES2013}. Each sequence represents the order in which the components failed, as well as the time of the failure. Each grid component is considered an event type, whereas a component failure is an event instance. This dataset includes 4492 cascade sequences and 565 distinct event types.

In the original dataset, each file, representing one blackout, has a list of the components that failed in that blackout. The original schema of the power grid data is as follows: $\langle$\textit{event indicator}, \textit{timestamp}, \textit{component id}$\rangle$. The event indicator can be one of 0, 1, and -1, which refer to an initiating event, a dependent event, and a stop event, respectively. There is always at least one \textit{initiating event} at the beginning of each component failure sequence (with $0$ as its starting time). Since these events are always at the beginning of the sequence, there is no inward edge towards them in the event precedence network. A \textit{dependent event} is the result of the initiating event or another dependent event. A blackout sequence always has at least one dependent event. We treat both an initiating event and a dependent event in the same way. On the other hand, a \textit{stop event} denotes the end of the blackout and is not a real event. Therefore, we ignore stop events. The \textit{timestamp} and the \textit{component id} are respectively the starting time of an event and the attribute that uniquely identifies a grid component.

To create an event stream, we modify the schema and mix the data from the files in random order while preserving the temporal order of the component failures in each blackout. The modified schema, 
\textit{$\langle$timestamp}, \textit{component id}, \textit{blackout id}, \textit{event indicator}$\rangle$, has an additional tag \textit{blackout id} to identify the blackout to which the component failure belongs to. So, the blackout id is the CRA in the power grid dataset.

\paragraph{\textbf{MSNBC.com web dataset}} This dataset consists of click-stream data of 989818 sequences obtained from the University of California, Irvine's machine learning repository~\cite{Heckerman_1999}. Each sequence reflects the browsing activities, arranged in temporal order, in one user session. The dataset gives a random sample of the length of visits of users browsing the msnbc.com web site on the whole day of September 28, 1999. The length of the visit is an estimate of the total number of clicks or pages seen by each user and is based on the ``Internet Information Server (IIS) logs for msnbc.com and news-related portions of msn.com''. A webpage category is an event type, and a webpage visit is an event instance. The session id of the visit is the CRA for its event instance.

The number of distinct event types is 17. That is, a sequence can have web activities related to 17 different webpage categories. These event types (i.e., webpage categories) are frontpage, news, technology, local, opinion, on-air, miscellaneous, weather, msn-news, health, living, business, msn-sports, sports, summary, bbs, and travel. The total number of event instances (i.e., page visits) is 4698795.

To create an event stream, we randomly mix the events while preserving the temporal order of the events for each session. The schema of the events is $\langle$\textit{time\-stamp}, \textit{webpage category}, \textit{session id}, $\Phi$$\rangle$, where $\Phi$ denotes an empty attribute set.

\subsubsection{Experiment Platform} \label{sec:platform}
The experiments are conducted on a 2.3 GHz Intel Core i5 machine with 4GB of memory, running Windows 7. The algorithms are implemented in Java 1.7.0.

\subsection{Experiment Results} \label{sec:results}
We perform two sets of experiments to evaluate the RSET and ES algorithms. One set of experiments is to evaluate the prediction accuracy, and the other set of experiments is to evaluate the runtime. There are two objectives in each set of experiments. The first objective is to compare the query processing with the run-time causal inference mechanism (of RSET and ES) against the query processing with the traditional causal inference mechanism (of the FCNI algorithm). For a fair comparison, as the goal is only to compare the causal inference mechanisms, the query processing mechanism of either RSET algorithm or ES algorithm can be used in the FCNI algorithm, and, in this experiment, we choose the RSET algorithm. The second objective is to compare the query processing mechanism between the RSET algorithm and the ES algorithm. In addition, the effects of \textit{k} on the proposed algorithms are studied in each set of experiments.

{We divide each real dataset into $70\%$ for training and $30\%$ for testing the proposed algorithms. The division of MSNBC dataset is done by session id, and the division of electric power grid dataset is by blackout id.} From the event stream of training dataset, EPN is constructed as an input to the RSET and ES algorithms, whereas a causal network is constructed as an input to the FCNI algorithm. {The window observation period \textit{T} is set to 10~msec for the MSNBC dataset and 10~sec for the electric power grid dataset.} The testing data simulates event stream. As soon as a new event arrives, it is added to the partitioned window. Then, the top-k prediction query execution is triggered in response to the most recent event at position $\delta$ (called the \emph{EOP index}) in the sequence of cause events in the same partition. Note that the RSET and ES algorithms perform query processing over the EPN whereas the FCNI algorithm does so over the causal network. Upon the arrival of a new event, the measurements of prediction accuracy and runtime are repeated and the calculated average accuracy and average runtime are reported.

\subsubsection{Accuracy}\label{sec:exp1}

\begin{figure*}[t!]
\centering
\subfigure[k =1]{\includegraphics[scale=0.43]{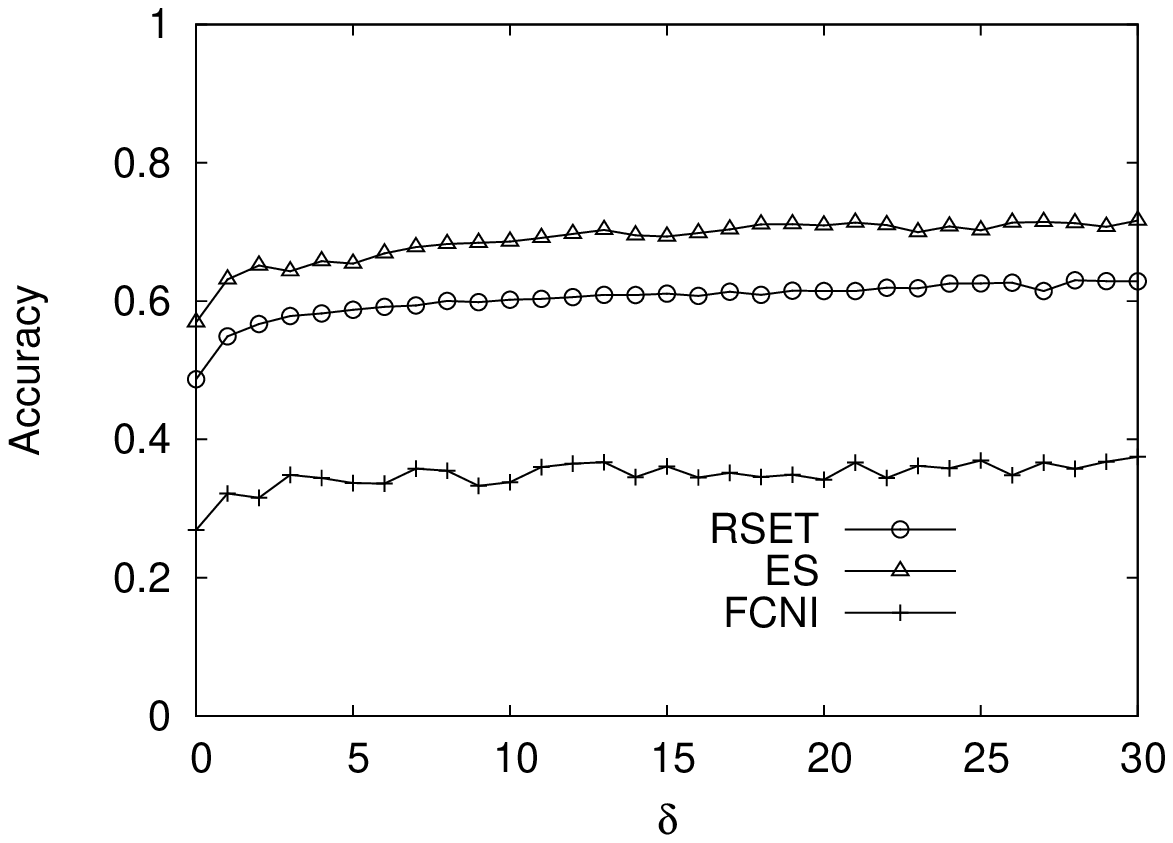}}
\quad
\subfigure[k = 3]{\includegraphics[scale=0.43]{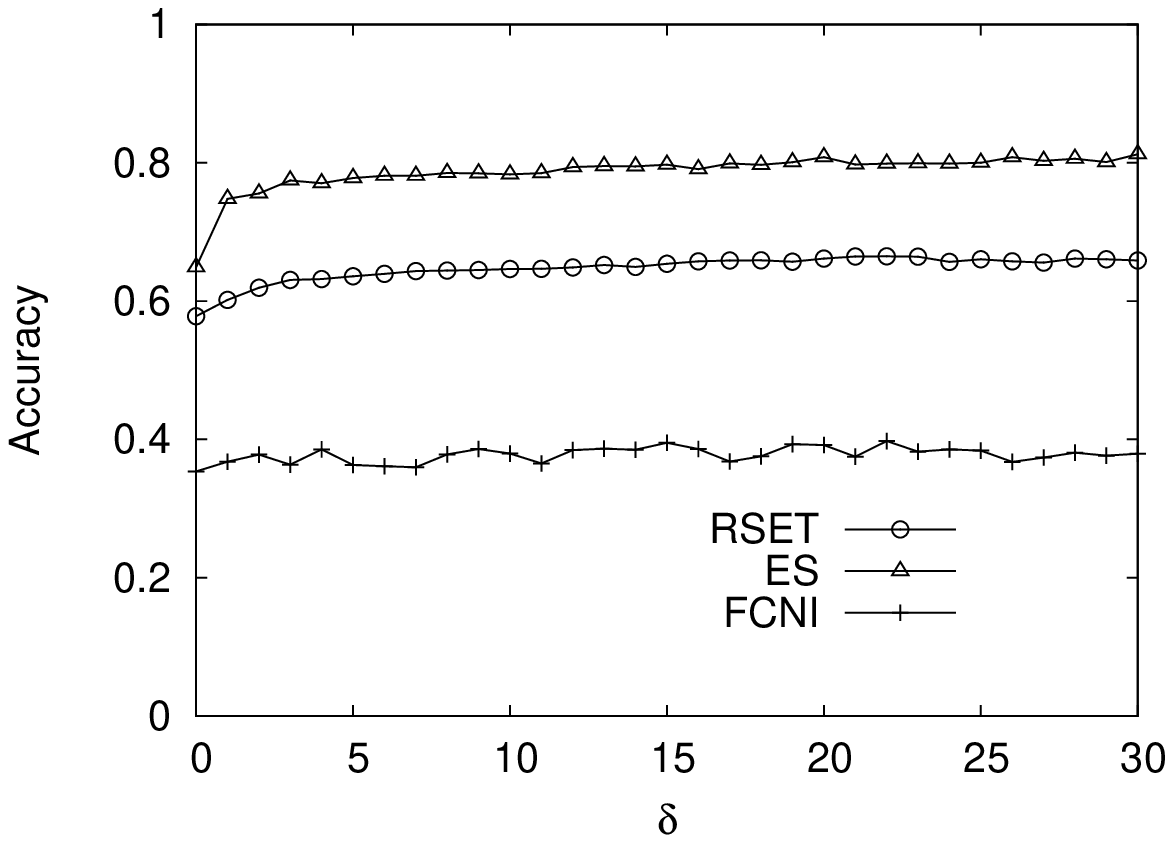}}
\quad
\subfigure[k = 5]{\includegraphics[scale=0.43]{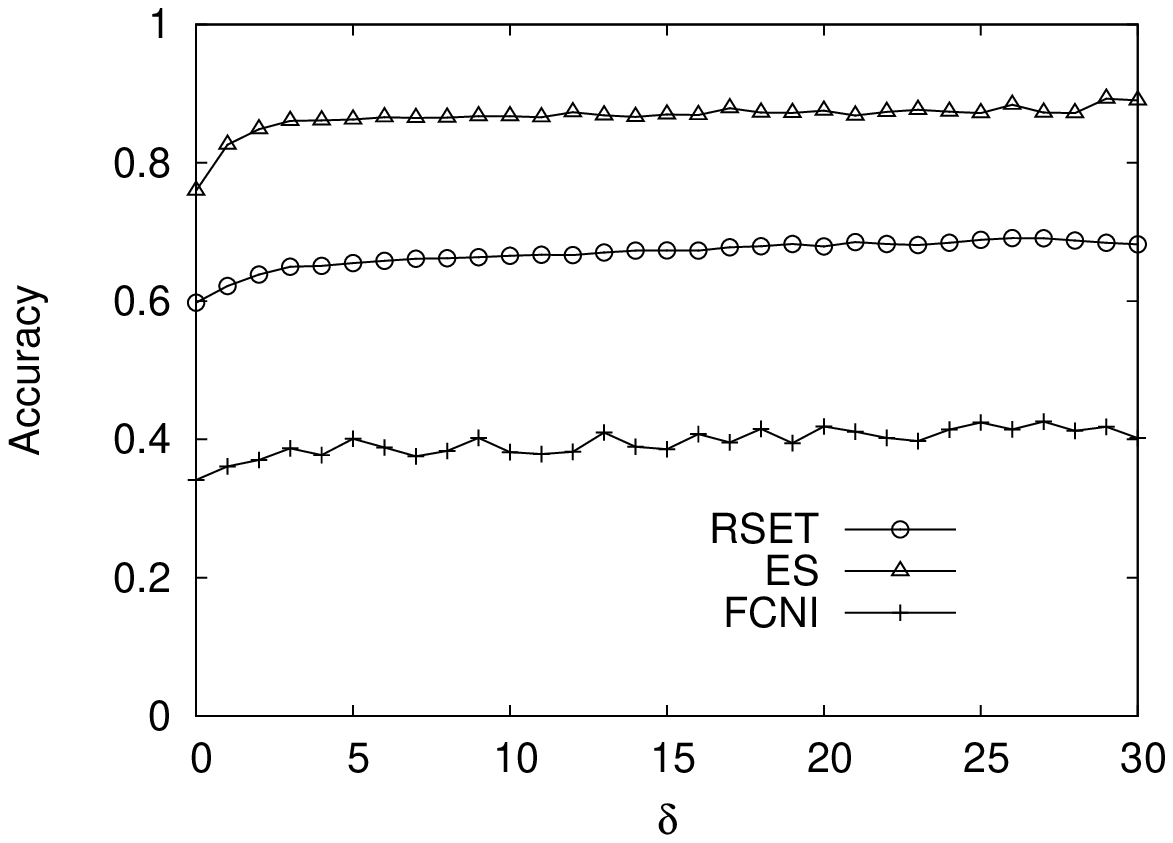}}
\quad
\subfigure[k = 7]{\includegraphics[scale=0.43]{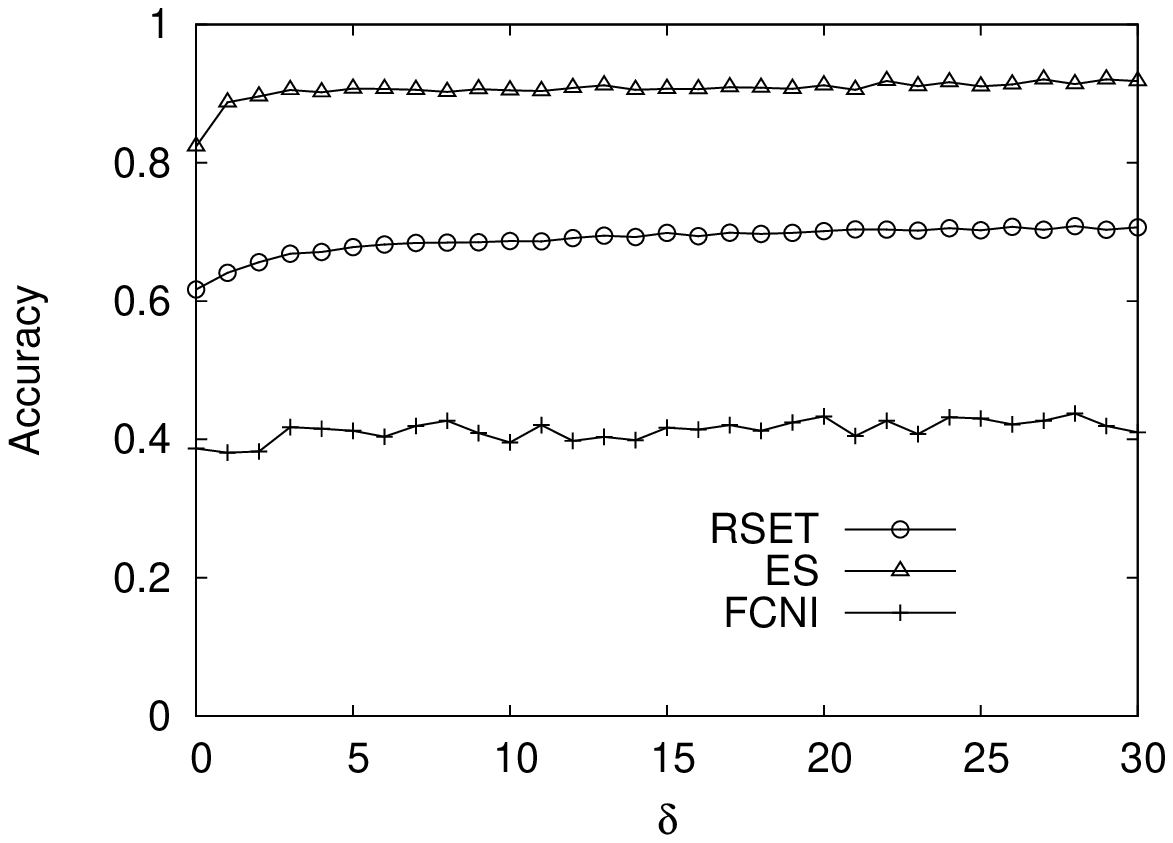}}
\quad
\subfigure[k = 9]{\includegraphics[scale=0.43]{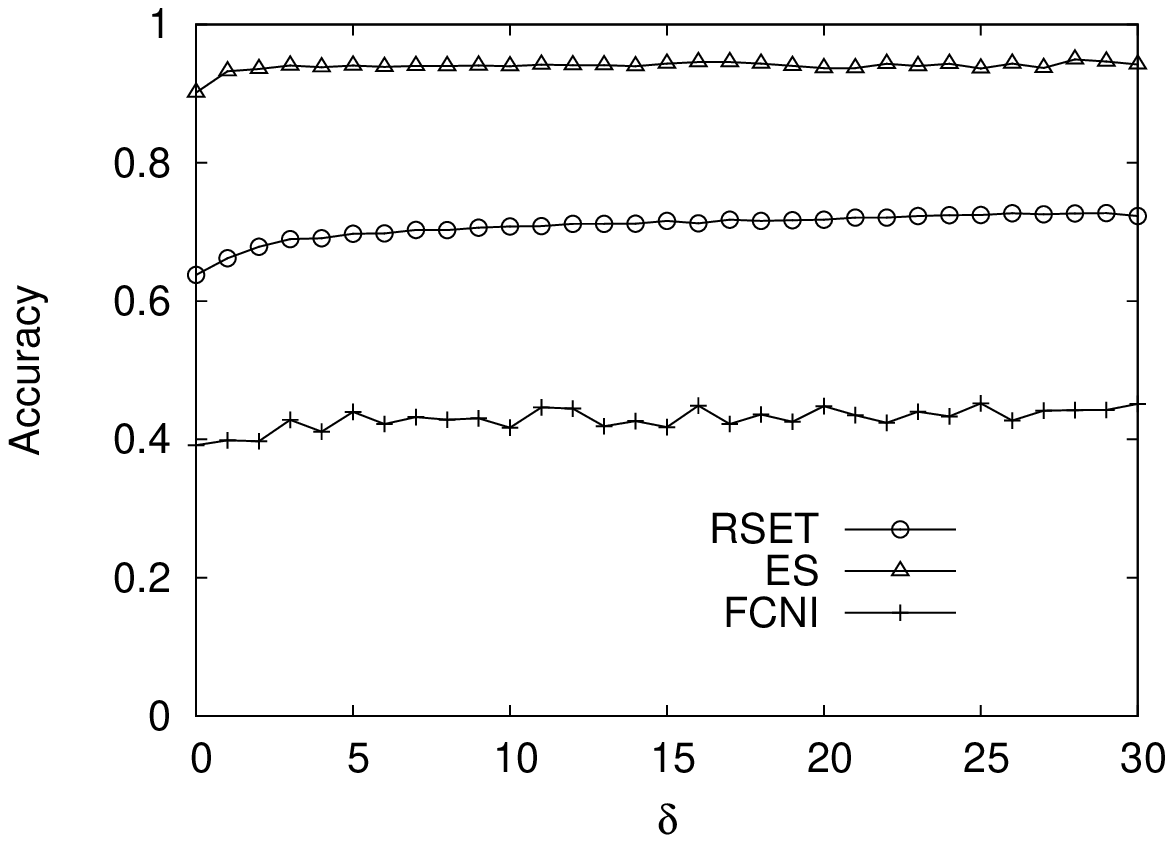}}
\caption{Hit-or-miss accuracies of the RSET, ES, and FCNI algorithms w.r.t EOP index ($\delta$) for MSNBC dataset.
\small{\emph{(Note that the EOP index $\delta$ is the position of the most recent event in the sequence of cause events in the same partition.)}}}
\label{fig:exp1_msnbc_0_accuracy}
\end{figure*}

\begin{figure*}[t!]
\centering
\subfigure[k =1]{\includegraphics[scale=0.43]{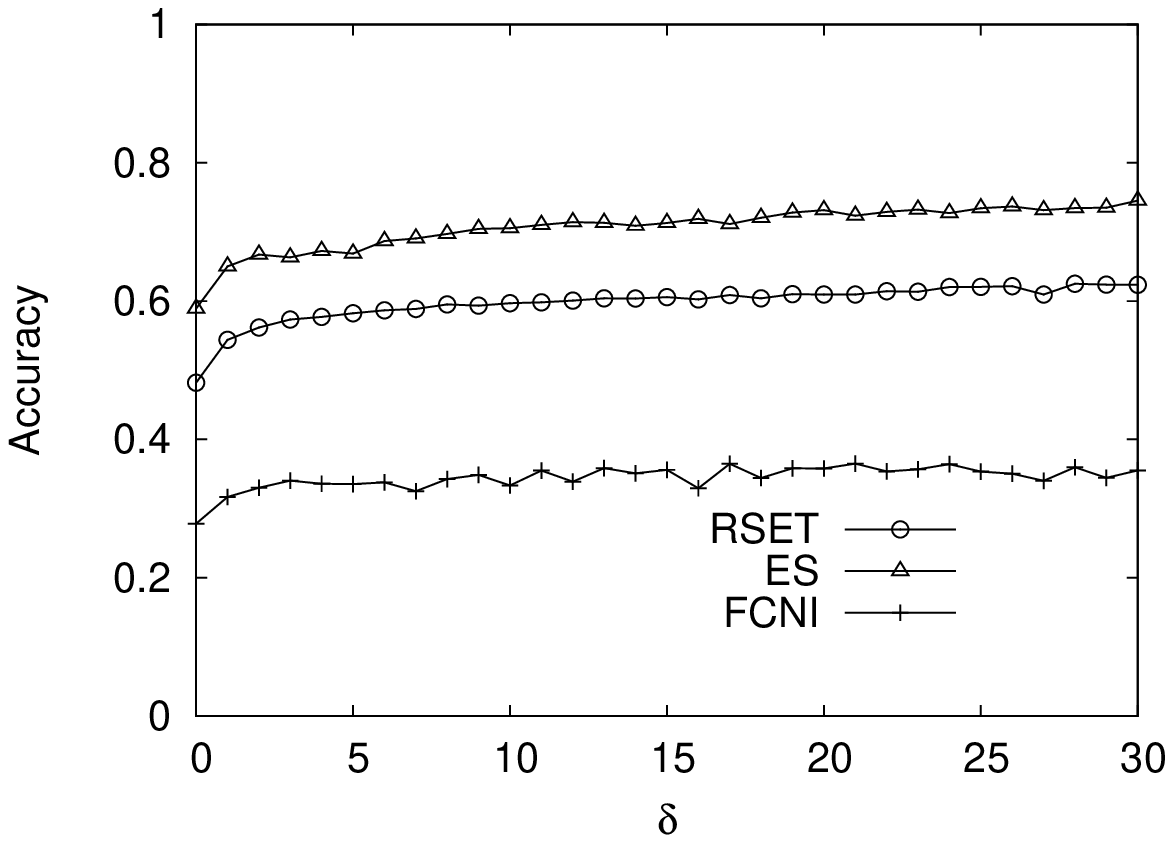}}
\quad
\subfigure[k = 3]{\includegraphics[scale=0.43]{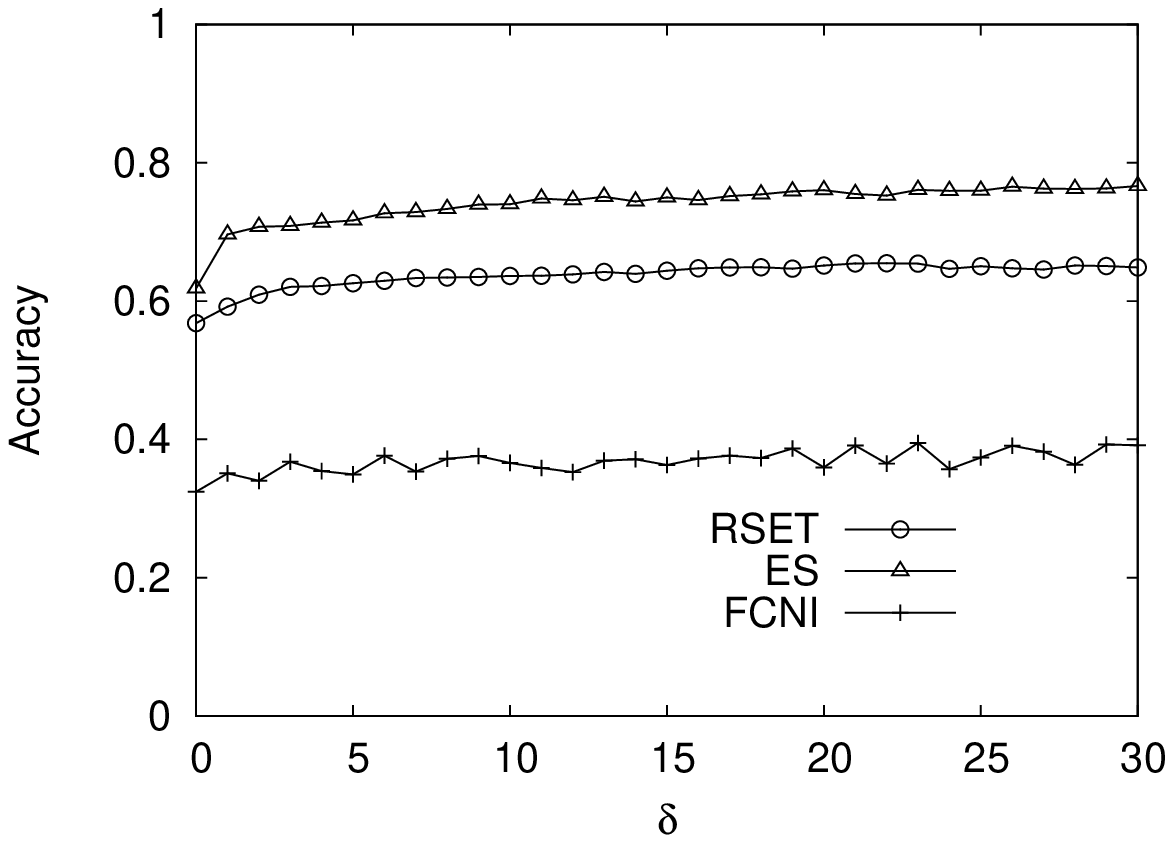}}
\quad
\subfigure[k = 5]{\includegraphics[scale=0.43]{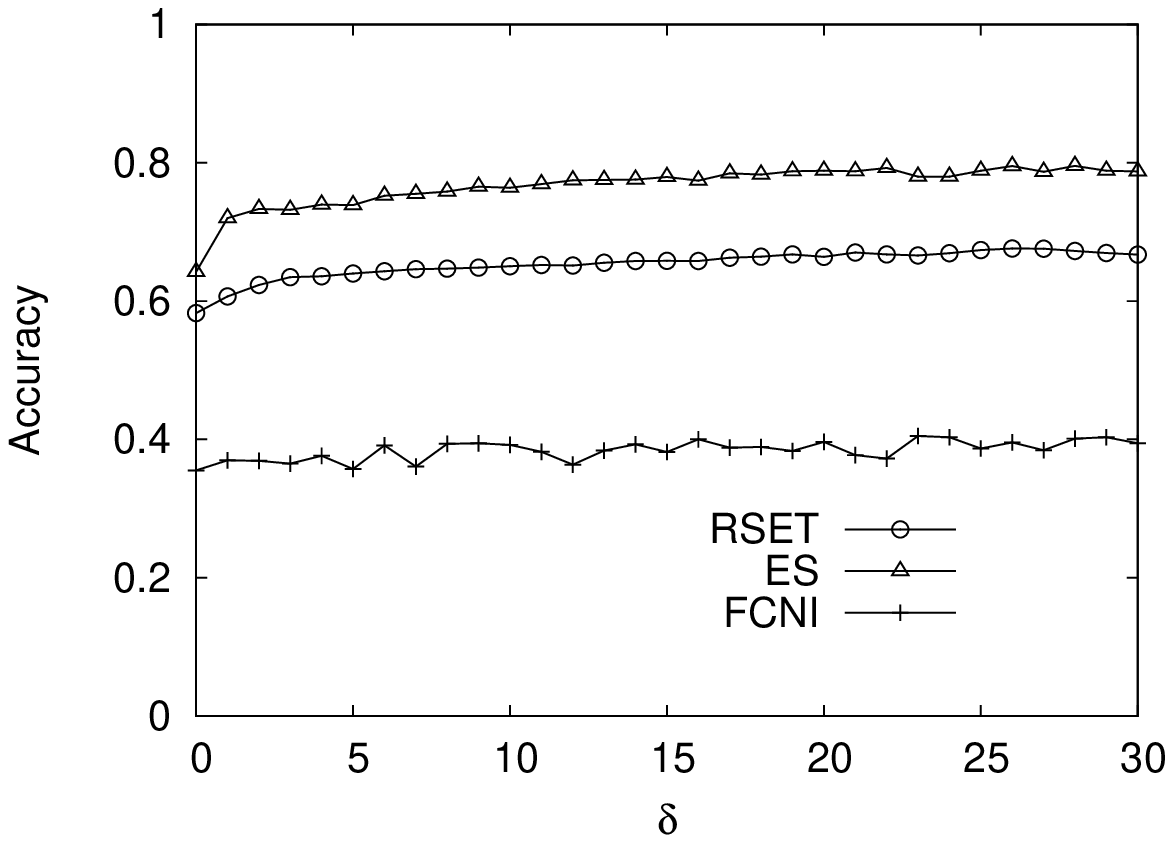}}
\quad
\subfigure[k = 7]{\includegraphics[scale=0.43]{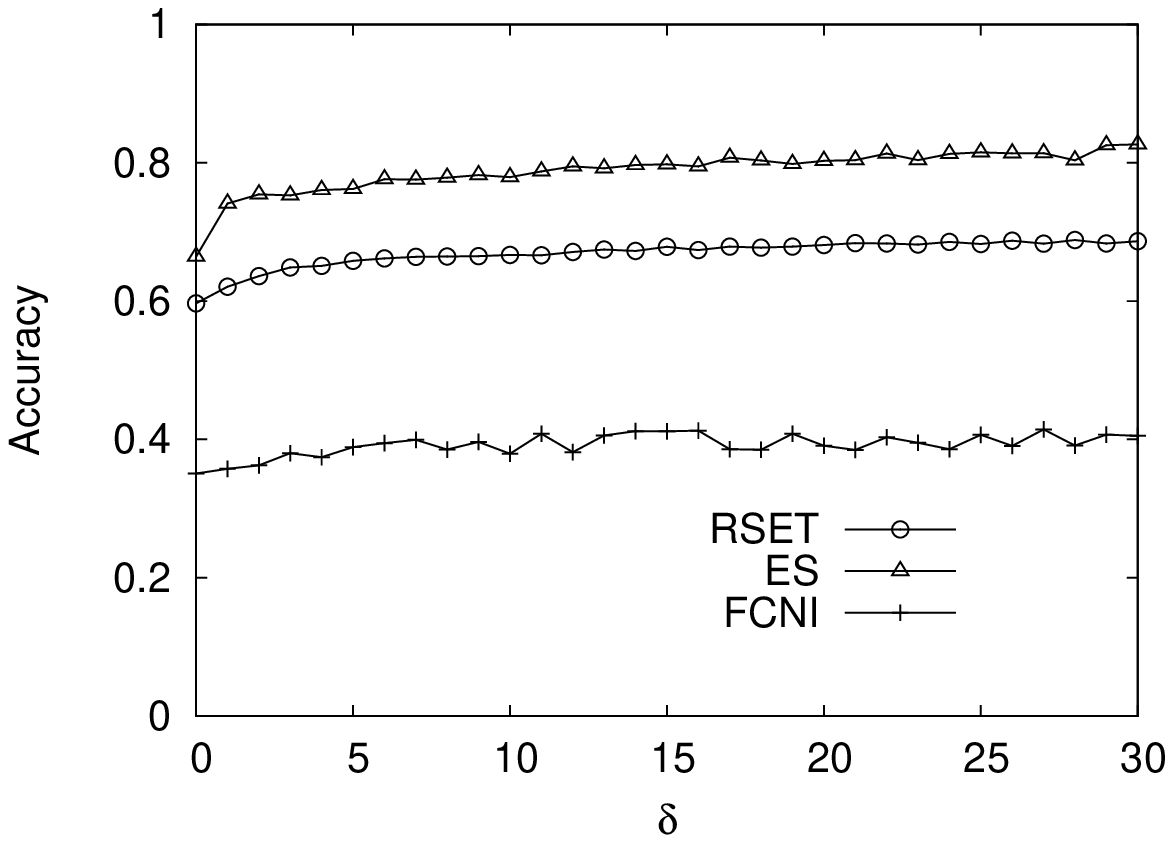}}
\quad
\subfigure[k = 9]{\includegraphics[scale=0.43]{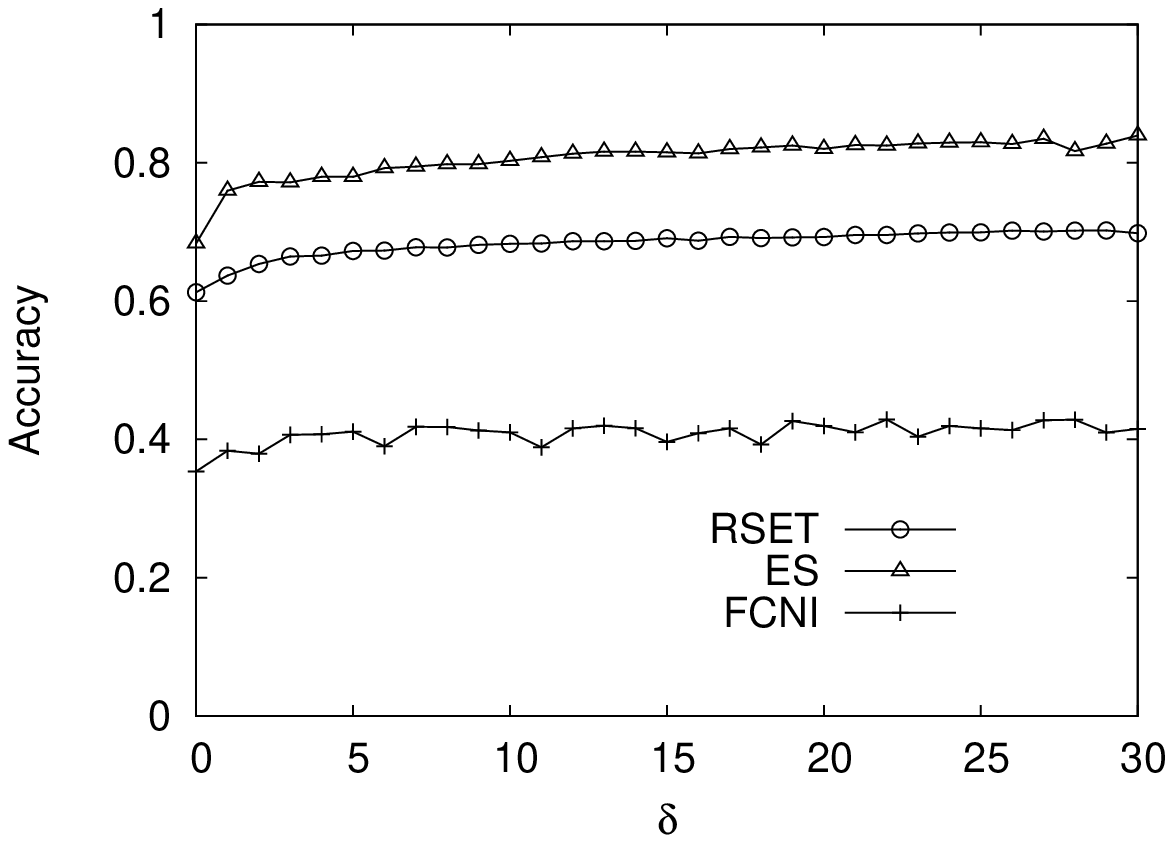}}
\caption{Weighted accuracies of the RSET, ES, and FCNI algorithms w.r.t EOP index ($\delta$) for MSNBC dataset.}
\label{fig:exp1_msnbc_1_accuracy}
\end{figure*}

\begin{figure*}[t!]
\centering
\subfigure[k =1]{\includegraphics[scale=0.43]{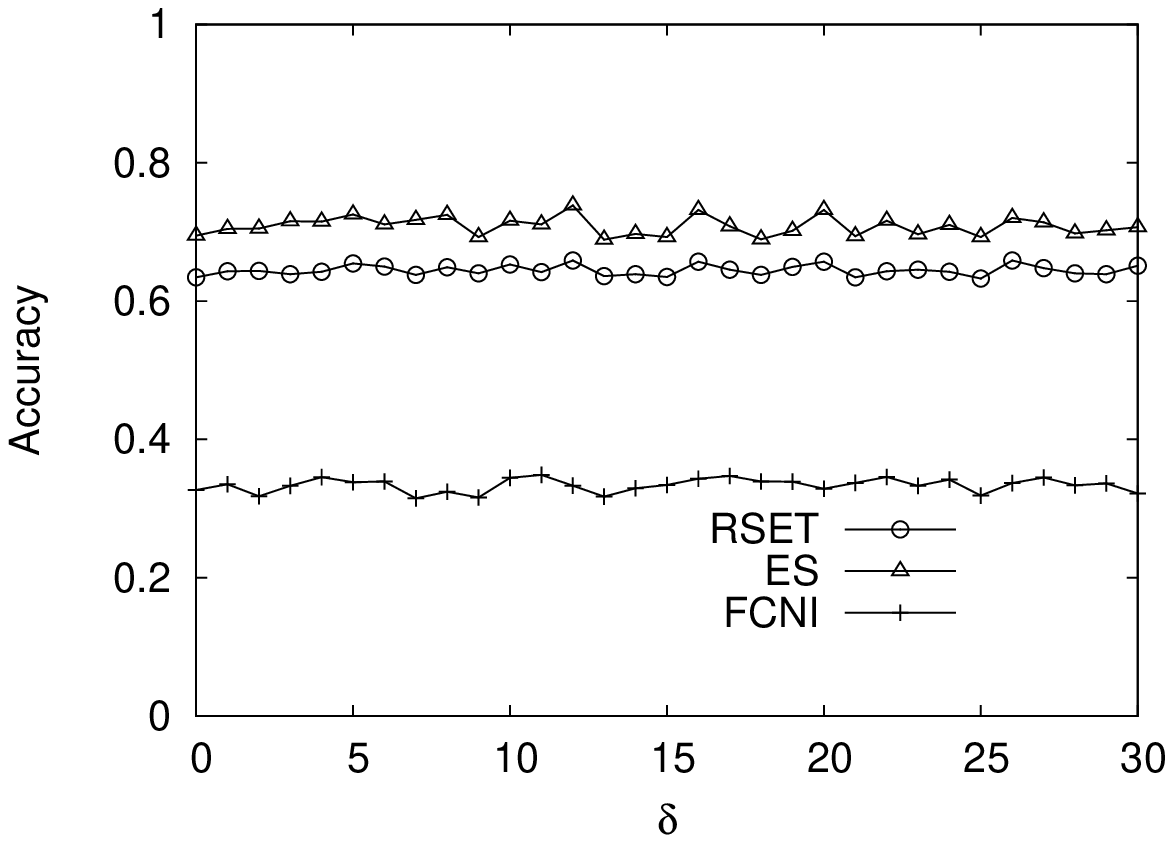}}
\quad
\subfigure[k = 5]{\includegraphics[scale=0.43]{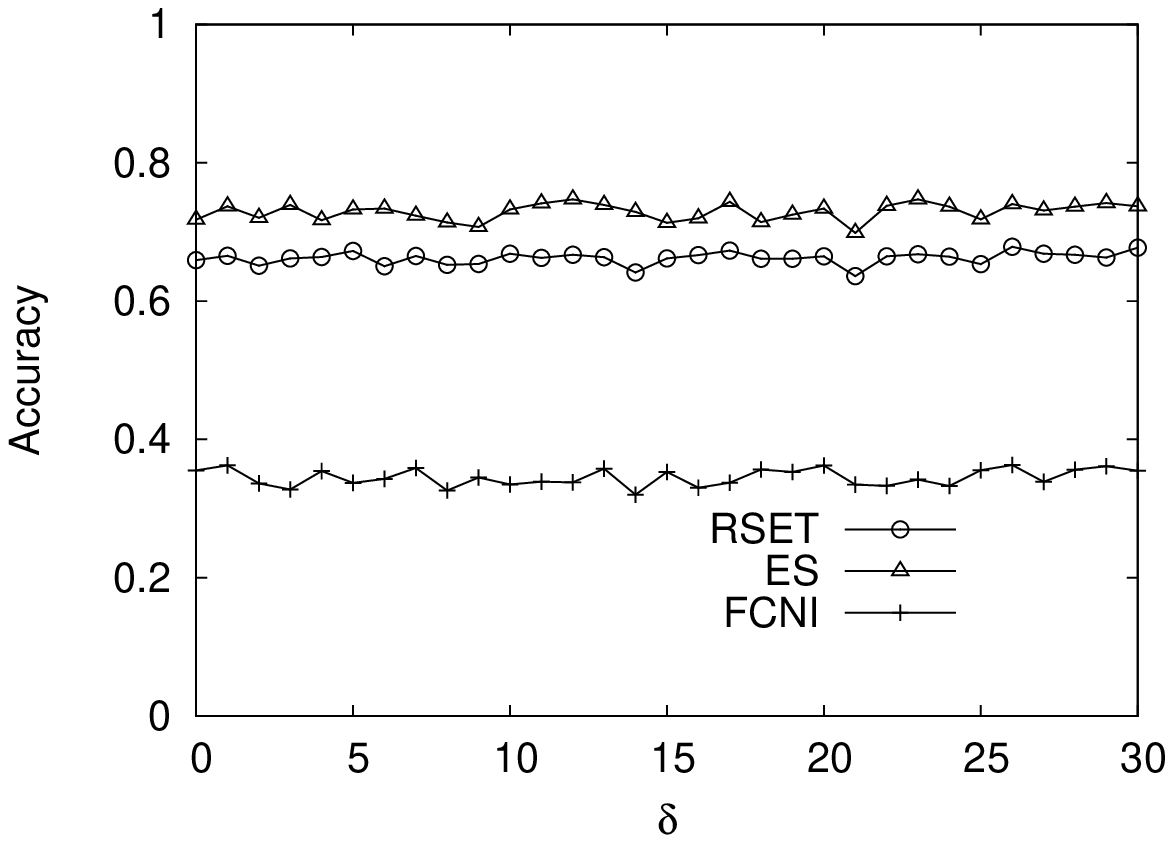}}
\quad
\subfigure[k = 10]{\includegraphics[scale=0.43]{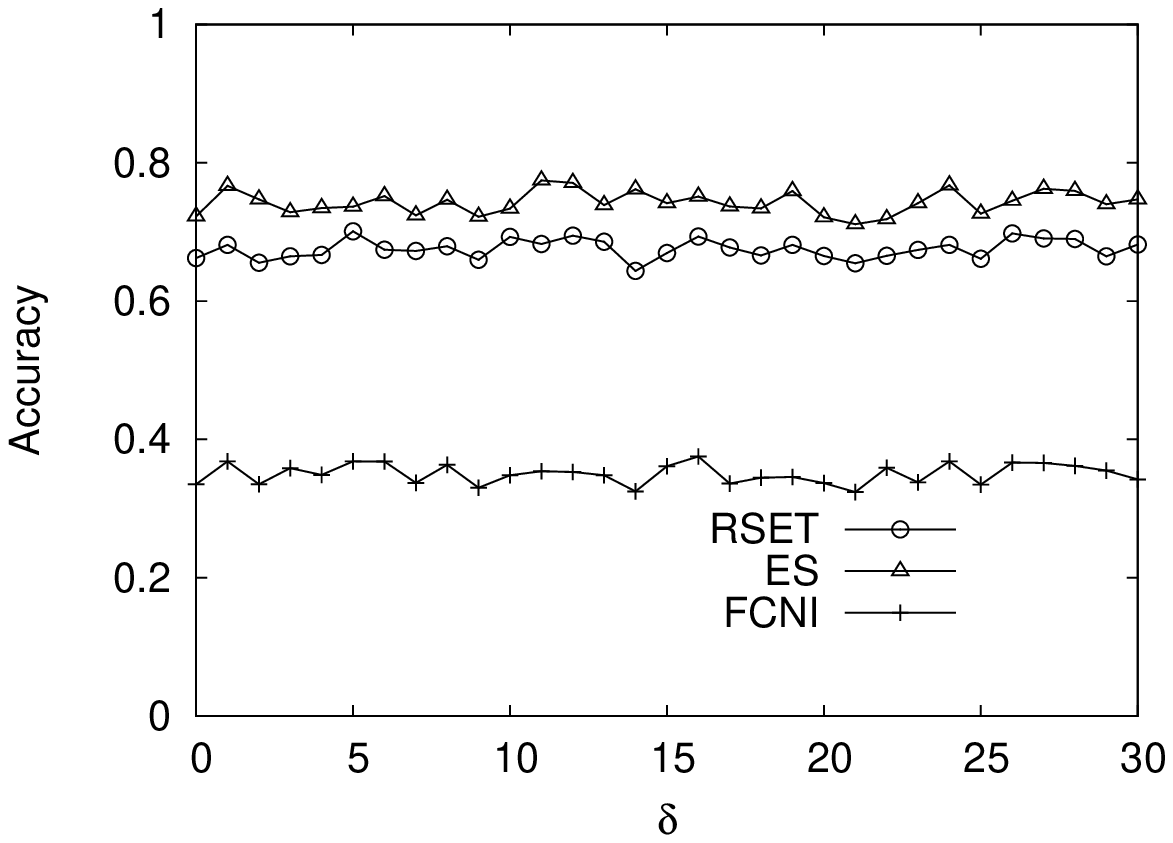}}
\quad
\subfigure[k = 15]{\includegraphics[scale=0.43]{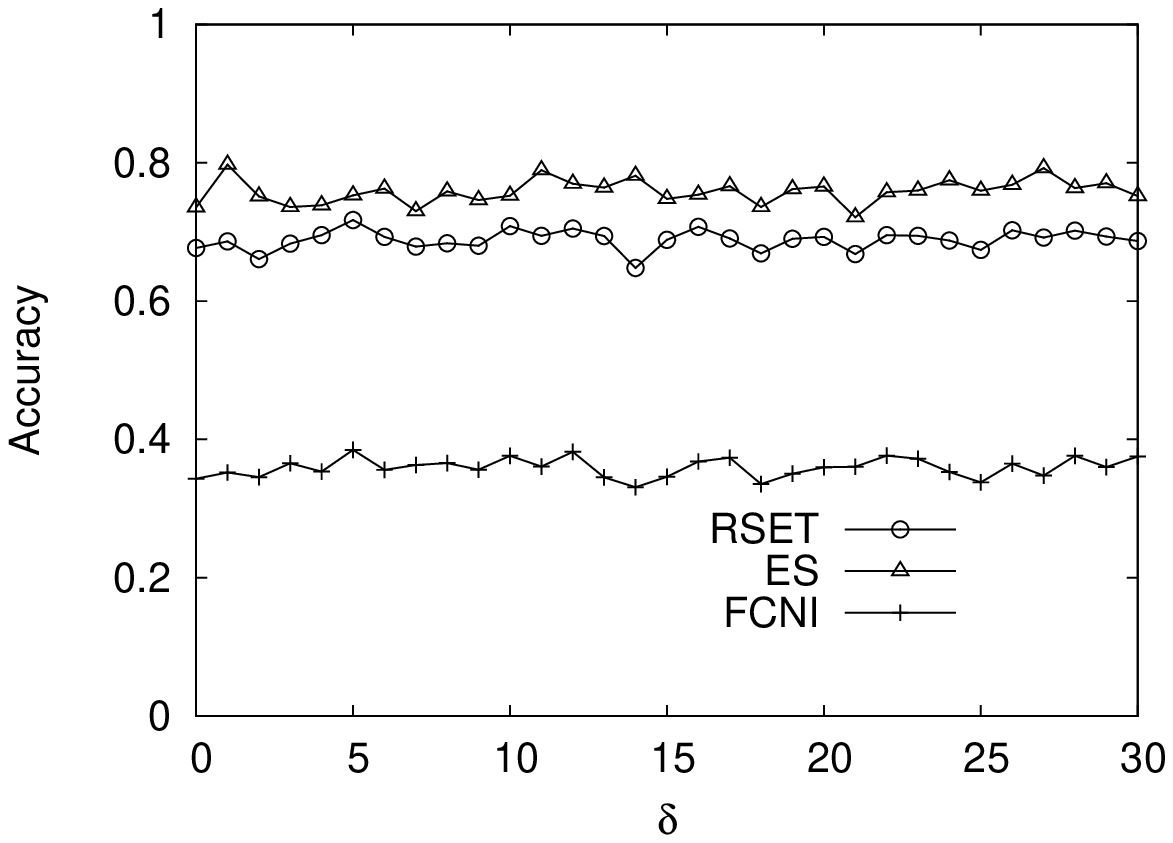}}
\quad
\subfigure[k = 20]{\includegraphics[scale=0.43]{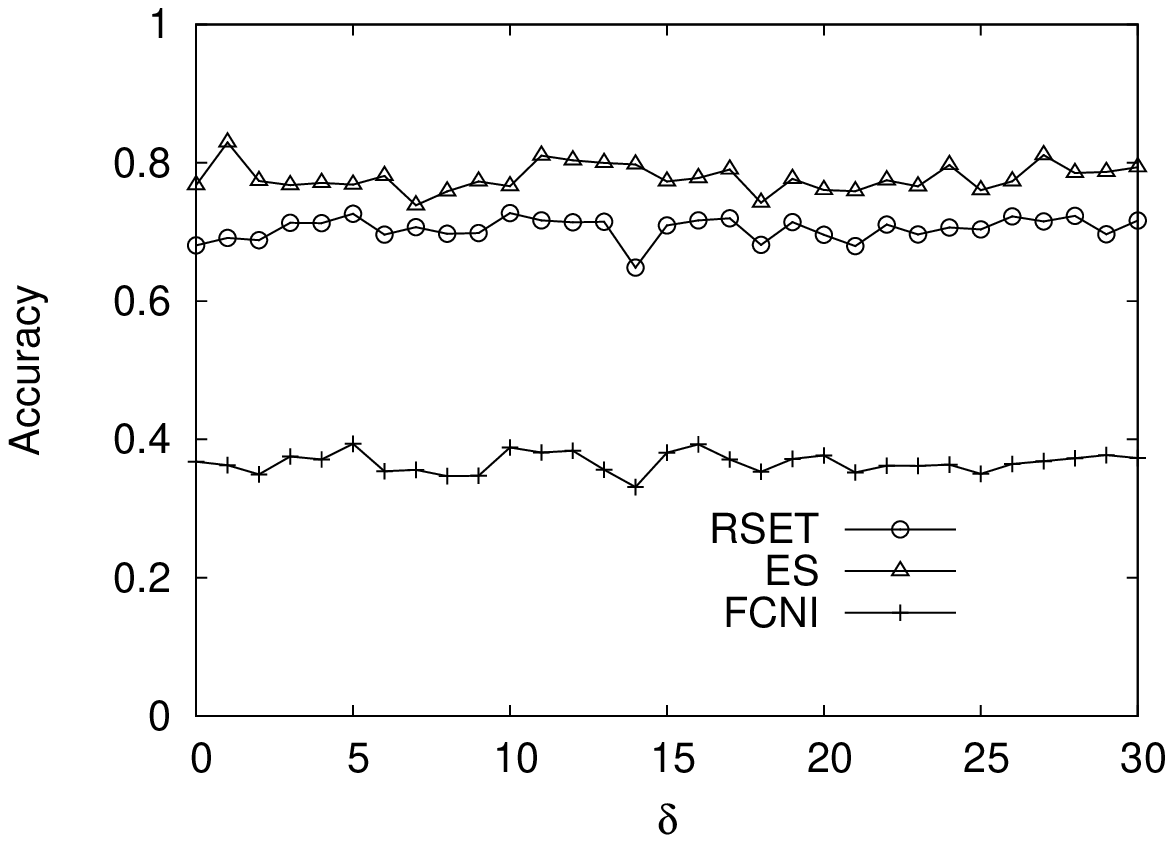}}
\caption{Hit-or-miss accuracies of the RSET, ES, and FCNI algorithms w.r.t EOP index ($\delta$) for the power grid dataset.}
\label{fig:exp1_powergrid_0_accuracy}
\end{figure*}

\begin{figure*}[t!]
\centering
\subfigure[k =1]{\includegraphics[scale=0.43]{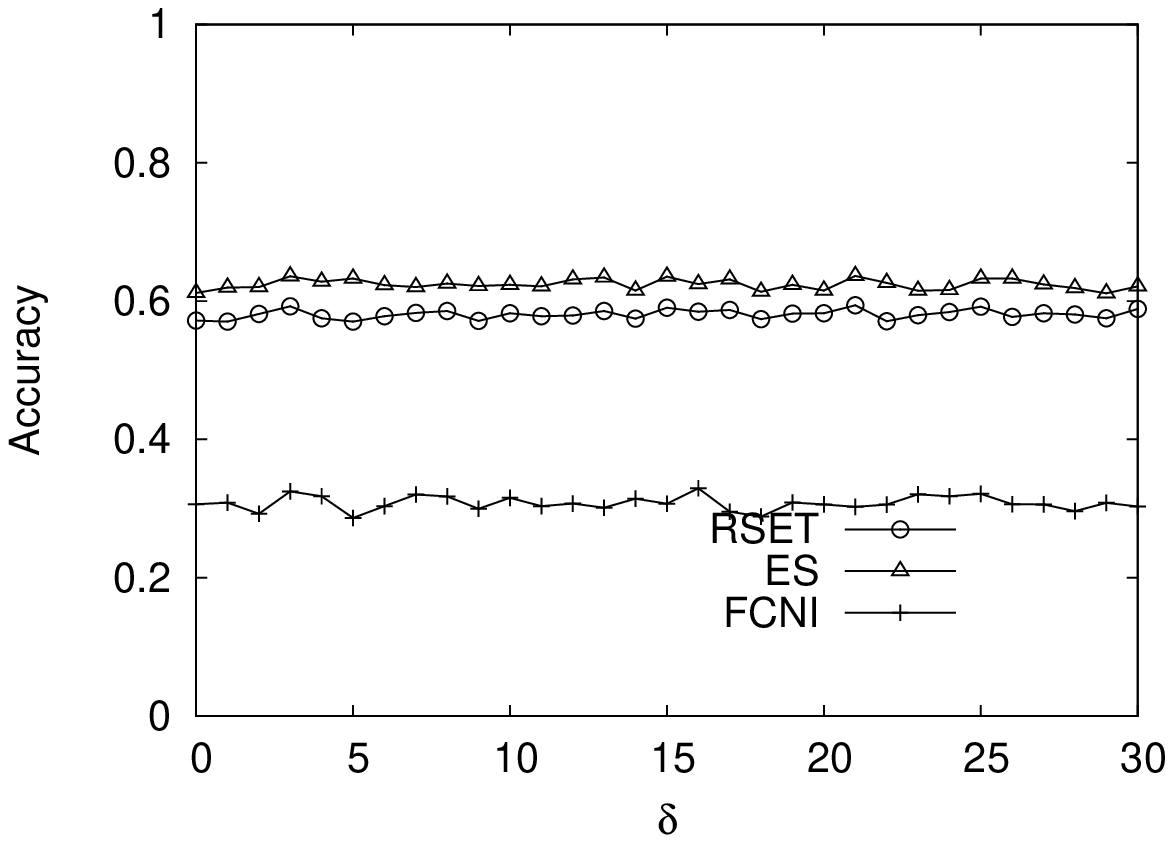}}
\quad
\subfigure[k = 5]{\includegraphics[scale=0.43]{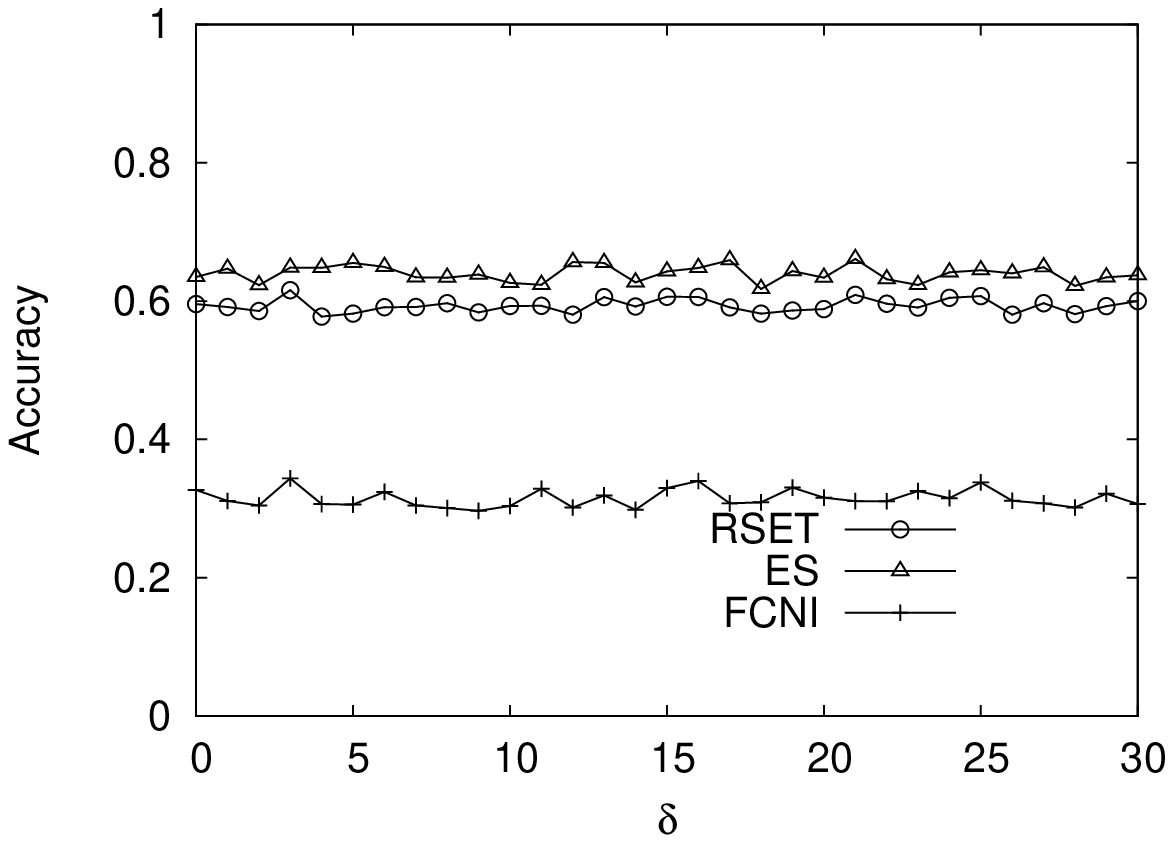}}
\quad
\subfigure[k = 10]{\includegraphics[scale=0.43]{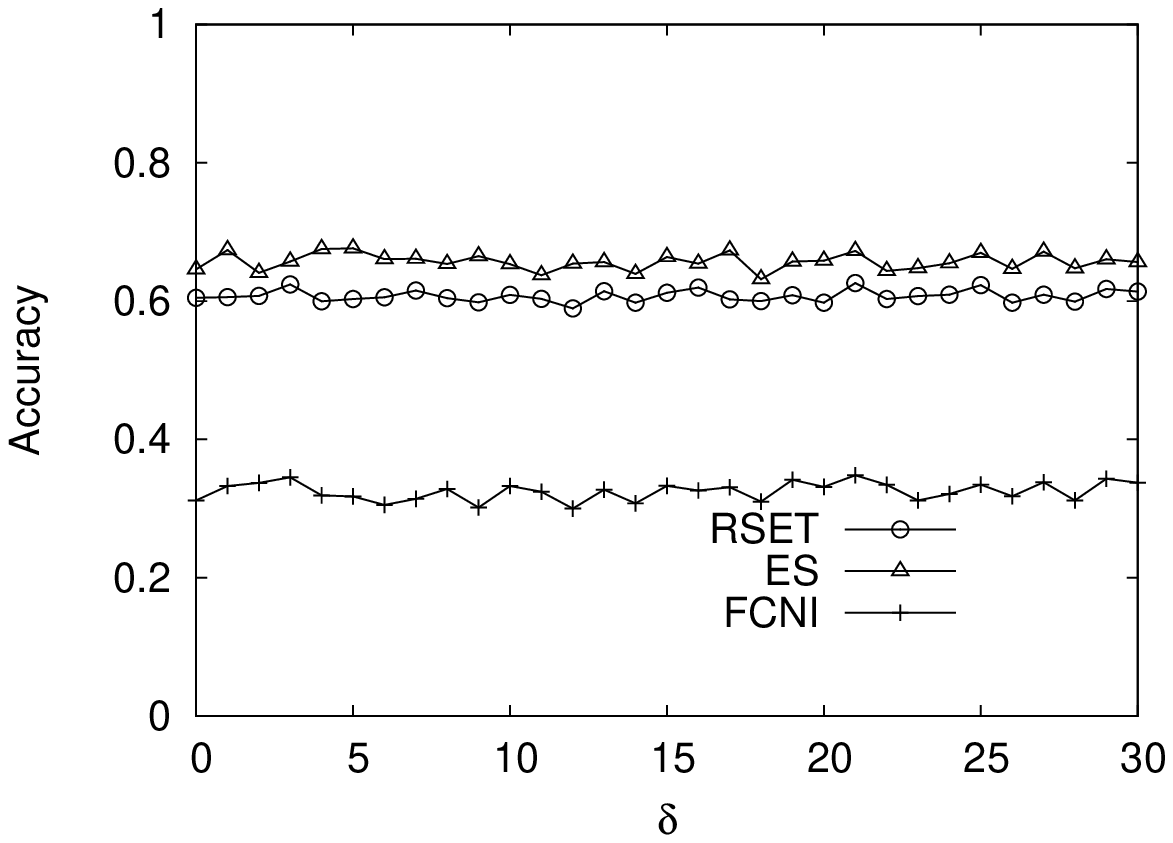}}
\quad
\subfigure[k = 15]{\includegraphics[scale=0.43]{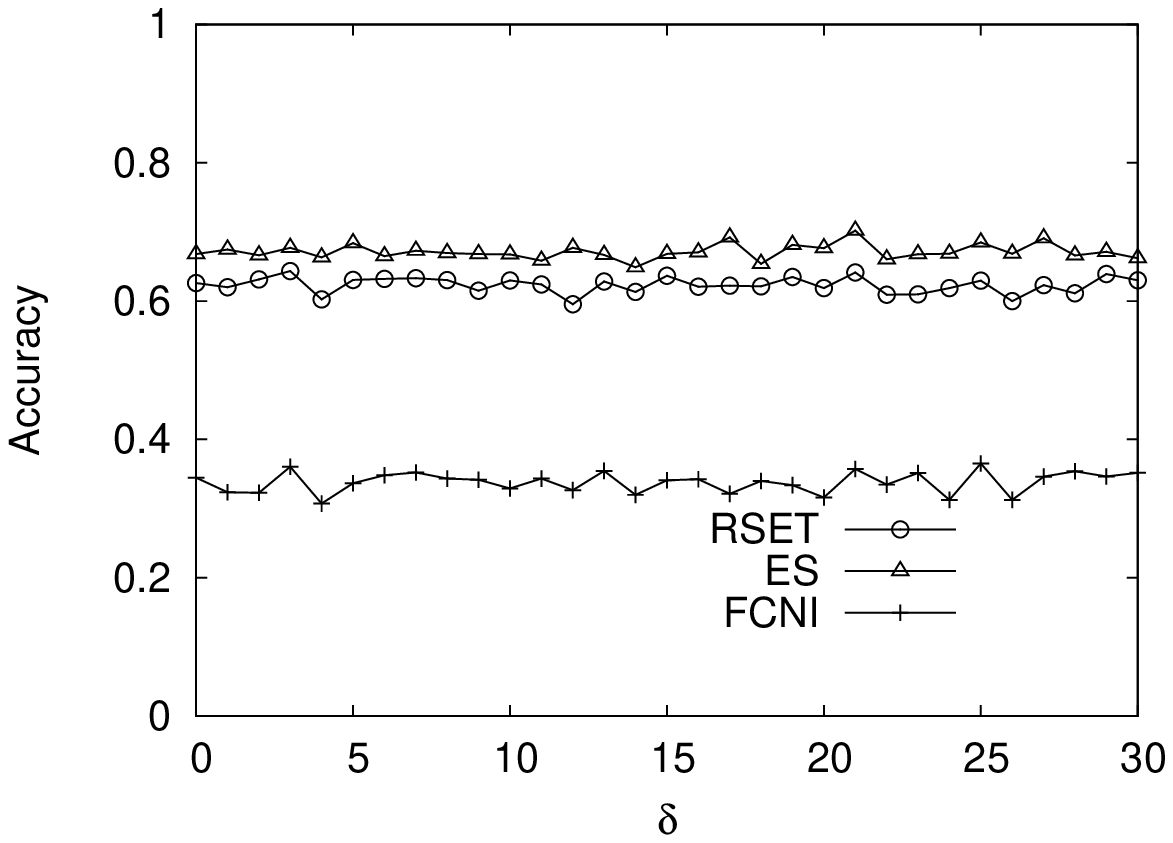}}
\quad
\subfigure[k = 20]{\includegraphics[scale=0.43]{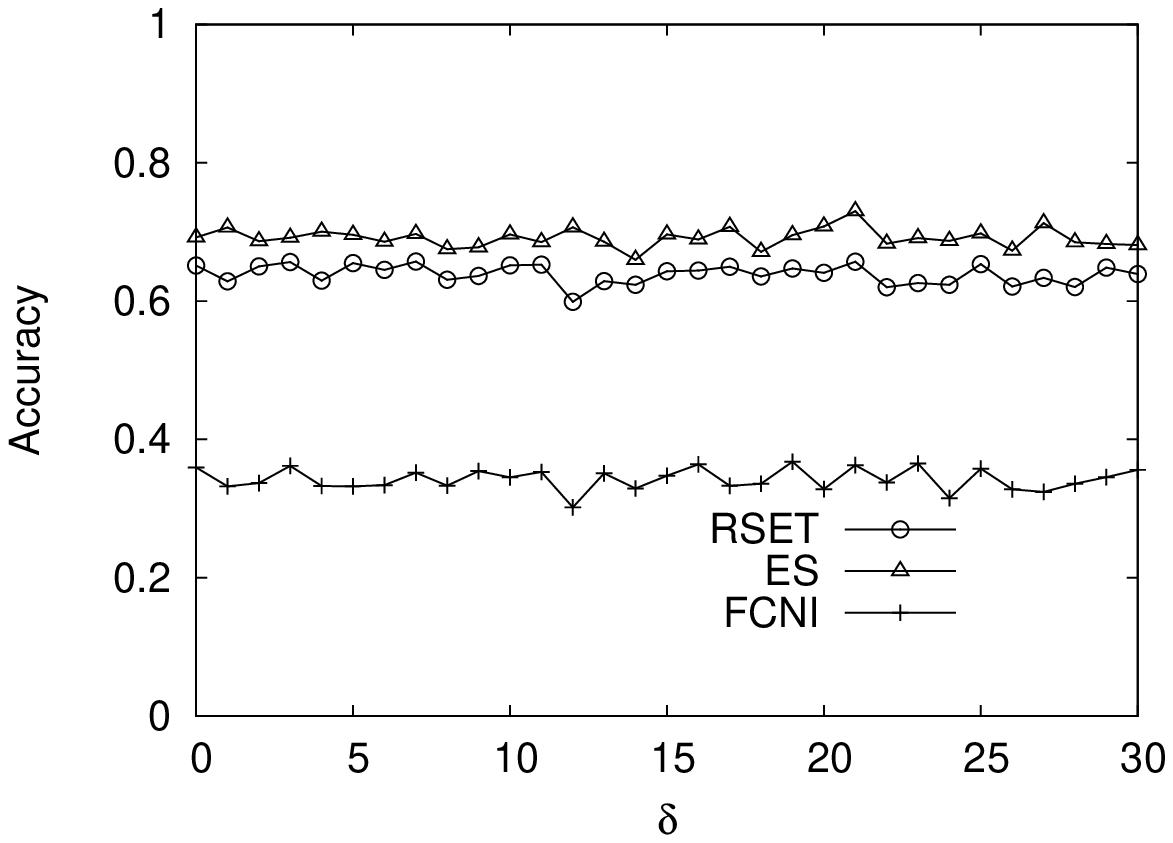}}
\caption{Weighted accuracies of the RSET, ES, and FCNI algorithms w.r.t EOP index ($\delta$) for the power grid dataset.}
\label{fig:exp1_powergrid_1_accuracy}
\end{figure*}

\begin{figure*}[t!]
\centering
\subfigure[hit-or-miss]{\includegraphics[scale=0.48]{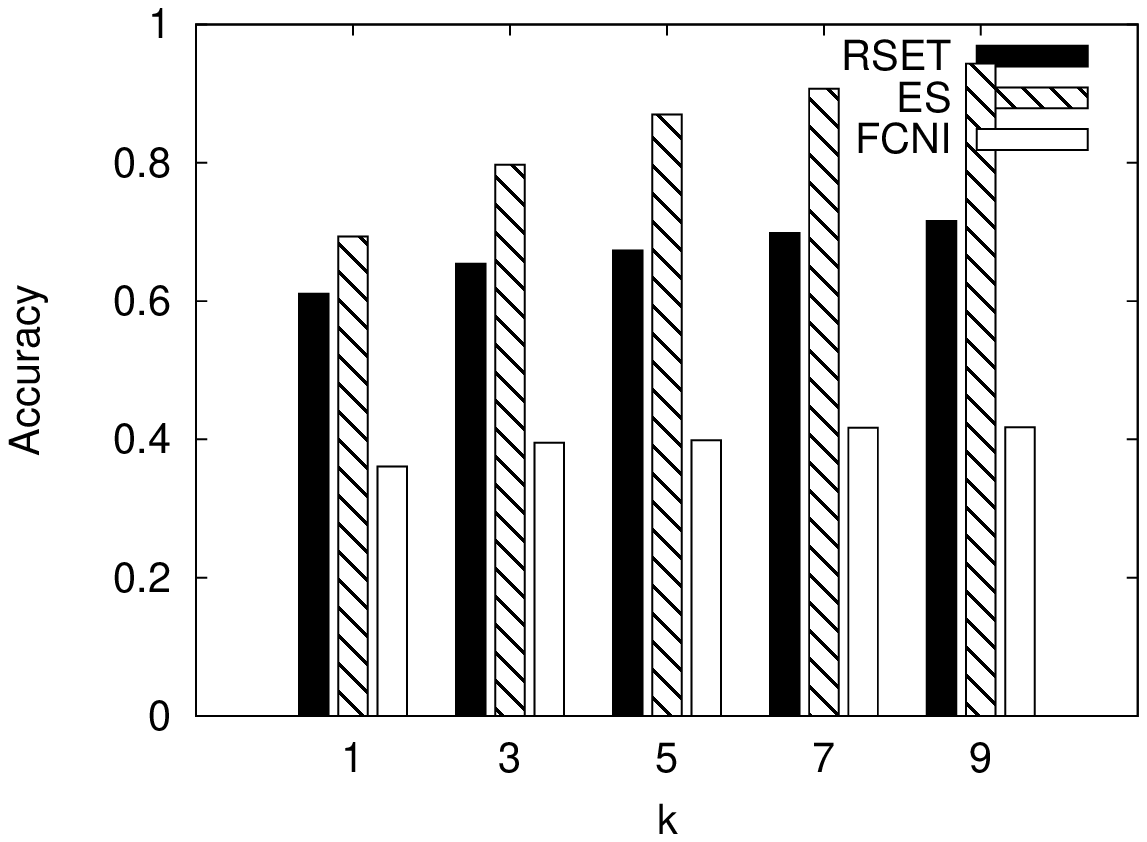}}
\subfigure[weighted]{\includegraphics[scale=0.48]{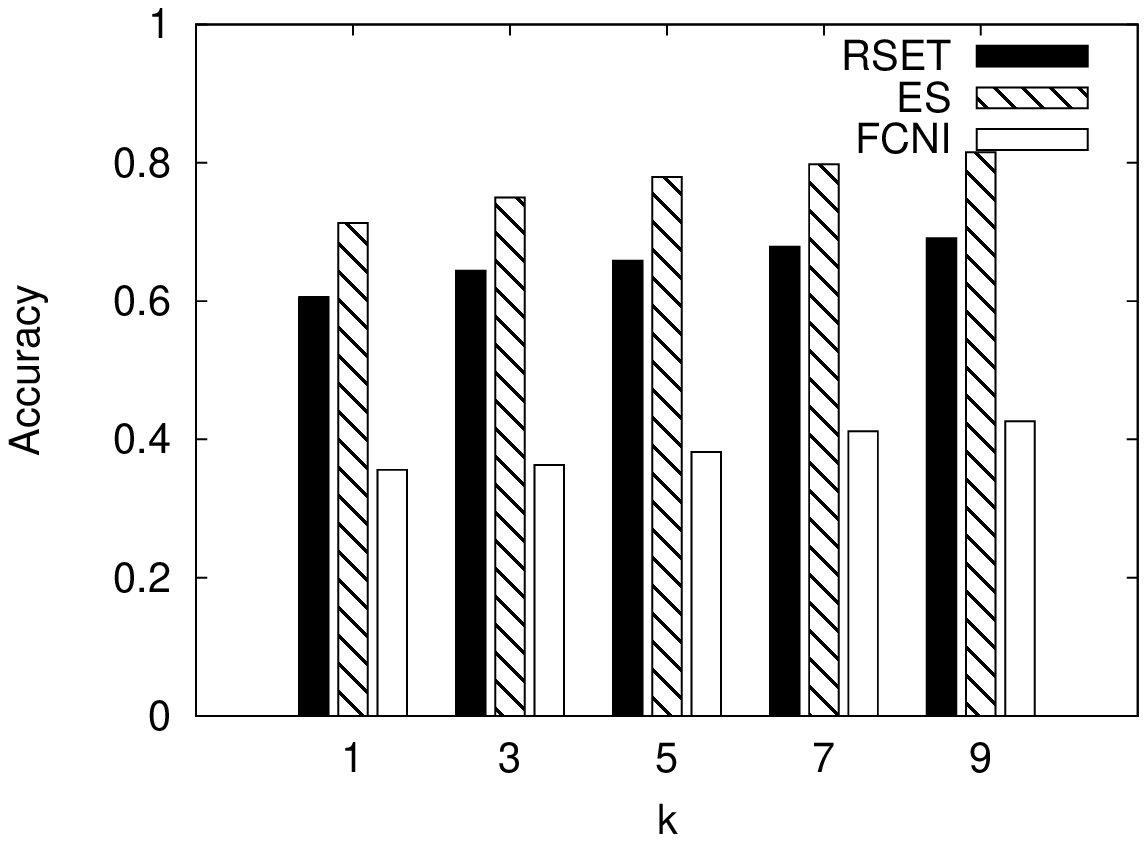}}
\caption{Accuracies of the RSET, ES, and FCNI algorithms w.r.t $k$ for MSNBC dataset.}
\label{fig:exp1_msnbc_k_accuracy}
\end{figure*}

\begin{figure*}[t!]
\centering
\subfigure[hit-or-miss]{\includegraphics[scale=0.48]{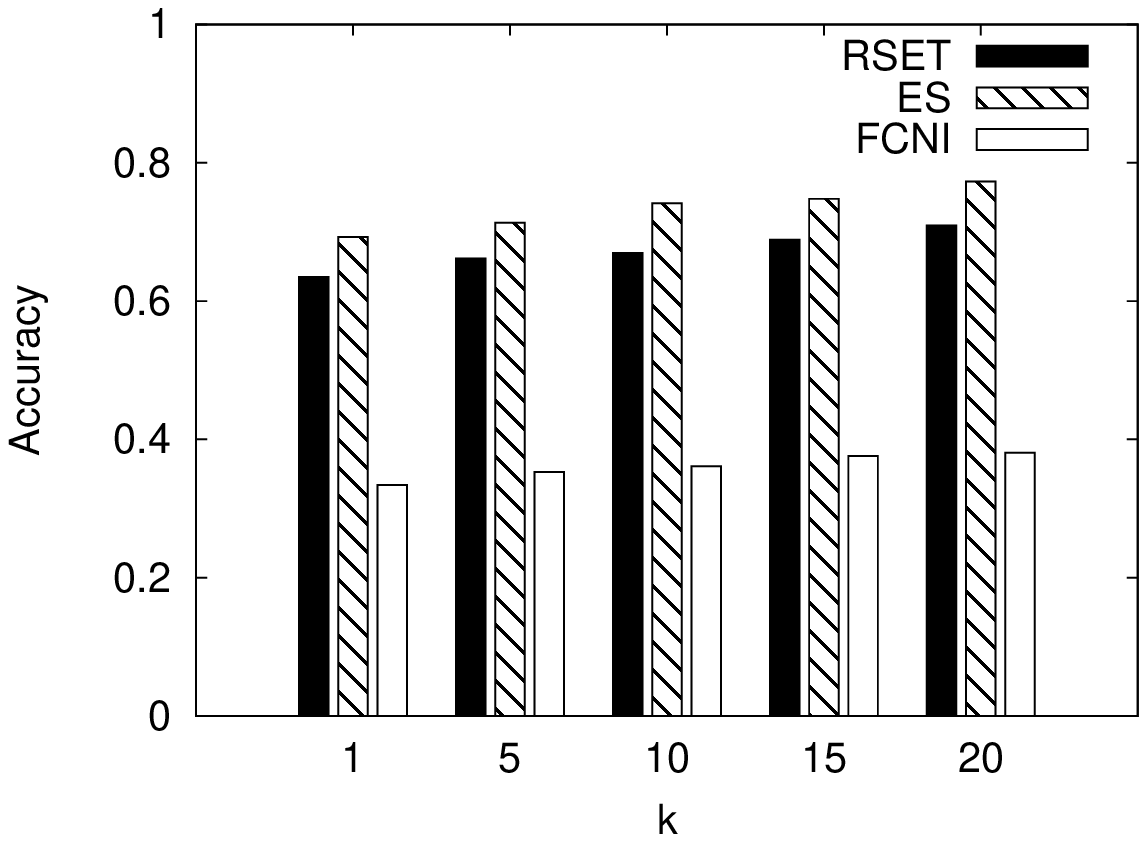}}
\subfigure[weighted]{\includegraphics[scale=0.48]{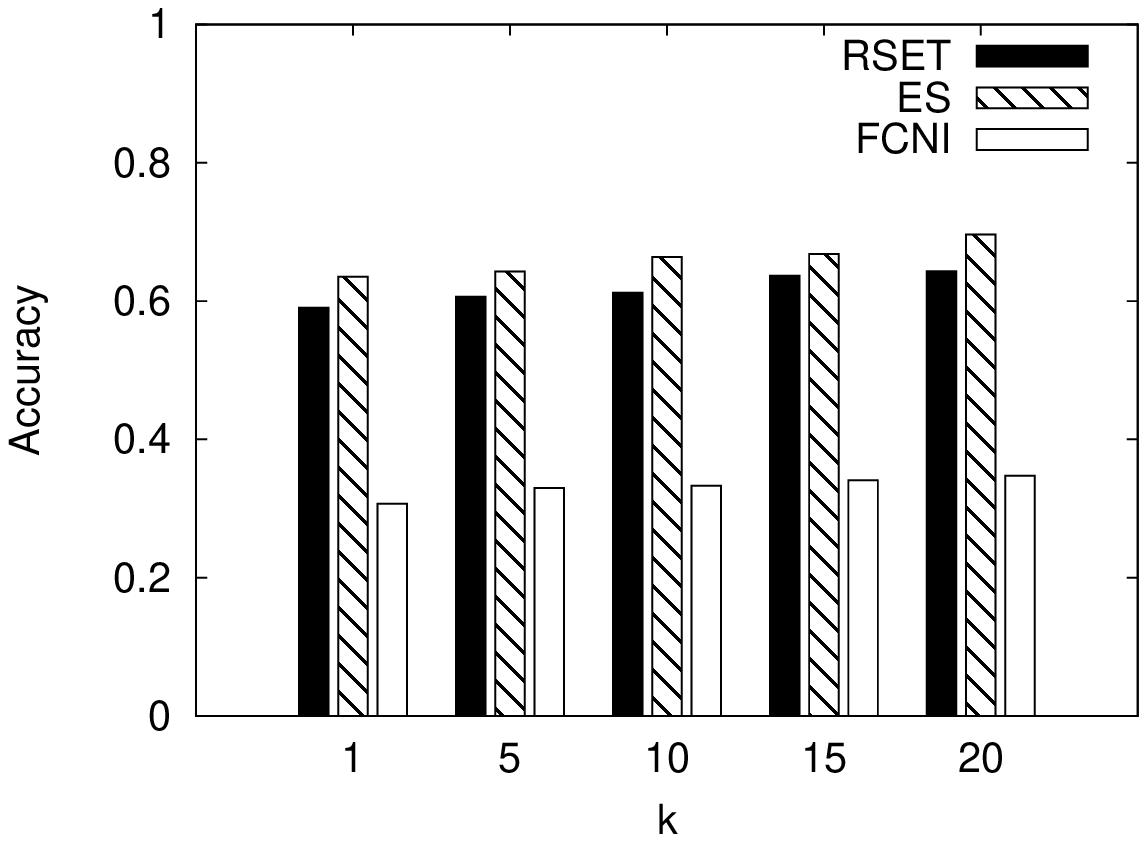}}
\caption{Accuracies of the RSET, ES, and FCNI algorithms w.r.t $k$ for the power grid dataset.}
\label{fig:exp1_powergrid_k_accuracy}
\end{figure*}

Figures~\ref{fig:exp1_msnbc_0_accuracy} and \ref{fig:exp1_msnbc_1_accuracy}, respectively, show the hit-or-miss and the weighted accuracies for the MSNBC dataset and Figures~\ref{fig:exp1_powergrid_0_accuracy} and \ref{fig:exp1_powergrid_1_accuracy} show them for the power grid dataset. In these figures, the accuracies of the RSET algorithm, the ES algorithm, and the FCNI algorithm are compared for different values of the EOP index ($\delta$) over the sequence of events in the condition set. In addition, Figures~\ref{fig:exp1_msnbc_k_accuracy} and~\ref{fig:exp1_powergrid_k_accuracy} show the average hit-or-miss and weighted accuracies for different values of \textit{k} in the MSNBC dataset and the power grid dataset, respectively.

As expected, for all cases, the hit-or-miss accuracy is never lower than the weighted accuracy. Clearly, a hit in the hit-or-miss accuracy always receives the score of $1$ while a hit in the weighted accuracy receives a score lower than $1$ unless the observed event type in the test data has the highest score in the ranked list. When \textit{k} is one, the size of the result ranked list is one, hence Equation~\ref{eq:2_weighted_accuracy} reduces to Equation~\ref{eq:hit_or_miss_accuracy}, and therefore the two accuracy measures give the same value.

\subsubsection*{\textbf{Comparison of the causal inference mechanisms}}

All results show that the prediction accuracies of both ES and RSET algorithms are significantly higher than that of the FCNI algorithm at every EOP index ($\delta$). This difference in accuracy comes from the difference in their causal models. (Recall that for fairness the FCNI algorithm uses the same query processing mechanism used in the RSET algorithm.) Thus, it confirms the expectation that the traditional causal model (of the FCNI algorithm) is so limited due to its lack of support for cyclic causality and the loss of causal information that the prediction accuracy is compromised significantly. On the other hand, the RSET and ES algorithms both use run-time causal inference which can handle cyclic causality and causal information loss, thereby achieving higher prediction accuracies.

The results also show that the accuracy of all three algorithms increases with the increase of \textit{k}. The reason for this is that more event types are considered as the probable next effects. Moreover, the accuracy in the ES and the RSET algorithms is always higher than in the FCNI algorithm. This indicates that the deficiencies of the FCNI algorithm (i.e., acyclic causality and causal information loss) leads to excluding many important causal relationships and, as a result, the accuracy is always much lower than that of the ES and the RSET algorithms regardless of the value of \textit{k}.

\subsubsection*{\textbf{Comparison of the query processing mechanisms}}

Three observations are made from the results.
First, all results show that the prediction accuracy of the ES algorithm is always higher than that of the RSET algorithm. This is evident from the fact that 
the ES algorithm performs an exhaustive search whereas the RSET algorithm performs only a partial search.

Second, the accuracy of the RSET algorithm is more comparable to that of the ES algorithm in the power grid dataset than in the MSNBC dataset. This can be explained as follows. When the ratio of $k$ to $N$ is larger, both algorithms have higher probabilities of making correct predictions but the gap between their accuracies is larger because ES performs exhaustive search while RSET performs partial. Note that $k/N$ is smaller in the power grid dataset (the largest $k/N$ considered is 0.035) than in the MSNBC dataset (the largest $k/N$ considered is 0.53), and therefore the gap is smaller for the power grid dataset.

Third, as discussed earlier, as the value of \textit{k} increases, the accuracies of all three algorithms increase. Here we add further evaluation on the ES algorithm and the RSET algorithm with a focus on their search mechanisms. As $k$ increases, the search space of the RSET algorithm increases, leading to a higher gain in the accuracy. In the case of the ES algorithm, the search space remains constant regardless of $k$, but the number of candidate effects from which the highest \textit{k} is selected increases and, consequently, the accuracy still increases. Intuitively, the rate of increase in the accuracy is higher for the hit-or-miss accuracy metric than the weighted accuracy metric in both algorithms.

\subsubsection{Runtime} \label{sec:exp2}

In this experiment, we compare the runtime among the three algorithms (RSET, ES, FCNI). In addition, we analyze the effects of \textit{k} on the runtime.

\begin{figure*}[t!]
\centering
\subfigure[k =1]{\includegraphics[scale=0.43]{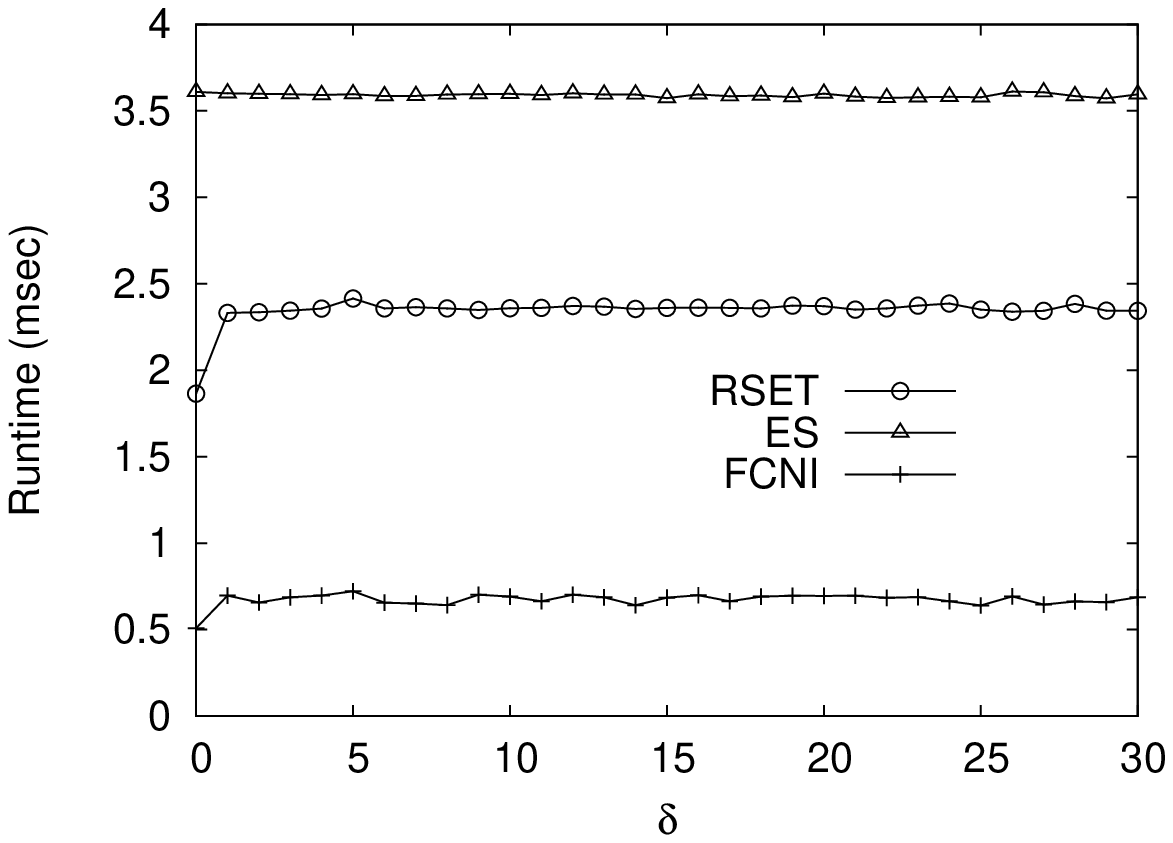}}
\quad
\subfigure[k = 3]{\includegraphics[scale=0.43]{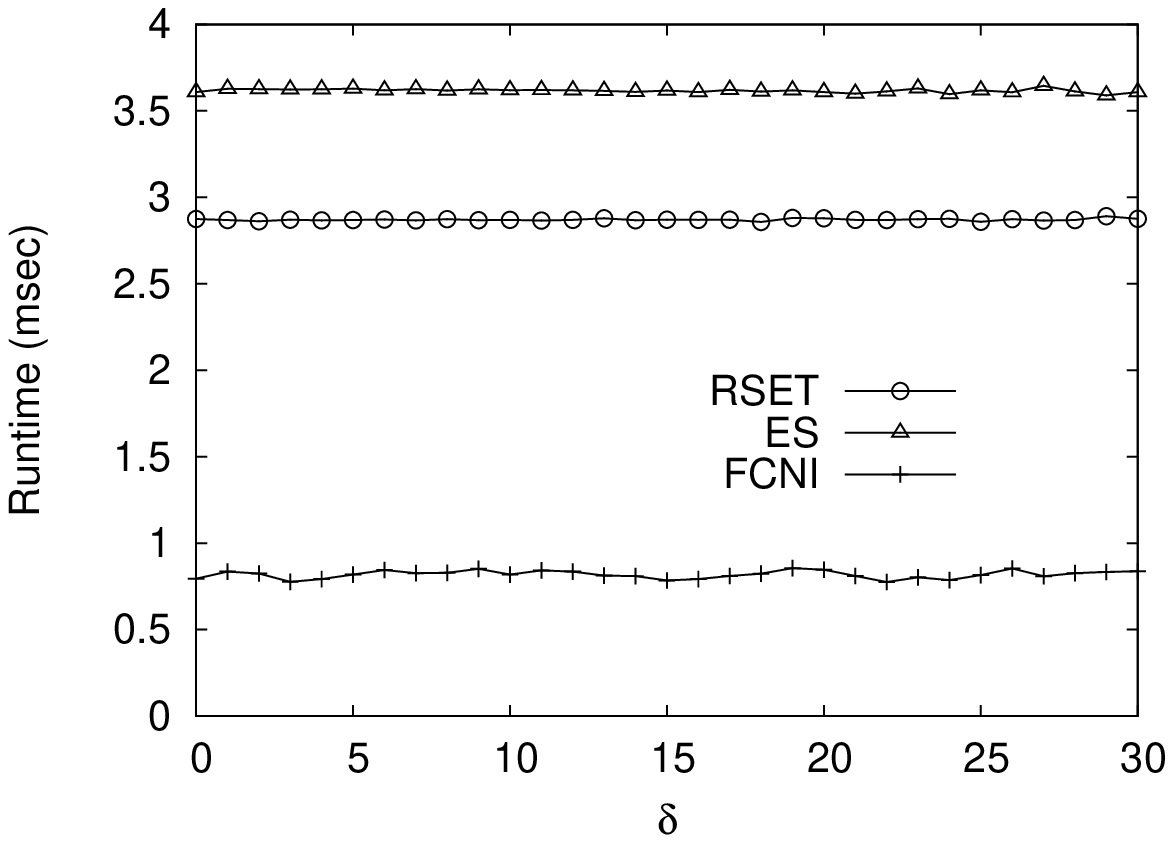}}
\quad
\subfigure[k = 5]{\includegraphics[scale=0.43]{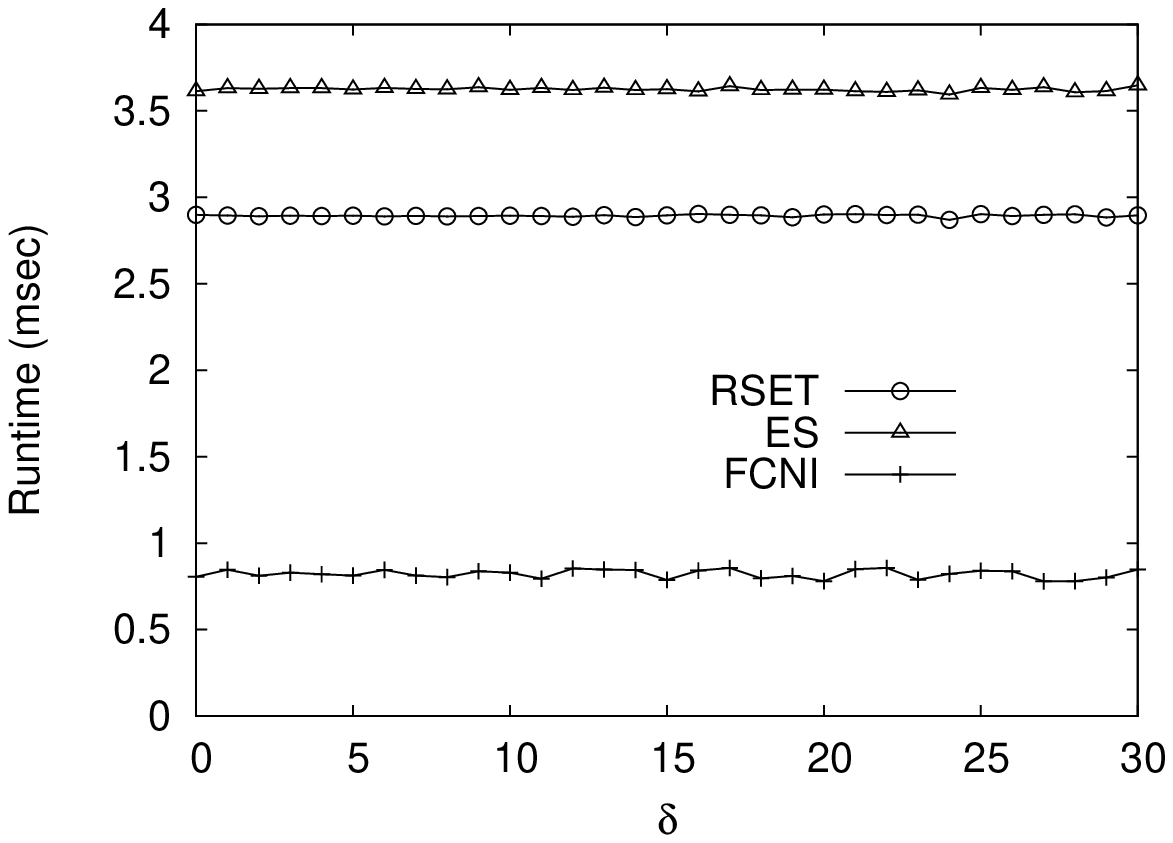}}
\quad
\subfigure[k = 7]{\includegraphics[scale=0.43]{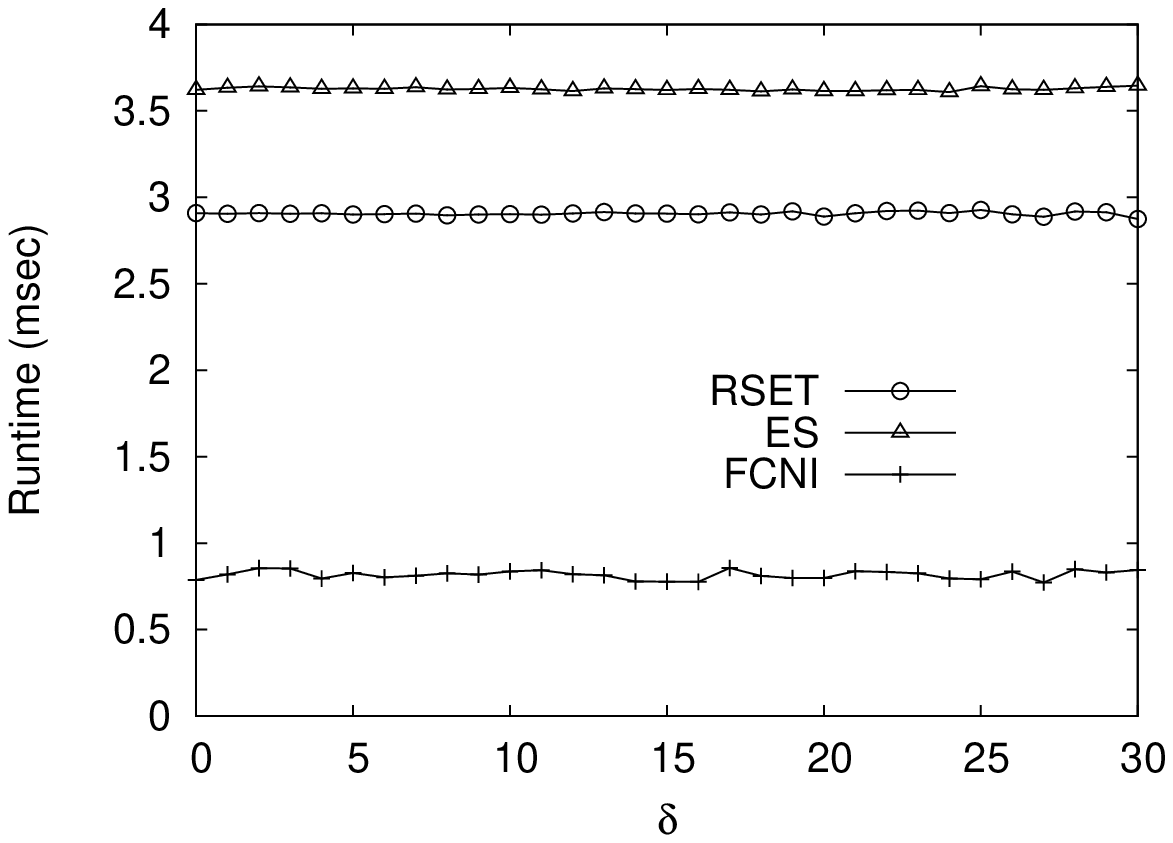}}
\quad
\subfigure[k = 9]{\includegraphics[scale=0.43]{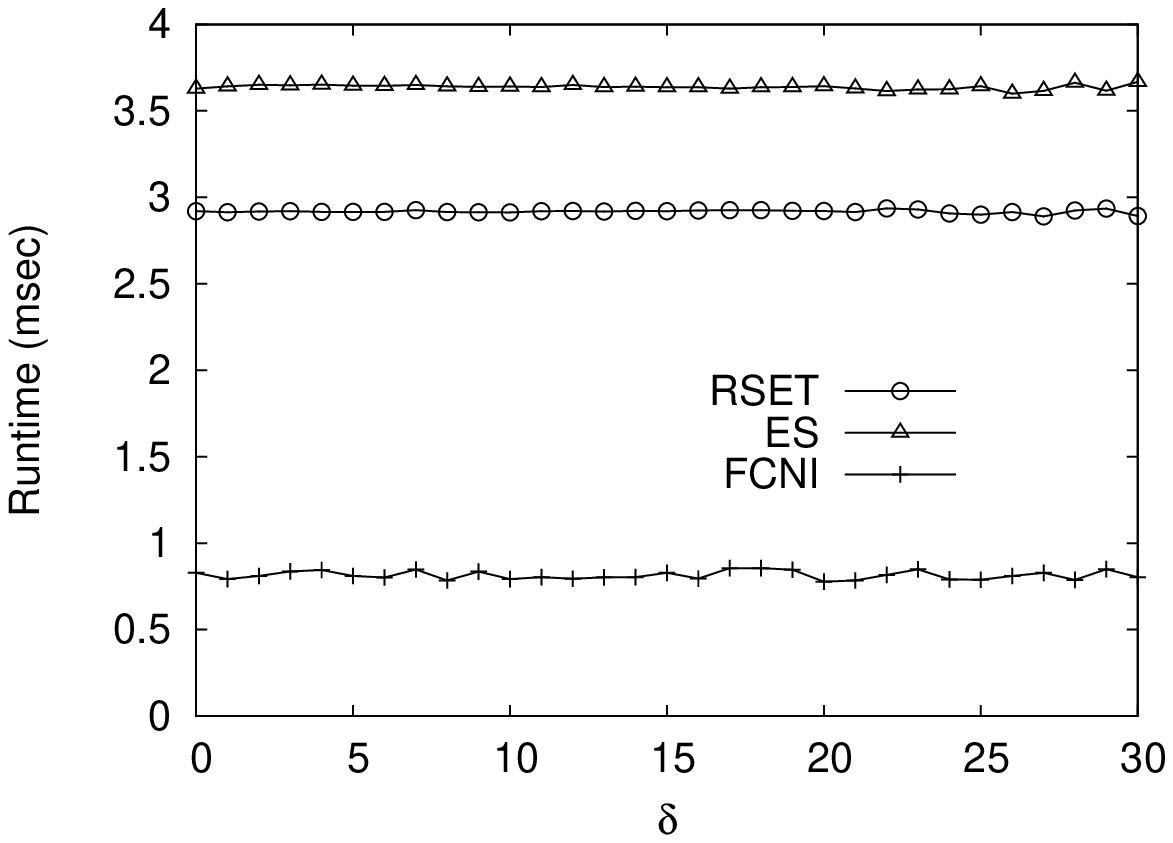}}
\caption{Runtime of the RSET, ES, and FCNI algorithms w.r.t EOP index ($\delta$) for MSNBC dataset.}
\label{fig:exp2_msnbc_1_runtime}
\end{figure*}

\begin{figure*}[t!]
\centering
\subfigure[k =1]{\includegraphics[scale=0.43]{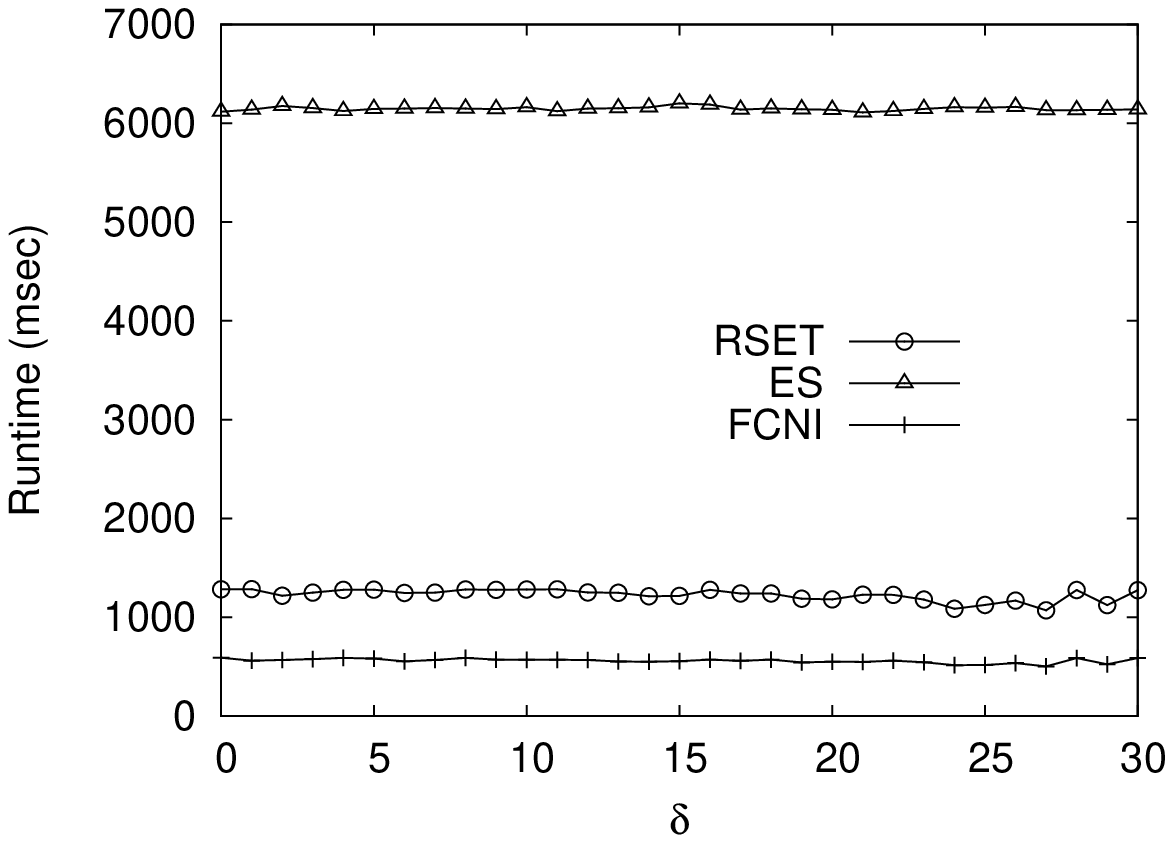}}
\quad
\subfigure[k = 5]{\includegraphics[scale=0.43]{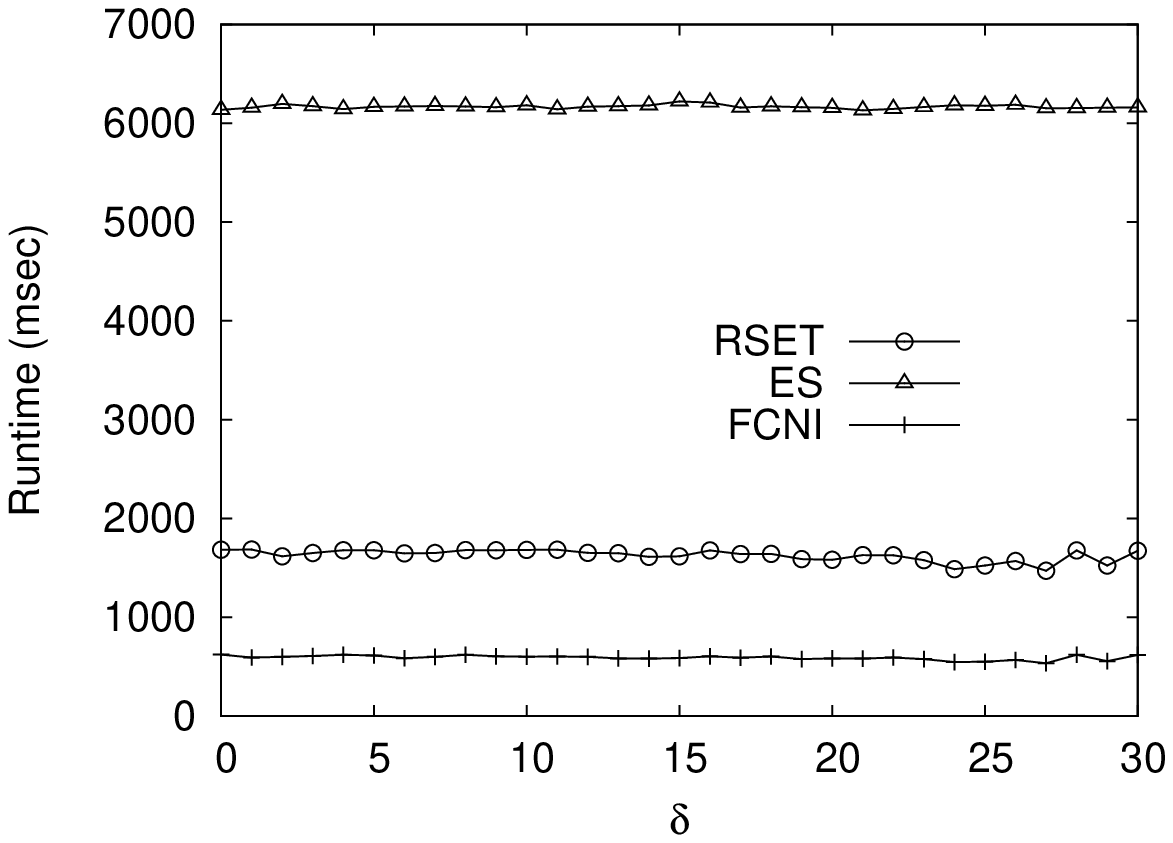}}
\quad
\subfigure[k = 10]{\includegraphics[scale=0.43]{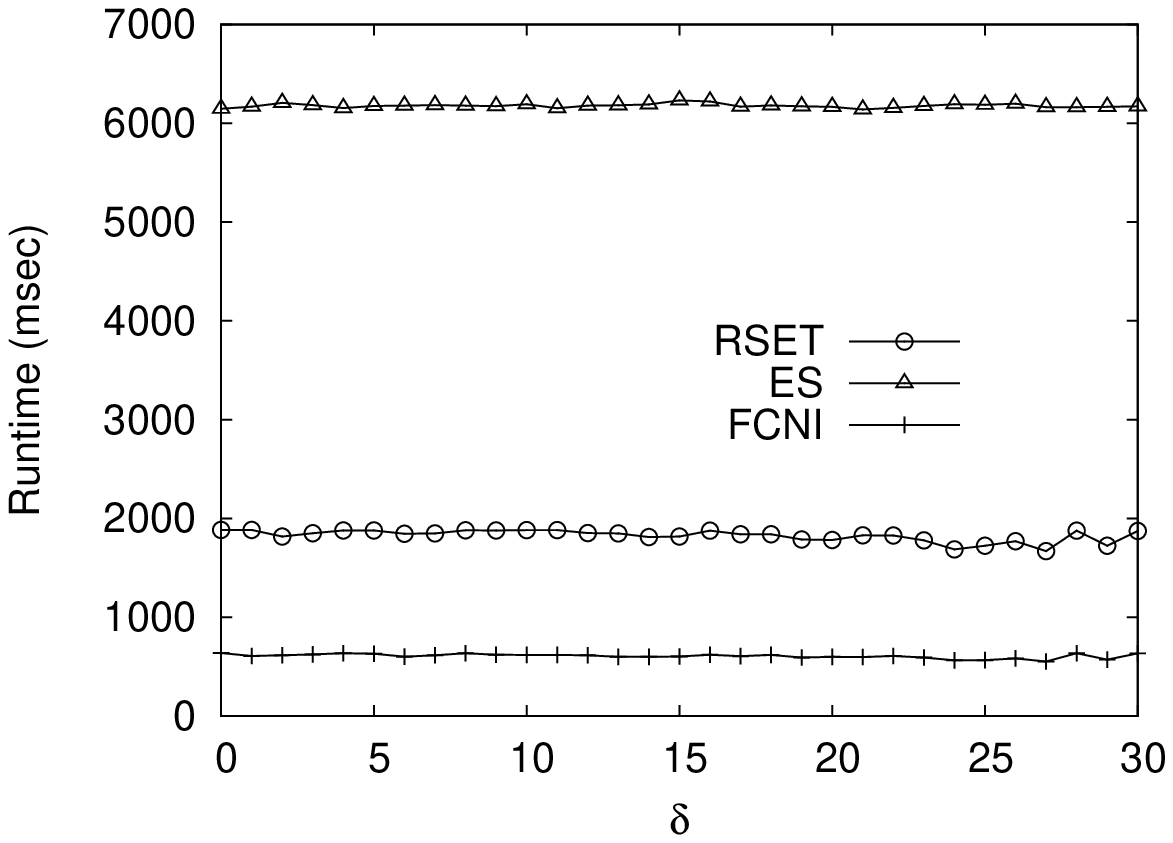}}
\quad
\subfigure[k = 15]{\includegraphics[scale=0.43]{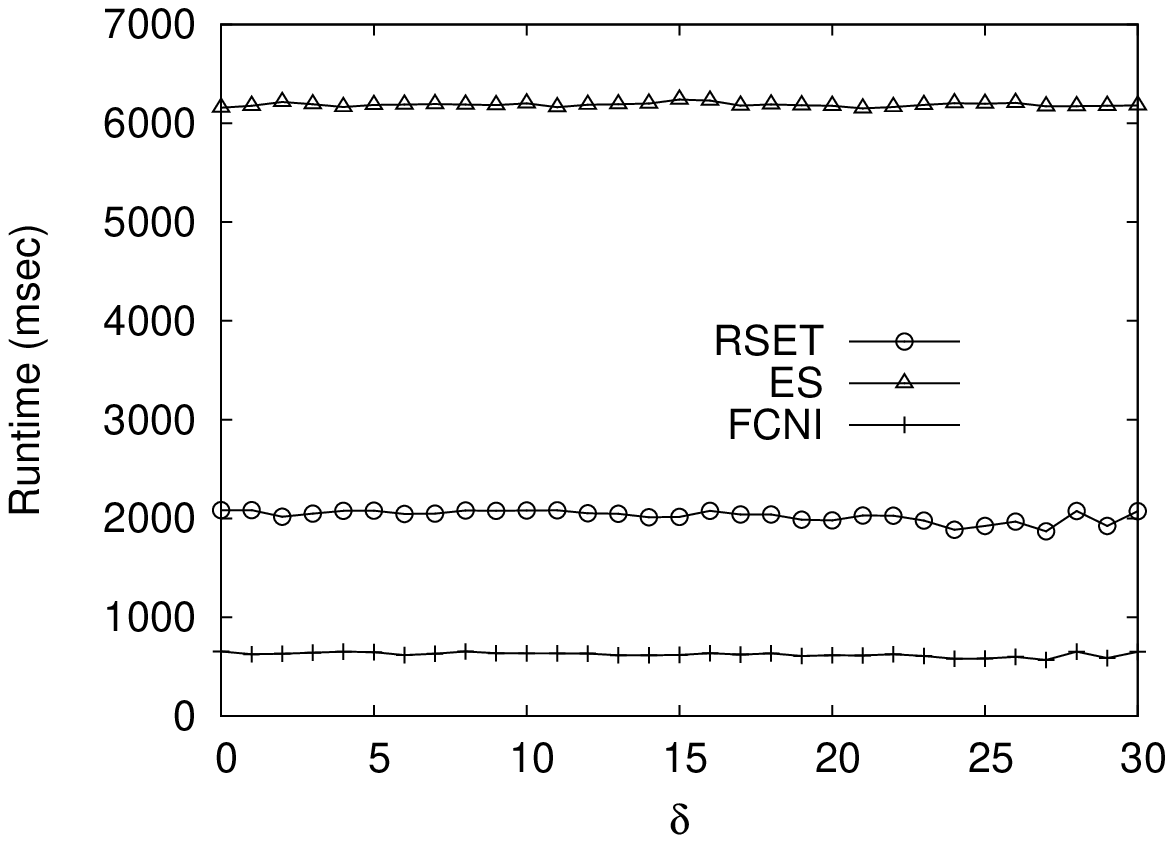}}
\quad
\subfigure[k = 20]{\includegraphics[scale=0.43]{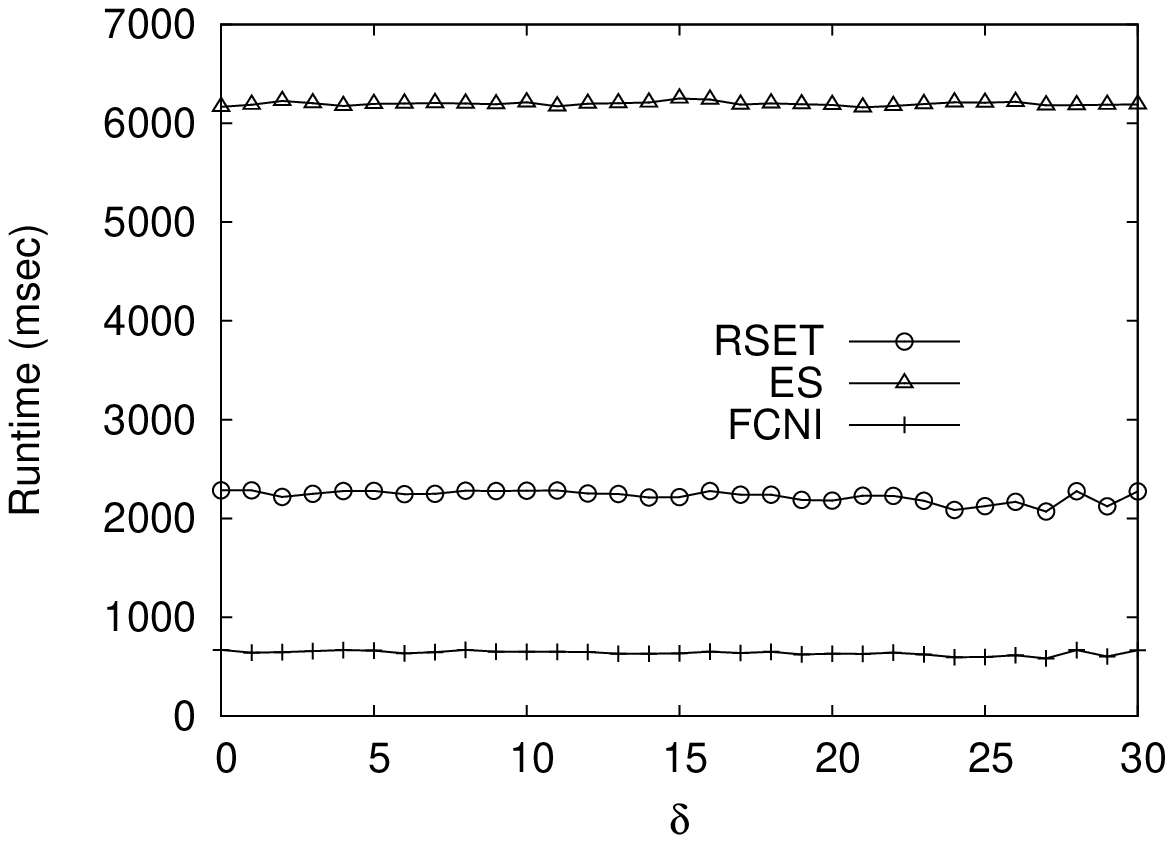}}
\caption{Runtime of the RSET, ES, and FCNI algorithms w.r.t EOP index ($\delta$) for the power grid dataset.}
\label{fig:exp2_powergrid_1_runtime}
\end{figure*}

\begin{figure*}[t!]
\centering
\subfigure[MSNBC]{\includegraphics[scale=0.48]{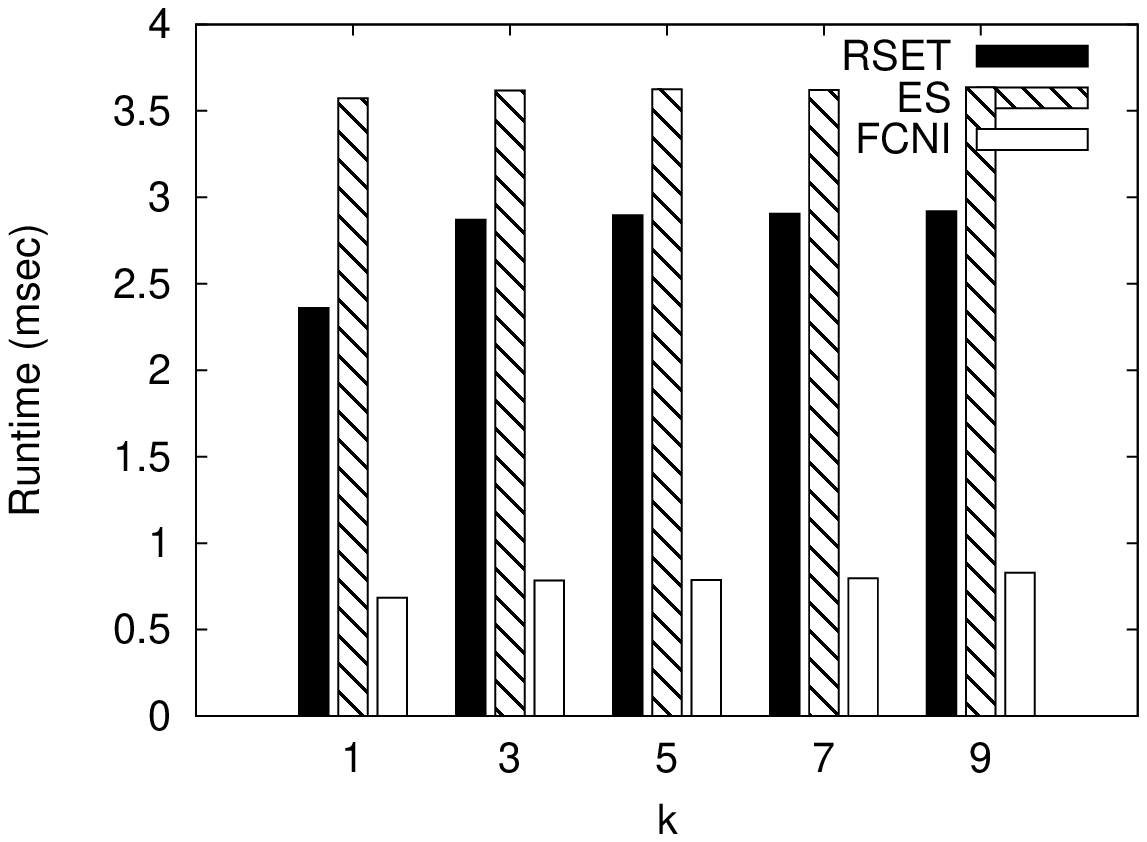}}
\subfigure[Powergrid]{\includegraphics[scale=0.48]{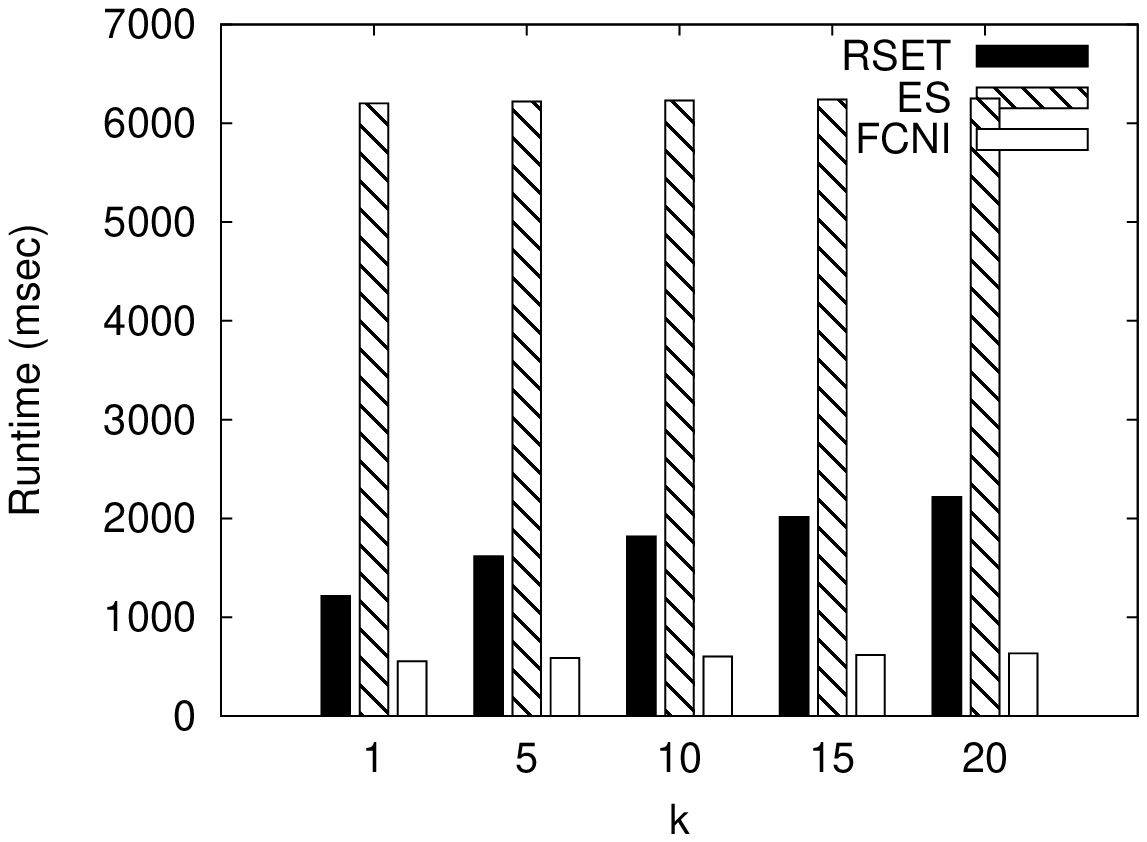}}
\caption{Runtime of the RSET, ES, and FCNI algorithms w.r.t $k$.}
\label{fig:exp2_k_runtime}
\end{figure*}

Figures~\ref{fig:exp2_msnbc_1_runtime} and \ref{fig:exp2_powergrid_1_runtime} show the runtime for the MSNBC dataset and the power grid dataset, respectively. In these figures, the runtime of the three algorithms are compared for different values of the EOP index ($\delta$) over the sequence of events in the condition set. In addition, Figure~\ref{fig:exp2_k_runtime} shows the runtime for different values of \textit{k} in the MSNBC and the power grid datasets.

\subsubsection*{\textbf{Comparison of the causal inference mechanisms}. }

The results show that the runtimes of the RSET and ES algorithms are longer than that of the FCNI algorithm at every EOP index for every value of $k$. As discussed in Section~\ref{sec:rset}, the RSET and ES algorithms have an overhead of the run-time causal inference during query processing while FCNI algorithm does not as it uses a pre-built causal network for prediction. Therefore, the runtimes for the ES and RSET algorithms are always longer than that of the FCNI algorithm.
Interestingly, the runtimes of the three algorithms are longer in the power grid dataset than the MSNBC dataset. This is due to a larger number of event types (i.e., N) in the power grid dataset.

\subsubsection*{\textbf{Comparison of the query processing mechanisms}. }
The results suggest three observations.
First, as expected, the runtime of the RSET algorithm is always shorter than the ES algorithm. The main reason is in the differen search scope (i.e., exhaustive in ES and partial in RSET) as discussed in Section~\ref{sec:query_processing}. In addition, there is an overhead in the ES algorithm, unlike the RSET algorithm, to calculate the causal search order.

Second, the runtime difference between the RSET algorithm and the ES algorithm is smaller for the MSNBC dataset (Figure~\ref{fig:exp2_msnbc_1_runtime}) than for the power grid dataset (Figure~\ref{fig:exp2_powergrid_1_runtime}). This is due to the larger network size (i.e., $N$) in the power grid dataset (than in the MSNBC dataset). That is, a larger network size results in a larger search space (which is typical over event streams) and thus requires longer runtime for query processing.

Third, the runtime of the ES algorithm does not change with an increase in \textit{k}. The ES algorithm always runs an exhaustive search, irrespective of the value of \textit{k}, and uses \textit{k} only to filter out the top-k event types out of $N$ event types at the end, which has insignificant effect on the overall runtime. On the other hand, the runtime of the RSET algorithm does increase with an increase in \textit{k}. The search space covered by the RSET algorithm increases with \textit{k}, leading to the increased runtime.

\subsubsection{Discussion of experiment results}

The proposed run-time causal inference mechanism, in the RSET and ES algorithms, handles cyclic causality and avoids the causal information loss, and thus improves prediction accuracy significantly. The FCNI algorithm, on the other hand, performs worse than both the RSET and ES algorithms as it cannot handle cyclic causality. The ratios of the number of cycles over the number of edges in the EPN are $0.69$ and $0.85$ for the power grid dataset and the MSNBC dataset, respectively. Intuitively, the accuracy of the FCNI algorithm would suffer increasingly more as the number of cycles increases. Thus, despite its reduced runtime, the FCNI algorithm is not suitable for top-k predictive query processing over event streams.

The accuracy of the RSET algorithm is comparable to the accuracy of the ES algorithm. Therefore, the RSET algorithm, as it is much faster than the ES algorithm, is more suitable for \emph{real-time} continuous top-k query processing over event streams whereas the ES algorithm is more suitable when the time is of lesser importance. {This will become increasingly evident for real-time applications with hundreds (or possibly thousands) of event types because then the pruning effect of reduced search and early termination becomes increasingly more significant.}

Between the two datasets, the runtime for the power grid dataset is much longer than the runtime for the MSNBC dataset. This is due to the difference in the numbers of event types in these two datasets. That is, the much larger number of event types in the power grid dataset leads to much more conditional independence tests during the run-time causal inference, thus resulting in slower query execution. In our work, the runtime measurements were made on a low-end laptop. The use of a more powerful computational setup (e.g., parallel processing) would further reduce the runtime.

\section{Conclusion and Future Work} \label{sec:conclusions}
This paper focused on the problem of continuous top-k prediction over event streams. We proposed a novel causal inference mechanism, called run-time causal inference, to support the cyclic causal relationships and overcome the causal information loss. Then, we proposed two query processing algorithms, called the \textit{Reduced Search with Early Termination (RSET)} algorithm and the \textit{Exhaustive Search (ES)} algorithm, which use run-time causal inference to predict top-k effects continuously. Finally, through experiments, we demonstrated that the proposed approach overcomes the two main limitations of the traditional causal inference approach -- acyclic causality and causal information loss. We showed that the proposed RSET and ES algorithms greatly improved the causal inference power for real data seen these days in various applications.

There are a number of issues we plan to address in the future work. First, in this paper we assume events in a stream are in the correct temporal order. If, however, events arrive out of order, erroneous relationships are introduced in the event precedence network, thereby degrading the accuracy of prediction. One idea to deal with such an out-of-order stream is to allow for ambiguity in the edge direction by introducing, for example, undirected edges and then allow the algorithm to resolve edge directions at query processing time. Second, in this paper we only support \emph{direct} causality and, therefore, one level of prediction, under the assumption that an event is the most likely effect of only its immediately preceding event. Extended from this, supporting \emph{indirect} causality between events, thus multiple levels of causal prediction through a chain of intermediate events, would be interesting. Since the mechanism to compute the propagation of probabilities through the event precedence network is already in place (see Examples~\ref{example:ES} and~\ref{example:RSET}) this extension is not conceptually difficult. However, the computational cost, which increases exponentially with the number of prediction levels, may be a challenge. Third, in this paper event types are assumed to be \emph{provided} by domain experts.
Some applications may require that the EPN constructor automatically identify and define event types from event streams.

\bibliographystyle{abbrv}
\bibliography{RSET}

\begin{thebibliography}{10}

\bibitem{AcharyaL_DaWaK2013}
S.~Acharya and B.~Lee.
\newblock Fast causal network inference over event streams.
\newblock In {\em Data Warehousing and Knowledge Discovery}, volume 8057 of
  {\em Lecture Notes in Computer Science}, pages 222--235. Springer Berlin
  Heidelberg, 2013.

\bibitem{Akdere_DCP2010}
M.~Akdere, U.~\c{C}etintemel, and E.~Upfal.
\newblock Database-support for continuous prediction queries over streaming
  data.
\newblock {\em Proc. VLDB Endow.}, 3(1-2):1291--1301, Sept. 2010.

\bibitem{BishopFH_DMA75}
Y.~M. Bishop, S.~E. Fienberg, and P.~W. Holland.
\newblock {\em Discrete Multivariate Analysis: Theory and Practice}.
\newblock MIT Press, 1975.

\bibitem{BowesNGC_AAI2000}
J.~Bowes, E.~Neufeld, J.~Greer, and J.~Cooke.
\newblock A comparison of association rule discovery and bayesian network
  causal inference algorithms to discover relationships in discrete data.
\newblock In H.~Hamilton, editor, {\em Advances in Artificial Intelligence},
  volume 1822 of {\em Lecture Notes in Computer Science}, pages 326--336.
  Springer Berlin Heidelberg, 2000.

\bibitem{ChengM_JAIR00}
J.~Cheng and M.~J. Druzdzel.
\newblock Ais-bn: An adaptive importance sampling algorithm for evidential
  reasoning in large {Bayesian} networks.
\newblock {\em J. Artif. Intell. Res. (JAIR)}, 13:155--188, 2000.

\bibitem{ChengD_UAI01}
J.~Cheng and M.~J. Druzdzel.
\newblock Confidence inference in {Bayesian} networks.
\newblock In {\em Proceedings of the Seventeenth conference on Uncertainty in
  artificial intelligence}, UAI'01, pages 75--82, San Francisco, CA, USA, 2001.
  Morgan Kaufmann Publishers Inc.

\bibitem{ChengGKD_02}
J.~Cheng, R.~Greiner, J.~Kelly, D.~Bell, and W.~Liu.
\newblock Learning {Bayesian} networks from data: an information-theory based
  approach.
\newblock {\em Artificial Intelligence}, 137(1-2):43--90, 2002.

\bibitem{Chickering_JMLR2002}
D.~M. Chickering.
\newblock Learning equivalence classes of {Bayesian}-network structures.
\newblock {\em J. Machine Learning Research}, 2:445--498, 2002.

\bibitem{deCampos_JMLR2006}
L.~M. de~Campos.
\newblock A scoring function for learning {Bayesian} networks based on mutual
  information and conditional independence tests.
\newblock {\em J. Machine Learning Research}, 7:2149--2187, 2006.

\bibitem{EllisW_JASA2008}
B.~Ellis and W.~H. Wong.
\newblock Learning causal {Bayesian} network structures from experimental data.
\newblock {\em J. American Statistics Association}, 103(482):778--789, 2008.

\bibitem{ElloumiZ_biological2013}
M.~Elloumi and A.~Y. Zomaya.
\newblock {\em Biological Knowledge Discovery Handbook: Preprocessing, Mining
  and Postprocessing of Biological Data}, volume~23.
\newblock John Wiley \& Sons, 2013.

\bibitem{EppsteinH_PES2013}
M.~Eppstein and P.~Hines.
\newblock A {"Random Chemistry"} algorithm for identifying collections of
  multiple contingencies that initiate cascading failure.
\newblock {\em IEEE Transactions on Power Systems}, 27(3):1698--1705, Aug 2012.

\bibitem{FriedmanLNP_UBN2000}
N.~Friedman, M.~Linial, I.~Nachman, and D.~Pe'er.
\newblock Using {Bayesian} networks to analyze expression data.
\newblock In {\em Proceedings of the Fourth Annual International Conference on
  Computational Molecular Biology}, RECOMB '00, pages 127--135, New York, NY,
  USA, 2000. ACM.

\bibitem{Clark_TCS2003}
C.~Glymour.
\newblock {Learning, prediction and causal Bayes nets.}
\newblock {\em Trends in cognitive sciences}, 7(1):43--48, Jan. 2003.

\bibitem{GuyonACEP_SS08}
I.~Guyon, C.~F. Aliferis, G.~F. Cooper, A.~Elisseeff, J.-P. Pellet, P.~Spirtes,
  and A.~R. Statnikov.
\newblock Design and analysis of the causation and prediction challenge.
\newblock In {\em IEEE World Congress on Computational Intelligence Causation
  and Prediction Challenge}, pages 1--33, 2008.

\bibitem{Haghani_MA2010}
P.~Haghani, S.~Michel, and K.~Aberer.
\newblock The gist of everything new: Personalized top-k processing over web
  2.0 streams.
\newblock In {\em Proceedings of the 19th ACM International Conference on
  Information and Knowledge Management}, CIKM '10, pages 489--498, New York,
  NY, USA, 2010. ACM.

\bibitem{Heckerman_UAI1995}
D.~Heckerman.
\newblock A {Bayesian} approach to learning causal networks.
\newblock In {\em Proceedings of the Eleventh conference on Uncertainty in
  artificial intelligence}, UAI'95, pages 285--295, San Francisco, CA, USA,
  1995. Morgan Kaufmann Publishers Inc.

\bibitem{Heckerman_1999}
D.~Heckerman.
\newblock {UCI} machine learning repository
  [http://archive.ics.uci.edu/ml/datasets/msnbc.com+\\anonymous+web+data],
  1999.

\bibitem{Henrion_UAI91}
M.~Henrion.
\newblock Search-based methods to bound diagnostic probabilities in very large
  belief nets.
\newblock In {\em Proceedings of the Seventh conference on Uncertainty in
  Artificial Intelligence}, UAI'91, pages 142--150, San Francisco, CA, USA,
  1991. Morgan Kaufmann Publishers Inc.

\bibitem{Jiang_DKE2005}
X.-R. Jiang and L.~Gruenwald.
\newblock Microarray gene expression data association rules mining based on
  {BSC}-tree and {FIS}-tree.
\newblock {\em Data $\&$ Knowledge Engineering}, 53(1):3 -- 29, 2005.

\bibitem{KemenyS_FMC1969}
J.~Kemeny and J.~Snell.
\newblock {\em Finite Markov chains}.
\newblock University series in undergraduate mathematics. VanNostrand, New
  York, repr edition, 1969.

\bibitem{KlopotekM_Studia2006}
M.~A. K{\l}opotek.
\newblock Cyclic {Bayesian} network--{Markov} process approach.
\newblock {\em Studia Informatica}, 1(2):7, 2006.

\bibitem{Kullback_book1968}
S.~Kullback.
\newblock {\em Information Theory and Statistics}.
\newblock Dover Publication, 2nd edition, 1968.

\bibitem{LiL_PAKDD2009}
G.~Li and T.-Y. Leong.
\newblock Active learning for causal {Bayesian} network structure with
  non-symmetrical entropy.
\newblock In {\em Proceedings of the 13th Pacific-Asia Conference on Advances
  in Knowledge Discovery and Data Mining}, PAKDD '09, pages 290--301, Berlin,
  Heidelberg, 2009. Springer-Verlag.

\bibitem{TsauYAC_Springer2008}
T.~Y. Lin, Y.~Xie, A.~Wasilewska, and C.-J. Liau, editors.
\newblock {\em Data Mining: Foundations and Practice}, volume 118 of {\em
  Studies in Computational Intelligence}.
\newblock Springer, 2008.

\bibitem{MacKay_LIGM99}
D.~J.~C. MacKay.
\newblock Introduction to {Monte} {Carlo} methods.
\newblock In {\em Learning in graphical models}, pages 175--204. MIT Press,
  Cambridge, MA, USA, 1999.

\bibitem{Mazlack_ACI2004}
L.~J. Mazlack.
\newblock Mining causality from imperfect data.
\newblock In {\em Proceedings of the sixth International FLINS Conference on
  Applied Computational Intelligence Proceedings}, Applied Computational
  Intelligence, pages 155--160, 2004.

\bibitem{MeganckLM_MDAI2006}
S.~Meganck, P.~Leray, and B.~Manderick.
\newblock Learning causal {Bayesian} networks from observations and
  experiments: a decision theoretic approach.
\newblock In {\em Proceedings of the Third international conference on Modeling
  Decisions for Artificial Intelligence}, MDAI'06, pages 58--69, Berlin,
  Heidelberg, 2006. Springer-Verlag.

\bibitem{MohammadN_SCI2010}
Y.~Mohammad and T.~Nishida.
\newblock Mining causal relationships in multidimensional time series.
\newblock In E.~Szczerbicki and N.~Nguyen, editors, {\em Smart Information and
  Knowledge Management}, volume 260 of {\em Studies in Computational
  Intelligence}, pages 309--338. Springer Berlin Heidelberg, 2010.

\bibitem{MoralDJ_SAC12}
P.~Moral, A.~Doucet, and A.~Jasra.
\newblock An adaptive sequential {Monte Carlo} method for approximate
  {Bayesian} computation.
\newblock {\em Statistics and Computing}, 22:1009--1020, 2012.

\bibitem{Pearl_B1995}
J.~Pearl.
\newblock Causal diagrams for empirical research.
\newblock {\em Biometrika}, 82:669--88, 1995.

\bibitem{Pearl_S1998}
J.~Pearl.
\newblock Graphs, causality, and structural equation models.
\newblock {\em Sociological Methods and Research}, 27:226--84, 1998.

\bibitem{Pearl_book2009}
J.~Pearl.
\newblock {\em Causality: Models, Reasoning and Inference}.
\newblock Cambridge University Press, 2nd edition, 2009.

\bibitem{MargolinH_2013Reasoning}
B.~M. Rottman and R.~Hastie.
\newblock Reasoning about causal relationships: Inferences on causal networks.
\newblock {\em Psychological Bulletin}, 140(1):109--139, January 2014.

\bibitem{RudinLSKM_COLT2011}
C.~Rudin, B.~Letham, A.~Salleb-Aouissi, E.~Kogan, and D.~Madigan.
\newblock Sequential event prediction with association rules.
\newblock In {\em Proceedings of the 24th Annual Conference on Learning
  Theory}, COLT '11, pages 615--634, 2011.

\bibitem{SchenkelCKMNP_2008}
R.~Schenkel, T.~Crecelius, M.~Kacimi, S.~Michel, T.~Neumann, J.~X. Parreira,
  and G.~Weikum.
\newblock Efficient top-k querying over social-tagging networks.
\newblock In {\em Proceedings of the 31st Annual International ACM SIGIR
  Conference on Research and Development in Information Retrieval}, SIGIR '08,
  pages 523--530, New York, NY, USA, 2008. ACM.

\bibitem{Shachter_UAI90}
R.~D. Shachter.
\newblock Evidence absorption and propagation through evidence reversals.
\newblock In {\em Proceedings of the Fifth Annual Conference on Uncertainty in
  Artificial Intelligence}, UAI '89, pages 173--190, Amsterdam, The
  Netherlands, 1990.

\bibitem{SilversteinBMU_DMKD2000}
C.~Silverstein, S.~Brin, R.~Motwani, and J.~Ullman.
\newblock Scalable techniques for mining causal structures.
\newblock {\em Data Mining and Knowledge Discovery}, 4(2-3):163--192, 2000.

\bibitem{SpiritesGS_book2000}
P.~Spirtes, C.~Glymour, and R.~Scheines.
\newblock {\em Causation, Prediction, and Search}.
\newblock MIT Press, 2nd edition, 2000.

\bibitem{SpirtesGS_PACSS90}
P.~Spirtes, C.~N. Glymour, and R.~Scheines.
\newblock Causality from probability.
\newblock In {\em Proceedings of the Conference on Advanced Computing for the
  Social Sciences}, ACSS, 1990.

\bibitem{TulupyevN_MICAI2005}
A.~L. Tulupyev and S.~I. Nikolenko.
\newblock Directed cycles in {Bayesian} belief networks: probabilistic
  semantics and consistency checking complexity.
\newblock In {\em Proceedings of the 4th Mexican international conference on
  Advances in Artificial Intelligence}, pages 214--223. Springer, 2005.

\bibitem{Blackout_2004d}
{US-Canada Power System Outage Task Force}.
\newblock {Final Report on the August 14, 2003 Blackout in the United States
  and Canada}.
\newblock Technical report, April 2004.

\bibitem{VaimanBCCDHPMZ_PS2012}
M.~Vaiman, K.~Bell, Y.~Chen, B.~Chowdhury, I.~Dobson, P.~Hines, M.~Papic,
  S.~Miller, and P.~Zhang.
\newblock Risk assessment of cascading outages: Methodologies and challenges.
\newblock {\em IEEE Transactions on Power Systems}, 27(2):631--641, 2012.

\bibitem{VelosoAGMM_LRQ2008}
A.~A. Veloso, H.~M. Almeida, M.~A. Gon\c{c}alves, and W.~Meira~Jr.
\newblock Learning to rank at query-time using association rules.
\newblock In {\em Proceedings of the 31st Annual International ACM SIGIR
  Conference on Research and Development in Information Retrieval}, SIGIR '08,
  pages 267--274, New York, NY, USA, 2008. ACM.

\bibitem{YoungSK_SIGN2011}
S.~S. Young and A.~Karr.
\newblock Deming, data and observational studies.
\newblock {\em Significance}, 8(3):116--120, 2011.

\bibitem{Zhang_ECI1996}
N.~L. Zhang and D.~Poole.
\newblock Exploiting causal independence in {Bayesian} network inference.
\newblock {\em J. Artif. Int. Res.}, 5(1):301--328, Dec. 1996.

\bibitem{ZhangP_94}
Zhang.N.L. and P.~D.
\newblock A simple approach to {Bayesian} network computations.
\newblock In {\em Proceedings of the Tenth Canadian Conference on Artificial
  Intelligence}, CCAA '94, pages 171--178, 1994.

\end{thebibliography}

\end{document}